\documentclass[useAMS,usenatbib]{mn2e}
\newcommand \kms{km s$^{-1}$}
\newcommand \zabs{z$_{\rm abs}$}

\newcommand \url{}

\usepackage{longtable}
\usepackage{graphicx}
\usepackage{lscape}
\usepackage{rotating}
\usepackage{amssymb}

\def \nodata{. . .}

\def \ndsphstar{280}
\def \Msolar{M$_{\odot}$}
\def \alphafe{[$\alpha$/Fe]}
\def \HI{H\textsc{i}}
\def \sion{\textsc{ii}}
\def \mnfe{[Mn/Fe]$_{\rm DC}$}
\def \nldla{341}
\def \nlstars{2818}

\begin{document}

\title[Chemistry of metal-rich DLAs II]{The chemistry of the most metal-rich damped Lyman $\alpha$ systems at $z\sim2$ II. Context with the Local Group}


\author[Berg et al.] {
\parbox[t]{\textwidth}{
Trystyn A. M. Berg$^1$,  Sara L. Ellison$^1$, J. Xavier Prochaska$^2$, Kim A. Venn$^1$, Miroslava Dessauges-Zavadsky$^3$}\\\\
$^1$ Department of Physics and Astronomy, University of Victoria, Victoria, British Columbia, V8P 1A1, Canada.\\
$^2$ Department of Astronomy and Astrophysics, University of California, Santa Cruz, Santa Cruz, CA, 95064, USA.\\
$^3$ Observatoire de Gen\`eve, 51 Ch. des Maillettes, 1290 Sauverny, Switzerland.}


\maketitle
\begin{abstract}
Using our sample of the most metal-rich damped Lyman $\alpha$ systems (DLAs) at \zabs{}$\sim2$, and two literature compilations of chemical abundances in \nldla{} DLAs and \nlstars{} stars, we present an analysis of the chemical composition of DLAs in the context of the Local Group. The metal-rich sample of DLAs at \zabs{}$\sim2$ probes metallicities as high as the Galactic disc and the most metal-rich dwarf spheroidals (dSphs), permitting an analysis of many elements typically observed in DLAs (Fe, Zn, Cr, Mn, Si, and S) in comparison to stellar abundances observed in the Galaxy and its satellites (in particular dSphs).  Our main conclusions are: (1) non-solar [Zn/Fe] abundances in metal-poor Galactic stars  and in dSphs over the full metallicity range probed by DLAs, suggest that Zn is not a simple proxy for Fe in DLAs and therefore not a suitable indicator of dust depletion. After correcting for dust depletion, the majority of DLAs have subsolar [Zn/Fe] similar to dSphs;  (2) at [Fe/H]$\sim-0.5$, a constant [Mn/Fe]$\sim-0.5$ and near-solar \alphafe{} (requiring an assumption about dust depletion) are in better agreement with dwarf galaxies than Galactic disc stars; (3)  [$\alpha$/Zn] is usually solar or subsolar in DLAs.  However, although low ratios of [$\alpha$/Fe] are usually considered more `dwarf-like' than `Milky Way-like', subsolar [Zn/Fe] in Local Group dwarfs leads to supersolar [$\alpha$/Zn] in the dSphs, in contrast with the DLAs.  Therefore, whilst DLAs exhibit some  similarities with the Local Group dwarf population, there are also notable differences.

\end{abstract}


\section{Introduction}

Quasar absorption line systems provide opportunities for observing the evolution of the universe; from the epoch of reionization to the structure of galaxies in the local universe. A commonly used probe to study the evolution of galaxies over this large period of time is the gas-phase properties of a class of quasar absorption lines called damped Lyman $\alpha$ systems \citep[DLAs;][]{Wolfe05}. Defined based on the strength of the Ly$\alpha$ absorption feature, DLA sightlines probe galactic gas \citep[][]{Wolfe95} with column densities of N(\HI{})$\geq 2\times{}10^{20}$ atoms cm$^{-2}$ \citep{Wolfe86}. With DLAs typically spanning a range of redshifts from $z\sim$0--5, they provide an excellent testbed for tracking galaxy evolution over a large portion of the history of the universe.

Much of the previous observational work on DLAs has focused on understanding the chemistry of the gas within galaxies independent of galaxy type \citep[e.g.][]{Pettini94,Lu98,Centurion00,Ledoux02,Wolfe03}.  As DLAs span a large range in redshift, they have been used for tracking the metal enrichment of the Universe  \citep{Pettini97,Pettini99,Prochaska03ApJ595,Rafelski12,Rafelski14}. However, using a more detailed analysis coupled with the unique origin and properties of each element, the study of the elements' abundance patterns in stars and gas can provide insight into the processes of galactic systems that lead to the observed metal enrichment \citep[such as the role of supernovae {[SNe]} and the star formation history; ][]{McWilliam03,Venn04,Tolstoy09}.  For metal-poor systems ([Fe/H]$\leq-2.5$), individual DLA abundance patterns have been compared to Galactic carbon-enhanced metal-poor stars to understand the origin of the first stars \citep[][]{Cooke11,Cooke14}. Large surveys of DLAs have also looked at many commonly observed elements in an attempt to differentiate the nucleosynthetic trends  from DLA properties (such as dust depletion) in the quasi-stellar object (QSO) sightlines \citep{Lu96,Prochaska02II,DLAcat30,DZavadsky06}. In addition, the physical nature of DLAs have profound effects on the gas-phase abundances and can be used to characterize galactic properties, such as the sources of ionization \citep{Dodorico07,Ellison10, Zafar14Ar} or the amount of dust \citep[e.g.][]{Pettini94,Kulkarni97,Akerman05,Ledoux02,Vladilo11}.

In concert with the compilation of abundances in high-redshift galaxies, considerable progress has been made in gathering large samples of stellar abundances, both in the Milky Way, and in other nearby galaxies. Within our Local Group, there have been many studies that have focused on different populations of stars, including the Galactic bulge, thin disc, thick disc, and halo, as well as satellite galaxies and globular clusters \citep[e.g.~high spectral resolution studies from][]{McWilliam03,Venn04, Bensby14,Hendricks14}. We have some understanding of the role of various astrophysical properties, such as star formation history and stellar populations, from the chemistry of these Galactic components \citep[][]{McWilliam97,Tolstoy09}.  With a detailed comparison between the chemistry of DLAs and stars, we can infer the processes of galaxy evolution back to when the universe emerged from reionization, and better understand how galaxies like our own Milky Way came into existence. However, such a comparison is complicated by the nature of DLA observations. Whereas stars within a galaxy (or population) can trace a large range of the chemical evolution of the system; each DLA only probes a sightline through a single galaxy, providing a snapshot of the galaxy of given morphology \citep[e.g.][]{Peroux11,Fumagalli15} at a particular epoch. 

In this paper, we provide a detailed analysis of a variety of elements observed in DLAs (Fe, Zn, Cr, Mn, S, and Si) and compare the nucleosynthetic patterns to what is seen in the stellar components of the Milky Way system. The last sizeable compilation of DLA abundances was presented over a decade ago by \cite{Prochaska01I} and {Prochaska02II} using 28 DLAs (studying O, Si, S, Al, Ar, Cr, Fe, Ni, Zn, and Co).  Moreover, although some works make comparisons between the general metallicity distribution of DLAs and stars \citep{Meyer90,Pettini00,Rafelski12}, or focus on a few elements \citep{Lu96,DZavadsky04,Nissen07}, comprehensive comparisons of chemical enrichment of DLAs with stars are rare. The past decade has seen incredible growth in abundance measurements for both DLAs and stars, such that a detailed comparison is timely. Such a comparison can provide insight into what environments the gas came from to form the stars we see in our own Galaxy.

\section{Samples}
The comparison of the chemical evolution of DLAs to Local Group environments requires selecting several samples with accurate metal abundances. For this work, we have compiled three samples: the most metal-rich DLAs at \zabs{}$>1.5$ (further called the MRDLA sample) from \citet[further referred to as Paper I]{Berg15}, and two literature compilations of metal abundances in both stars and DLAs.  The properties of these samples are described and compared below. All abundances (for both stars and DLAs) have been converted to the \cite{Asplund09} meteoritic solar scale, unless otherwise stated.

\subsection{MRDLA Sample}

The MRDLA sample from Paper I contains 44 DLAs (of which 31 were newly observed) that were selected as candidate metal-strong DLAs\footnote{A metal-strong DLA is defined by \cite{HerbertFort06} as having large metal column densities; log$N$(Zn\sion{}) $\geq13.15$ or log$N$(Si\sion{})$\geq 15.95$.} from the \cite{HerbertFort06} catalogue (see Paper I for more details). These systems were chosen in order to study the most chemically evolved DLAs at \zabs$\sim2$, and look for exotic elements \citep[such as boron; see][]{Berg13}. These 44 DLAs were observed using the High Resolution Echelle Spectrometer \citep[HIRES;][]{Vogt94} on the Keck I telescope on Mauna Kea,  for a total of 83.8 h on our 31 new targets, with a minimum of 1 h on each sightline. The signal-noise ratio for the spectra span a range of 2--52 pixel$^{-1}$; with the typical spectrum having an SNR of $\sim10$ pixel$^{-1}$ (see Table 1 in Paper I). The MRDLA sample spans a large range in N(\HI{}) ($20.3\leq$log$N$(\HI{})$\leq22$) and is predominately within the redshift range $2\lesssim z_{\rm abs} \lesssim 3$. The bulk of the MRDLAs have a metallicity between $-1.5\leq {\rm [M/H]}\leq0.5$\footnote{Where the metallicity tracer M is selected following the method outlined by \cite{Rafelski12}. In brief, the metallicity indicator is chosen (in order of decreasing preference) from S, Si, Zn, and Fe. If Fe is adopted, a $+0.3$ dex correction is included to account for the apparent discrepancy between Fe and S, Si, and Zn.}, and only $\sim45\%$ are true  metal-strong DLAs as defined by \cite{HerbertFort06}. However this entire sample probes the upper 50$\%$ of DLA metallicties (see upper left panel of Figure \ref{fig:Festars} in Section \ref{sec:Mdist}); and therefore covers the metallicity range of the most metal-rich DLAs. For a summary of the column densities and the metallicities of this MRDLA sample, see Tables 35 and 36 in Paper I.

In addition to the metal abundances published in Paper I, we have measured column densities and $3\sigma$ upper limits for Ti\textsc{ii} in the MRDLA sample (see Table \ref{tab:DLAlit}). The column densities for Ti\textsc{ii} detections were determined with the apparent optical depth method \citep{Savage91}, using the same velocity limits as in Paper I. Upper limits were calculated using the SNR of the spectrum at the given line and the full width half-maximum of the strongest absorption feature in a detected metal line. The atomic data for Ti \textsc{ii} 1910 line were obtained from \cite{Morton03}.

\subsection{DLA Literature Sample}
\label{sec:DLAlit}
The literature on DLA chemistry spans nearly four decades of work, using many different telescopes and spectrographs. The first surveys searching for DLAs \citep{Wolfe86, Sargent89,Lanzetta91} used low resolution spectrographs to identify quasars with DLAs and measure the \HI{} column densities to study the evolution of \HI{}. The first systematic studies of metals were done with the Hale, William Herschel, and the Anglo-Australian telescopes \citep[][respectively]{Pettini90,Pettini94,Pettini97} focusing on weak lines that were unlikely to be saturated.  Observations with high-resolution spectrographs on 4-m class telescopes permitted a more detailed view of the kinematic structure in DLAs and opened the door to studying a wider range of elements \citep[e.g.][]{Carswell87,Bergeron91,Savaglio94,Roth95,Pettini95,Meyer95}. With the advent of HIRES on the Keck I telescope and Ultraviolet and Visual Echelle Spectrograph (UVES) on the Very Large Telescope (VLT), higher resolution observations could resolve the Ly$\alpha$ forest and metal lines clearly and thus provide more accurate abundances \citep[e.g.][]{Lu96,Pettini99,Centurion00,Molaro01,Prochaska02II,DZavadsky04,Akerman05,	DZavadsky06,Ledoux06,DLAcat70,Ellison10}. Follow-up observations of the initial DLA catalogues and the Sloan Digital Sky Survey (SDSS), in addition to targeting fainter background QSOs, has led to enormous databases of both \HI{} column densities and metal abundances in DLAs \citep[e.g.][]{Prochaska01I,Prochaska03ApJS147,Penprase10, Noterdaeme12}.

We have compiled a catalogue of metal column densities for all DLAs published between 1994 and 2014 which have had high-resolution (R$>10000$, but typically R$\sim40000$; the error in column densities are approximately $\pm0.1$ dex) observations completed. The high-resolution requirement selects data for which most velocity profile components are resolved to check for blending and saturation. With the necessity of using 8--10m class telescopes to obtain quality data for typical QSO magnitudes within a reasonable amount of time, a significant portion of the catalogue is limited to Keck/Echellete Spectrograph and Imager (ESI), Keck/HIRES, VLT/UVES,  or VLT/XSHOOTER. The catalogue contains \nldla{} DLAs, with column densities for a variety of elements (O, Mg, Al,  Si, S, Ca, Ti, Fe, Zn, Mn, Cr, Co, and Ni) and ionization states (e.g. Mg \textsc{i}, Mg \sion{}, Al \sion{}, Al \textsc{iii}); making this compilation the largest catalogue of DLA metal abundances currently available.  Table \ref{tab:DLAlit} contains our entire DLA literature compilation, including the MRDLA sample data for ease of access.

The properties of the DLA literature sample are described in Appendix \ref{sec:AppDLALit}. Overall, the high-resolution literature DLA sample spans a large range in redshift ($0 \lesssim z \lesssim 5$) and follows the \HI{} distribution seen for the large \HI{}-only surveys \citep[e.g.][]{Noterdaeme12}. With the large range in metallicity ($-3$ to $0.5$), the DLA literature sample provides a sufficient breadth to trace the chemical evolution of galactic gas over the metallicity regimes of the Local Group, and is not significantly biased in metallicity or redshift compared to other carefully chosen DLA samples such as \cite{Rafelski12}. 

\subsection{Stellar Literature Sample}

To compare DLA chemistry to the abundance patterns seen in the Local Group, we have compiled a catalogue of stellar abundances for \nlstars{} stars from the stellar halo, the disc, and a selection of satellite galaxies of the Milky Way (Large Magellanic Cloud (LMC), Fornax, Sagittarius, Carina, and Sculptor). Stellar abundances are useful as they offer insight into the chemistry of the interstellar medium (ISM) at the epoch of the stars' formation. Our literature compilation of stellar abundances is drawn from a variety of papers that use high-resolution observations of stars to obtain accurate abundances (R$>10000$, but typically R$\sim40000$; the error in [Fe/H] is approximately $\pm0.1$ dex). These include previous literature compilations \citep{Venn04,Frebel10, North12} as well as individual studies \citep{Reddy03,Shetrone03,Bensby05,Geisler05, Reddy06,Sbordone07,Carretta10,Letarte10,Tafelmeyer10,Venn12,Starkenburg13, Bensby14,Hendricks14,Skuladottir15}. Table \ref{tab:samplitstars} presents each paper with the number of stars from each Galactic component (halo, disc, and satellite galaxies), and the elements with the abundances used in the rest of this paper. Elements which include hyperfine structure corrections\footnote{Hyperfine structure corrections are required to correct the abundances of odd atomic number elements (such as Mn, Co, or Cu) from atomic line splitting, broadening the absorption line. These corrections are of the order $\sim0.1$ dex \citep{Prochaska00Mn} in stars.} in their abundances are flagged. A detailed explanation of the stellar literature sample is provided in Appendix \ref{sec:AppStelLit}, with the full literature compilation given in Table \ref{tab:stellar}.

\begin{table*}
\begin{center}
\caption{Summary of stellar literature sample}
\label{tab:samplitstars}
\begin{tabular}{lccc}
\hline
Paper & Galactic population & $N_{\rm stars}$ & Elements used \\
\hline
\cite{Reddy03} 		& Thin disc 	& 148 & Mg, Si, S, Cr, Mn$^{\rm H}$, Fe, Zn\\
\cite{Venn04} 		& Thin disc 	& 328$^{\star}$ & Mg, Fe(, Si, Cr)$^{\rm S}$\\
		--			& Thick disc 	& 165$^{\star}$ & Mg, Fe(, Si, Cr)$^{\rm S}$\\
		--			& Halo 			& 174$^{\star}$ & Mg, Fe(, Si, Cr)$^{\rm S}$\\
\cite{Bensby05}		& Thin disc		& 58$^{\star}$ & Mg, Si, Cr, Fe, Zn\\
		--			& Thick disc 	& 40$^{\star}$ & Mg, Si, Cr, Fe, Zn\\
\cite{Reddy06} 		& Thick disc 	& 94 & Mg, Si, S, Cr, Mn$^{\rm H}$, Fe, Zn\\
\cite{Frebel10} 	& Halo 			& 865$^{\star}$ & Mg, Si, Cr, Mn, Fe, Zn\\
\cite{Bensby14}		& Thin disc		& 427$^{\star}$ & Mg, Si, Cr, Fe, Zn\\
		--			& Thick disc 	& 249$^{\star}$ & Mg, Si, Cr, Fe, Zn\\
		--			& Halo 			& 38$^{\star}$ & Mg, Si, Cr, Fe, Zn\\
\cite{Shetrone03} 	& Satellite (Carina, Sculptor)	& 13 & Mg, Si, Cr, Mn$^{\rm H}$, Fe, Zn\\
\cite{Geisler05} 	& Satellite	(Sculptor)	& 4 & Mg, Si, Mn$^{\rm H}$, Fe, Zn\\
\cite{Sbordone07} 	& Satellite	(Sagittarius)	& 12 & Mg, Si, Cr, Mn$^{\rm H}$, Fe, Zn\\
\cite{Pompeia08} 	& Satellite	(LMC)	& 59 & Mg, Si, Cr, Fe\\
\cite{Carretta10} 	& Satellite	(Sagittarius)	& 27 & Mg, Si, Cr, Mn$^{\rm H}$, Fe\\
\cite{Letarte10} 	& Satellite (Fornax)		& 81 & Mg, Si, Cr, Fe, Zn\\
\cite{Tafelmeyer10} & Satellite (Fornax)		& 5 & Mg, Si, Cr, Mn, Fe\\
\cite{North12} 		& Satellite (Carina, Fornax, Sculptor)		& 172 & Mn$^{\rm H}$, Fe\\
\cite{Venn12} 		& Satellite (Carina)		& 9 & Mg, S, Cr, Mn$^{\rm H}$, Fe, Zn\\
\cite{Starkenburg13} & Satellite (Sculptor) 	& 7 & Mg, Cr, Fe\\
\cite{Hendricks14} & Satellite (Fornax) & 190 & Mg, Si, Fe\\
\cite{Skuladottir15} & Satellite (Sculptor) & 85 & Mg, S$^N$, Fe\\

\hline
\end{tabular}
  \\
Notes. $^{\rm H}$ -- Hyperfine structure corrections included.\\
$^{\rm N}$ -- Non-local thermodynamic equilibrium corrections are not included for consistency, but are available in \cite{Skuladottir15}.\\
$^{\star}$ -- Large literature compilation, likely containing a variety of different analyses.\\
$^{\rm S}$ -- Supplemented with abundances from original work \citep{Edvardsson93,Fulbright00,Stephens02}.\\
\end{center}
\end{table*}

\subsection{Metallicity distribution comparisons}
\label{sec:Mdist}
\begin{figure*}
\begin{center}

\includegraphics[width=\textwidth]{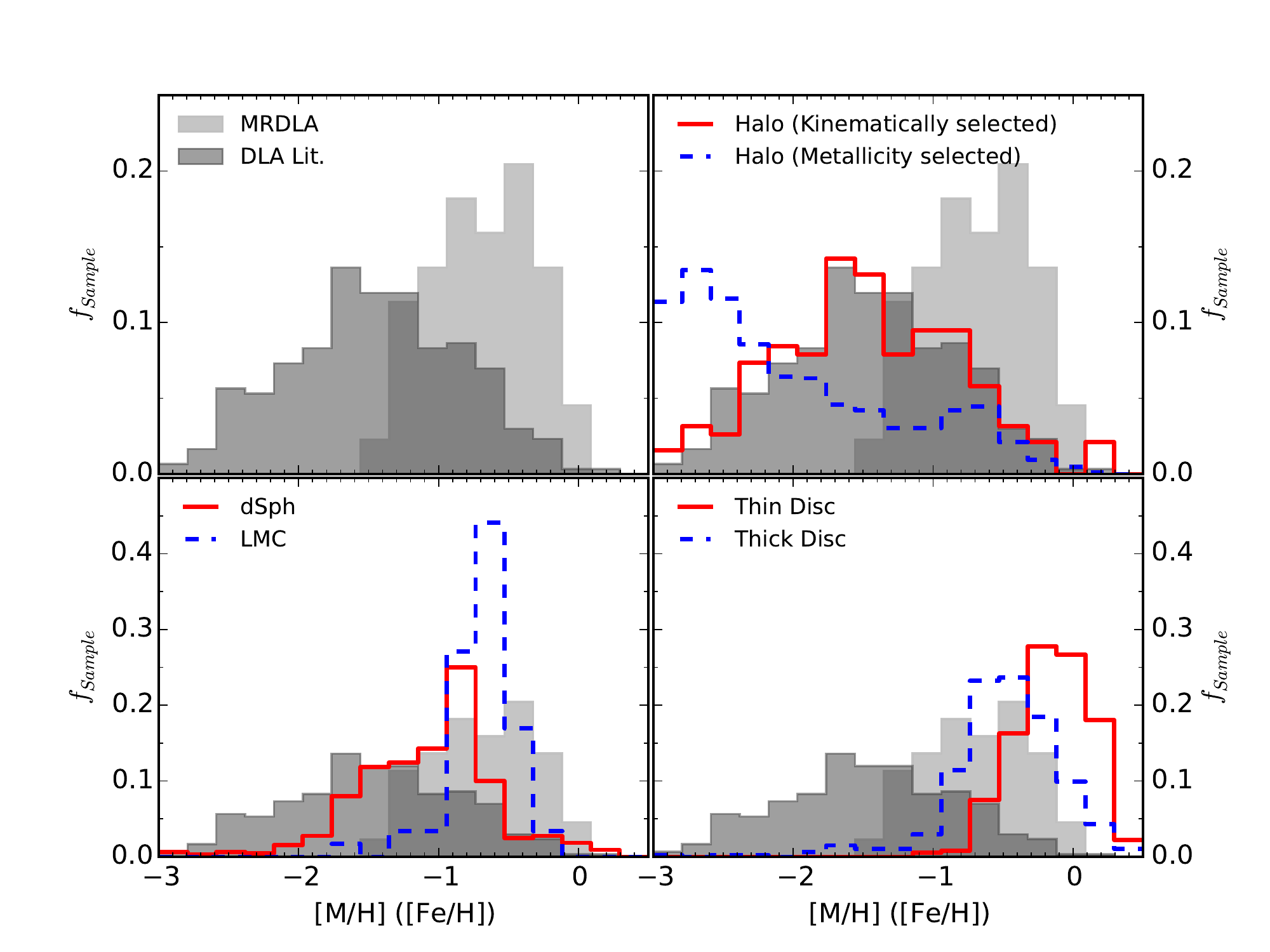}
\caption{Metallicity distributions for the samples of stars ([Fe/H]) and DLAs ([M/H]). Upper left: distributions for the MRDLAs  (light grey) and literature DLAs (dark grey). These are the same distributions shown in every panel. Upper right: stellar halo subsamples. The distributions for the kinematically selected \citep{Venn04,Bensby14} and metallicity selected \citep{Frebel10} halo stars are shown as the solid red and  blue dashed lines (respectively). Lower left: LMC (blue dashed line) and dSphs (solid red line) subsamples. Lower right: thin (solid red line) and thick (dashed blue line) disc stellar samples.}
\label{fig:Festars}
\end{center}
\end{figure*}

Figure \ref{fig:Festars} compares the metallicity distribution for each stellar subsample (using Fe as the metallicity tracer in stars)  to our MRDLA and literature DLA samples. As the DLA and stellar metallicity distributions represent different observed quantities (DLAs provide metallicities weighted by the cross-section of a galaxy, whereas stars represent a volume-limited sample), the comparison of their metallicity distributions \emph{does not} provide any information on the types of galaxies probed. However, a comparison of metallicity distributions will provide an approximate idea of what metallicity regimes DLAs typically span in context of the chemical environments of the Local Group samples selected. The metallicity distributions for the MRDLAs and literature DLA samples are shown in all panels as the solid light and dark histograms (respectively). We remind the reader that the metallicities derived following \cite{Rafelski12} scheme are based on $\alpha$-elements, rather than an Fe-peak element. However, this should only shift the DLA distributions by at most $-0.3$ dex relative to the stellar distributions.

Due to the nature of selecting halo stars by their kinematics or metallicity (see Appendix \ref{sec:AppHalo} for more details), we show the metallicity distributions from using a kinematically selected sample \citep{Venn04,Bensby14} or a metallicity selected sample \citep{Frebel10} separately in the top right panel of Figure \ref{fig:Festars}. From selecting halo stars in different ways, both stellar halo samples are biased and may not reflect the true nature of the Galactic halo stars. The metallicity-selected sample is generally too metal-poor to match the abundances of the DLAs. However the local, kinematically selected halo stars span the entire metallicity range of DLAs in the literature \citep[$-3\lesssim${[M/H]}$\lesssim-0.5$ in the upper right panel of Figure \ref{fig:Festars}; as previously seen in][]{Pettini97,Rafelski12}. However halo stars do not probe the high-metallicity regime present in the MRDLA sample (in fact, the MRDLA sample only probes the upper half of the kinematically selected halo metallicity distribution). However, the higher metallicity literature DLAs and MRDLA sample is probed by the metal-rich satellites (lower left panel) and Galactic disc stars (lower right panel).

The vast range in metallicity of the combined literature DLA and MRDLA samples spans the various chemical environments selected in the Local Group.  However, we cannot distinguish how DLAs evolve chemically in the context of these various local environments using metallicity distributions. We must look at the differences of individual elements to attempt to understand the chemical evolution of galactic gas at high redshifts.

\section{Element Comparison}
In order to identify similarities between the chemical enrichment of local stellar populations and DLAs (both MRDLAs, as well as other DLAs in the literature), this section compares the trends of various metals (Zn, Cr, Mn, S, and Si) in both the stellar and DLA samples\footnote{Both DLA samples contain systems close to their host quasar. These proximate DLAs have been shown to have slight differences in their abundances \citep{Ellison10} due to ionization differences. Systems within 3000 \kms{} of their host quasar are included in the analysis, but are flagged in the Figures in this section as points with black outlines.} described above. Such a comparison is complex, due to the complications of dust depletion in DLAs and atmospheric modelling of stars\footnote{
In the Figures presented in this section, non-local thermodynamic equilibrium (non-LTE) corrections are represented by magenta arrows. The arrows point in the direction in which the abundances should be adjusted, while its length represents the typical size of the correction. If multiple arrows are present, the size of the correction is metallicity dependent.}. To aid in the analysis, a summary of the origin of each element, the caveats in measuring their abundances in both stars and DLAs, and a brief summary of previous work is provided in Appendix \ref{sec:AppElem}.  The following analysis is two-fold: to better understand the most metal-rich DLAs in terms of their chemistry and amount of dust depletion, and to understand which stellar population is most similar to the typical DLA.

\subsection{Zn, Fe, and Cr}
\label{sec:ZnFeCr}

Many of the Fe-peak elements are heavily depleted onto dust in DLAs, and do not yield an accurate measurement of the Fe-peak abundance \citep[e.g.][see Appendices \ref{sec:AppFe}--\ref{sec:AppCr}]{Pettini94,Vladilo02a}. However, Zn is relatively undepleted in DLAs \citep[][]{Pettini97,Vladilo02a}, and roughly traces Fe in disc stars as seen in \cite{Sneden88} and \cite{Nissen07}. As a result, Zn has been adopted as a tracer for the iron peak by the DLA community.

\begin{figure*}
\begin{center}
\includegraphics[width=\textwidth]{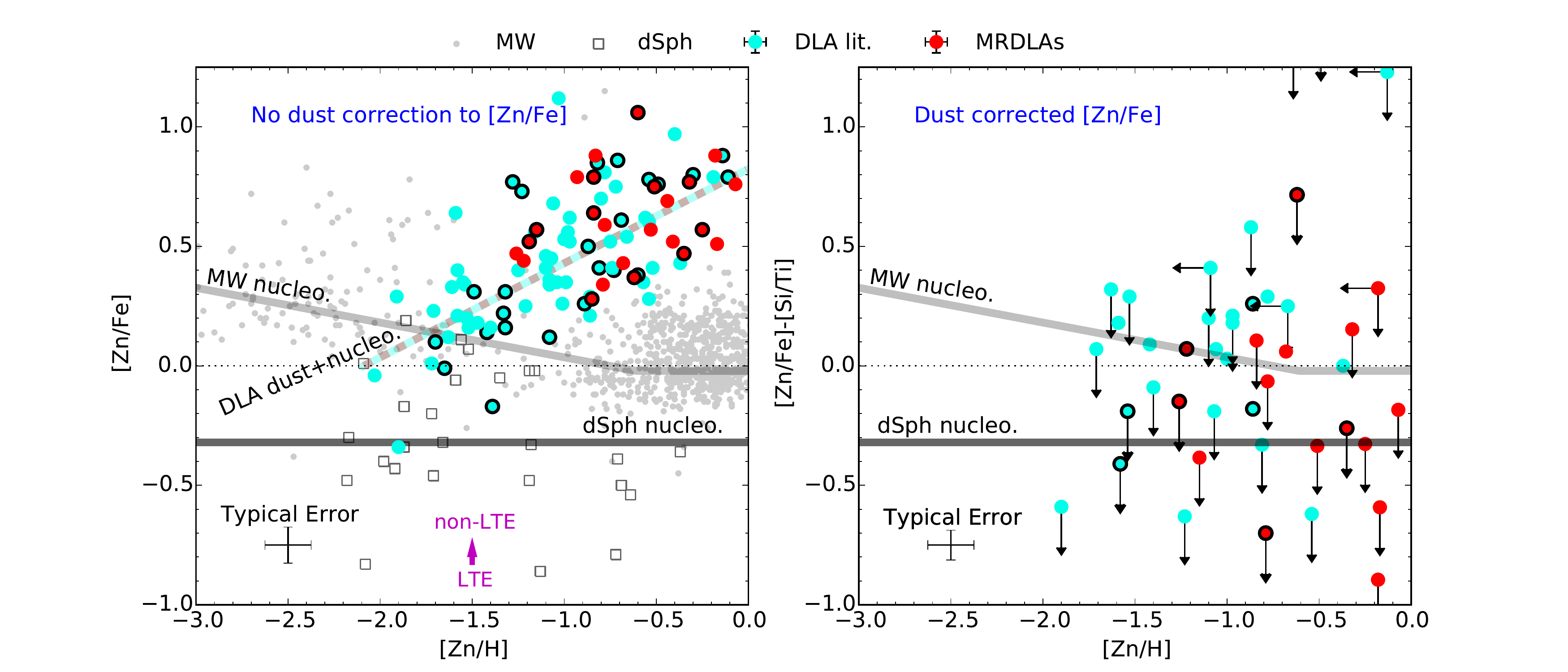}
\caption[{[Zn/Fe]} as a function of metallicity in stars and DLAs]{Left-hand panel:  [Zn/Fe] as a function of [Zn/H] in the DLA samples (circles; DLA literature sample in blue, MRDLAs in red) and stellar literature (unfilled dark grey squares for satellites and filled light grey circles for Galactic data). Only detections are plotted from all samples. The typical errorbar in DLAs is shown, while the typical error in stellar data is reflected by the scatter. DLAs within 3000 \kms{} of the host quasar are outlined by a black circle to flag systems where ionization corrections may be required. The magenta arrow shows the typical magnitude (0.1 dex) and direction of the stellar non-LTE corrections required for Zn in stars at all metallicities. The increase in [Zn/Fe] towards lower metallicities in Milky Way stars is attributed to a combination of non-LTE effects and nucleosynthetic origins in the first stars. Cartoon lines show the observed nucleosynthetic pattern in the Milky Way \citep[light grey;][]{Nissen07}, dSphs (dark grey), and DLAs (red-blue dashed line; also includes dust depletion). The relative difference between the DLA and stellar cartoon lines should provide a coarse estimate of the dust depletion. Right-hand panel: dust-corrected [Zn/Fe] in DLAs as a function of metallicity. The dust correction assumes that [Si/Ti] solely measures dust depletion, and is equivalent to [Zn/Fe] depletion to within $-0.1$ dex. The symbols and cartoon lines are the same as the left-hand panel, although upper limits are included.}
\label{fig:genZn}
\end{center}
\end{figure*}

Using Zn as an Fe-peak tracer in DLAs  is purely empirical. The assumption of Zn tracing Fe over all DLA metallicities has been previously called into question by several authors as some stellar populations manifest non-solar Zn/Fe abundances with variations of up to $\pm0.5$ dex. Such examples include: supersolar [Zn/Fe] in metal-poor stars \citep[][]{Nissen04,Nissen07}, Zn behaving similar to an $\alpha$-element \citep{Nissen11, Rafelski12}, and subsolar [Zn/Fe] in the bulge \citep{Barbuy15}. These observations have emphasized that the nucleosynthetic origin of Zn has multiple formation sites (see Appendix \ref{sec:AppZn} for more details). 

Figure \ref{fig:genZn} shows the trend of [Zn/Fe] for both stars and DLAs. Solely focusing on the stellar points in the left-hand panel, in Galactic stars above [Zn/H]$>-1.5$, [Zn/Fe] is indeed roughly solar, with a large scatter \citep[light grey circles;][]{Takeda05,Nissen07}. However, at lower metallicities, [Zn/Fe] is supersolar in most Galactic stars, which suggests that the dominant nucleosynthetic site of Zn production at low metallicities is different than at high metallicities \citep[also noted by][see Appendix \ref{sec:AppZn}]{Umeda02,Nissen07}. Thus, Zn is only a good tracer of Fe in the Milky Way at [Zn/H]$>-1.5$.

[Zn/Fe] abundances in the Galactic satellites are in stark contrast with the Milky Way ratios. [Zn/Fe] tends to be subsolar in dSphs  (dark grey squares) for all metallicities, with a large scatter in [Zn/Fe] for each individual galaxy. At low metallicities, subsolar [Zn/Fe] in dSphs has been attributed to a lack of retention of hypernovae ejecta with supersolar [Zn/Fe] \citep{Shetrone03,Sbordone07,Venn12}.

The DLAs in the left-hand panel of Figure \ref{fig:genZn} show increasing [Zn/Fe] with metallicity (from both the MRDLAs and literature samples; large red and blue circles, respectively). It has been a subject of debate whether the origin of increasing [Zn/Fe] is completely from dust depletion or also includes a nucleosynthetic component  \citep[e.g.][]{Lu96,Prochaska02II, Ledoux02}. Traditionally, the amount of Fe depletion has been assumed as the difference between the DLA and solar [Zn/Fe] typically seen in the Milky Way for [Fe/H]$>-1.5$ (light grey line in Figure \ref{fig:genZn}\footnote{The light grey line results from a fit of the data from \cite{Gratton03}, \cite{Cayrel04}, \cite{Nissen07}, and the Milky Way stars in the stellar sample. The data were binned by metallicity to remove any bias to the larger number of disc stars in comparison to the metal-poor stars. The best-fitting two-component model is [Zn/Fe]=$-0.15$[Fe/H] $-$ $0.11$ until [Zn/H]$=-0.61$, where [Zn/Fe] remains constant at [Zn/Fe]$\sim0$.}). The dSphs data (with the average dSph [Zn/Fe] given by the dark grey line) demonstrate that \emph{we cannot simply assume that Zn traces Fe in DLAs even at [Zn/H]$>-1.5$}. Moreover, the subsolar [Zn/Fe] ratios in dSphs indicate that even DLAs with [Zn/Fe]$\sim0$ may still be suffering from significantly depleted Fe.

The observation that, at [Zn/H]$<-1.5$, [Zn/Fe] DLAs are approaching the solar ratio has led to claims that dust depletion is minimal in low-metallicity DLAs \citep[][]{Pettini97,Akerman05}.  However, it is important to note that the Milky Way itself has slightly supersolar [Zn/Fe] at these metallicities.  Nonetheless, since the majority of the DLAs are consistent with the scatter of the Galactic data ([Zn/Fe]$=0.2\pm0.2$), based on Milky Way data alone it would seem plausible that dust does not play a significant role at [Zn/H]$<-1.5$.  However, the subsolar [Zn/Fe] abundances in dSphs calls this interpretation into question.  If the true [Zn/Fe] ratios in DLAs are more similar to dSphs than the Milky Way, then observed ratios of [Zn/Fe]$\sim0$ may still be consistent with significant (up to 1 dex) depletion. There are indeed two DLAs with subsolar [Zn/Fe] at low metallicities seen in Figure \ref{fig:genZn}, with another DLA in the literature sample at [Zn/H]$<-1.6$ with  [Zn/Fe]$<-0.4$. These metal-poor DLAs challenge the current understanding of dust depletion at low metallicities. We also note that MRDLAs and other high metallicity DLAs have depletions nearly twice as large ([Fe/Zn]$\gtrsim 0.6$ dex at [Zn/H]$\sim-0.5$) relative to the typical DLA ($\gtrsim 0.2$ dex at [Zn/H]$\sim-1.5$).

To estimate the contribution from dust depletion on [Zn/Fe], another pair of elements are needed which form in similar nucleosynthetic sites but have a difference in dust depletion. Si and Ti are both $\alpha$-elements that trace eachother \citep{Pritzl05, McWilliam13}, but Ti is more refractory than Si in the Milky Way \citep{Savage96}. By removing the contribution of dust in [Zn/Fe],  [Zn/Fe]$-$[Si/Ti] should only trace the nucleosynthetic differences between [Zn/Fe]. The precision of this correction is limited by the difference in relative dust depletion differences of [Zn/Fe] to [Si/Ti]. Fortunately, the dust depletion of [Zn/Fe] is at most 0.1 dex smaller than [Si/Ti] in Milky Way sightlines \citep[e.g. see][]{Savage96}, therefore  [Zn/Fe]$-$[Si/Ti] should be an accurate tracer of [Zn/Fe] nucleosynthesis in DLA sightlines. The size of these dust corrections is consistent with recent observations of the depletion of Zn in the ISM of the LMC and Small Magellanic Cloud \citep{Tchernyshyov15}, where  Zn is depleted in these satellites by up to 0.8 dex. The right-hand panel of Figure \ref{fig:genZn} shows [Zn/Fe] in DLAs after correcting for dust depletion (i.e. [Zn/Fe]$-$[Si/Ti]), where the $3\sigma$ limits are driven by Ti non-detections. Our dust depletion correction demonstrates that DLAs can be either consistent with subsolar [Zn/Fe] seen in dSphs or can exhibit [Zn/Fe]$\sim0$ as in the Milky Way. However, the large spread in dSphs [Zn/Fe] ($-0.9\lesssim$[Zn/Fe]$\lesssim0.2$) implies that all DLAs might be consistent with dSphs. Furthermore, this result indicates that [Zn/Fe] is not a perfect indicator of dust depletion, and that [Zn/Fe]$\sim0$ in DLAs does not mean the absorber is `dust-free'.

\begin{figure}
\begin{center}
\includegraphics[width=0.5\textwidth]{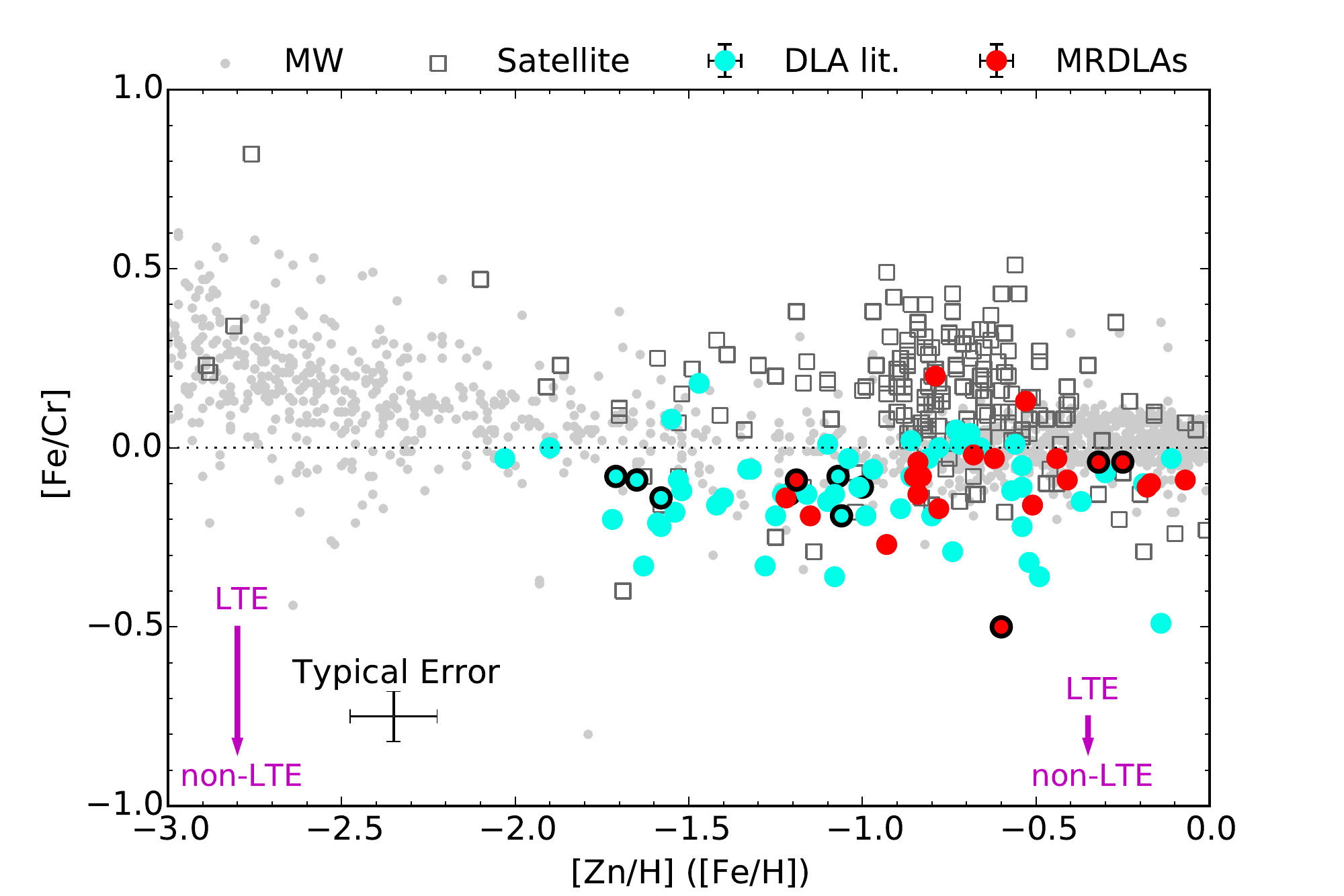}
\caption[{[Fe/Cr]} as a function of metallicity in stars and DLAs.]{[Fe/Cr] as a function of metallicity ([Zn/H] in DLAs; [Fe/H] in stars). The symbols are the same as Figure \ref{fig:genZn}. Non-LTE corrections as a function of metallicity for the stellar data are shown as the magenta arrows. [Fe/Cr]  is constant with metallicity in DLAs, suggesting there is no significant difference in the relative depletion between the two species.}
\label{fig:genCr}
\end{center}
\end{figure}

Like Fe, [Zn/Cr] has  also been used as a dust indicator in DLAs \citep[e.g.][see Appendix \ref{sec:AppCr} for more details]{Pettini97}. As we have argued above for [Zn/Fe], solar [Zn/Cr] is also not necessarily an indicator of zero depletion.  Nonetheless, [Fe/Cr] may reveal further insights into dust depletion. As Fe has a slightly higher condensation temperature than Cr \citep{Savage96}, [Fe/Cr] in DLAs is slightly subsolar \citep{Lu96,Prochaska02II}. In addition, one might expect to see a decrease in [Fe/Cr] with increasing metallicity if dust depletion is strongly metallicity dependent. To assess the amount of dust depletion of Fe relative to Cr in DLAs (the MRDLA sample in particular), Figure \ref{fig:genCr} shows the trend of [Fe/Cr] as a function of metallicity. In stars, [Fe/Cr] remains solar for all metallicities once non-LTE corrections have been included\footnote{[Fe/Cr] is overestimated by at most $+0.35$ dex at the low metallicities and $+0.1$ at high metallicities \citep{Bergemann10}.  The magenta arrows on Figure \ref{fig:genCr} demonstrate the approximate non-LTE correction at low and high metallicities. The surprising discrepancy between the bulk of the dSph [Cr/Fe] data and other dSphs and the Galaxy is likely a result of systematic discrepancies for deriving Cr abundances in Fornax \citep[see discussion in][]{Letarte10}.}, suggesting that any deviations from the solar value in DLAs will result from dust depletion. However, [Fe/Cr] is roughly constant in all DLAs, suggesting that any difference in the differential depletion of Fe and Cr onto dust is too subtle to observe.

In summary, based on recent measurements of Zn in both the stars and gas in Local Group dwarfs, we caution against the standard assumption that [Zn/Fe]$\sim0$ is  indicative of a dust-free sightline. Moreover, Cr does not provide any additional information on dust depletion that can not be obtained from Fe. Dust-corrected [Zn/Fe] show that DLAs show consistent [Zn/Fe] with both Milky Way and dSph nucleosynthesis.

\subsection{Si and S}
\label{sec:SSi}

\alphafe{} is a commonly used diagnostic to probe the nucleosynthetic differences between satellite and Milky Way stars. S and Si are the most commonly measured $\alpha$-elements in DLAs. Both Si and S are formed during O burning in stars, tracing each other over all metallicities \citep{Chen02}.  Although S has the advantage of being relatively undepleted onto dust, it is less commonly measured in DLAs as the lines of interest are often located within the Ly$\alpha$ forest. Si is easily measured in DLAs \citep{Prochaska02II}, yet is slightly depleted onto dust \citep{DZavadsky06,Vladilo11}. However stellar abundances of Si and S suffer from systematic errors (see Appendices \ref{sec:AppS} and \ref{sec:AppSi}), making a comparison to DLA abundances challenging. 

\begin{figure}
\begin{center}
\includegraphics[width=0.5\textwidth]{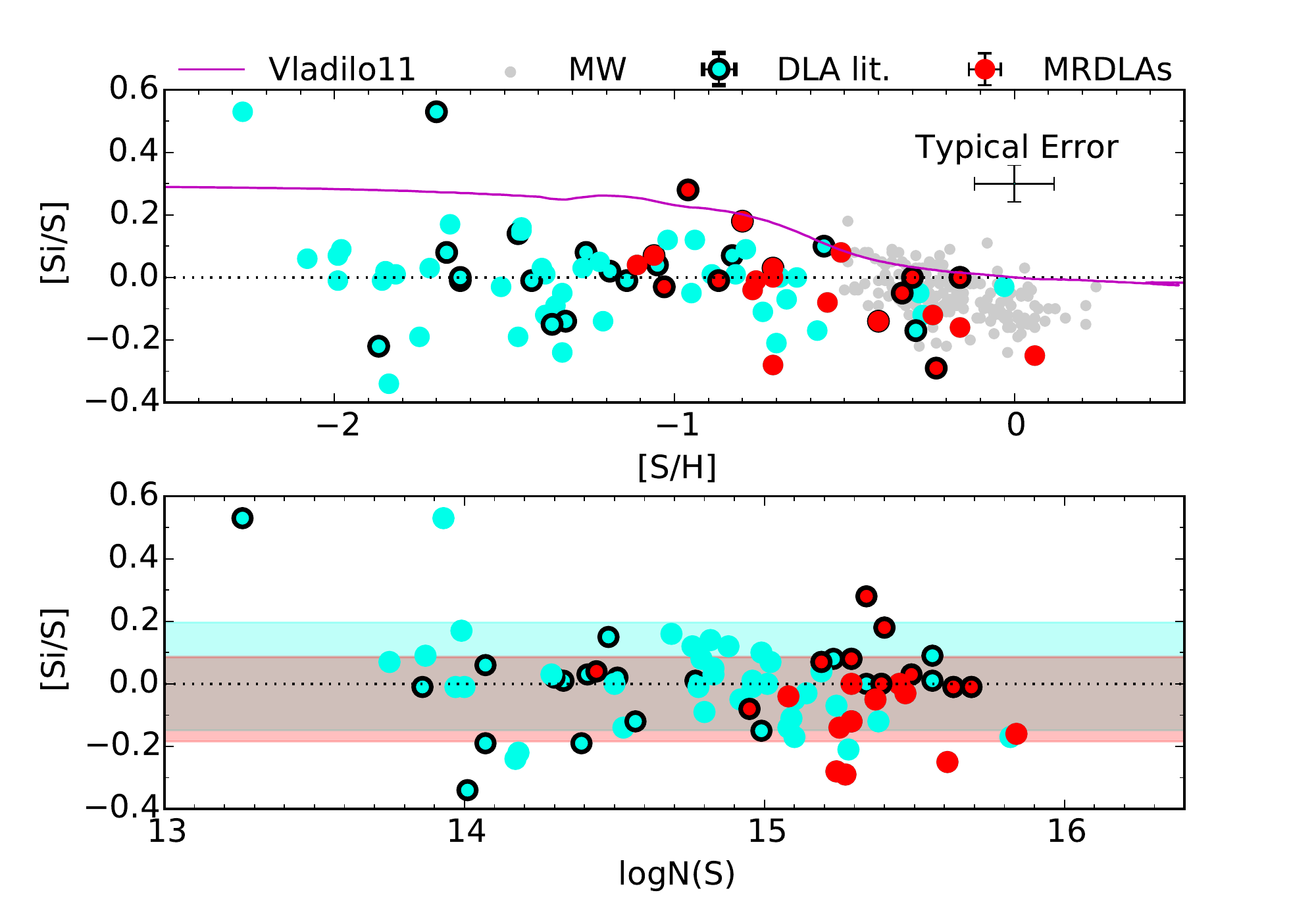}
\caption{[Si/S] as a function of metallicity ([S/H]; top panel) and metal column density (log$N$(S); bottom panel). The symbols are the same as in Figure \ref{fig:genZn}. The magenta line shows the undepleted [Si/S] prediction from \citet{Vladilo11} based on chemical evolution models of dwarf galaxies. [Si/S] remains consistent with the stellar data over all metallicities and column densities. The average values of [Si/S] and 1$\sigma$  scatter are shown by the blue and red shaded regions in the bottom panel.}
\label{fig:genSiS}
\end{center}
\end{figure}

To understand how well Si and S trace each other in DLAs, Figure \ref{fig:genSiS} shows the ratio [Si/S] as a function of metallicity (top panel) and log$N$(S) (bottom panel). In stars, [Si/S] indeed  appears to be approximately solar (with an average value of [Si/S]$\sim-0.05$), and is constant over the range of metallicities in the Galactic disc ($-0.5\lesssim$[S/H]$\lesssim0.2$). [Si/S] in DLAs is consistent with the scatter in the stars. 
 
If there is significant dust depletion of Si, we might expect either panel of Figure \ref{fig:genSiS} to show a decrease in [Si/S] with increasing metallicity or column density. However, both panels in Figure \ref{fig:genSiS} show a consistently solar [Si/S] in DLAs, with a mean value of $\langle{\rm[Si/S]_{MRDLA}}\rangle=-0.05\pm0.13$ and $\langle{\rm[Si/S]_{DLA}}\rangle=0.02\pm0.17$ for the MRDLAs and DLAs\footnote{ The two DLAs with [Si/S]$\sim0.5$ are outliers from the typical distribution. One system is close to the background quasar and likely suffers from ionization effects \citep{Ellison10}. The data for the other DLA show no sign of contamination, and appear to be an anomalously high [Si/S] system. Both DLAs are included in determining $\langle{\rm[Si/S]_{DLA}}\rangle$.} (respectively; shown as the shaded regions in the bottom panel of Figure \ref{fig:genSiS}).

The amount of Si depletion is much lower than that found by \cite{Vladilo11} who claimed that Si is depleted by an average $0.27\pm0.16$ dex in all DLAs. \cite{Vladilo11} used S and Zn abundances as input for a best-fitting chemical evolution model to predict [Si/S] without dust depletion, which assumes the same chemical evolution model of a single dwarf galaxy applies to all DLAs. With respect to the \cite{Vladilo11} model (the magenta line in Figure \ref{fig:genSiS}), the addition of the MRDLA data show a lack of Si depletion at higher metallicities. This discrepancy with the model prediction of increasing dust depletion as a function of metallicity suggests that the average depletion calculated by \cite{Vladilo11} does not represent the true amount of Si depletion in some systems. Furthermore, the lack of DLAs with [Si/S]$>0$ at [S/H]$>-0.5$ may reflect mild dust depletion of Si only at high metallicities.

\begin{figure*}
\begin{center}
\includegraphics[width=\textwidth]{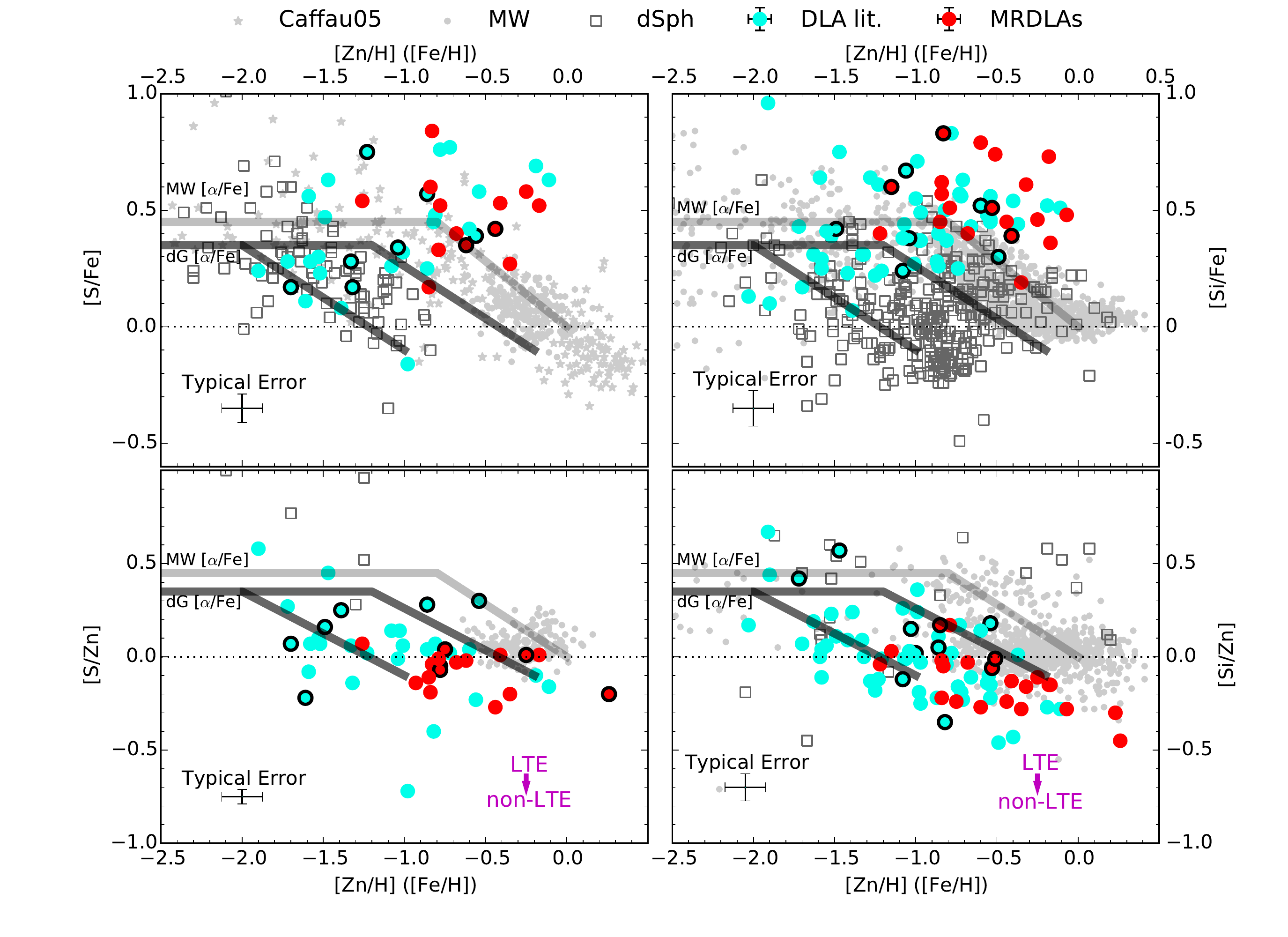}
\caption[ {[$\alpha$/Zn]} and {\alphafe{}} vs.~metallicity in stars and DLAs]{[S/Fe] (top left panel), [Si/Fe] (top right panel), [S/Zn] (bottom left panel), and [Si/Zn] (bottom right panel) as a function of  [Zn/H] in the DLA literature and MRDLA samples. The thick dark and light grey lines represent a cartoon of the \alphafe{}-metallicity trend to illustrate the difference between dwarf galaxies and the Milky Way \citep[respectively; based on the trends seen in][]{Tolstoy09}. The two dSph grey lines highlight the range in mass of dSphs, and the effect of mass on the \alphafe{} trend. [Fe/H] is used in place of [Zn/H] for the stars as the metallicity indicator. The symbols are the same as in Figure \ref{fig:genZn}, however data from \citet[star symbols]{Caffau05} were added to complete the trend of [S/Fe] in the Galactic population due to the lack of sulphur abundances in our stellar literature sample. DLAs have roughly constant \alphafe{}$\sim+0.5$ and [$\alpha$/Zn]$\lesssim0$ dex at the \alphafe{} knee of the Milky Way. If Zn is a tracer of Fe above [Zn/H]$-1.5$ dex in DLAs, the true \alphafe{} reaches a solar value at lower metallicities than the Milky Way but is similar to \alphafe{} in dSphs. However [$\alpha$/Zn] in DLAs disagrees with [$\alpha$/Zn] in dSphs. The magnitude and direction of both S and Si non-LTE corrections vary from line to line (as reflected in the scatter of the Galactic data), therefore the magenta arrows only show the non-LTE corrections for Zn. For S, the non-LTE correction can be as low as $-0.2$ dex \citep{Takeda05}, while corrections for Si can range between $\sim-0.02$ and $\sim0.25$ \citep{Shi09}. }
\label{fig:genalphaZn}
\end{center}
\end{figure*}

To assess the nucleosynthetic trend of \alphafe{} with metallicity in DLAs relative to the stellar literature sample, Figure \ref{fig:genalphaZn} shows [S/Fe] and [Si/Fe] (top left and right panels; respectively) and [S/Zn] and [Si/Zn] (bottom left and right panels; respectively). Due to the lack of S abundances in the literature sample, the data from \cite{Caffau05} have been included to supplement the Galactic stellar data\footnote{The \cite{Caffau05} data are excluded from the literature sample as it has only measured S and Fe abundances for a variety of stars, and is not focused on a particular Galactic population.}. In the Galactic data, \alphafe{} shows the well-known supersolar plateau at \alphafe{}$=+0.3$ at low metallicities, with a gradual decrease to solar \alphafe{} (the `knee') after [Fe/H]$>-1$ \citep{McWilliam97,Venn04}. The metallicity of the knee in dSphs depends on the mass of the galaxy, with lower mass dSphs having lower metallicity knees \citep{Tolstoy09}. To provide a representative trend of \alphafe{} in each panel of Figure \ref{fig:genalphaZn} for the Galaxy and its satellites, we  plot cartoon lines showing the typical trend of \alphafe{} for the two populations \citep[as interpreted from data in][light and dark grey lines, respectively]{Tolstoy09}. The spread in mass (and therefore metallicities of the \alphafe{} knee) in dSphs is shown by the two dark grey lines  representing the lowest and highest knees from dSphs in \cite{Tolstoy09}.

The bulk of the dSph Si abundances in Figure \ref{fig:genalphaZn} come from two of the more massive galaxies \citep[Sagittarius and Fornax; see][ and references therein]{Mcconnachie12}, and therefore show [Si/Fe] dropping to solar ratios at [Fe/H]$\lesssim-0.5$.  Nonetheless, [Si/Fe] is typically lower in dSphs than the Milky Way at a given metallicity and the dSphs can exhibit solar (or even subsolar) [Si/Fe] at the same metallicities at which the Milky Way shows an average enhancement of +0.3 dex\footnote{The light grey line in Figure \ref{fig:genalphaZn} plateau is offset slightly for display purposes.}. In contrast, [S/Fe] (primarily from the lower mass Sculptor dSph) has a knee at a lower metallicity of [Fe/H]$\sim-2$.

In the lower panels of Figure \ref{fig:genalphaZn} we show the Si and S abundances now relative to Zn.  Note that the \alphafe{} cartoon lines are the same as those from the upper panels, and are repeated in the lower panels for reference. The Milky Way [Si/Zn] data show a similar trend to the Galactic \alphafe{}\footnote{An identical trend with Galactic [Mg/Zn] is seen in Figure \ref{fig:MgZnstars}.}. In contrast to \alphafe{}, [$\alpha$/Zn] in dSphs is relatively high which is to be expected as [Zn/Fe] is typically subsolar in dSphs (see Figure \ref{fig:genZn}). The comparison between \alphafe{} and [$\alpha$/Zn] in the dSphs therefore epitomizes the challenge of using Zn as an Fe-peak tracer in DLAs. That is, even if \alphafe{} is solar (or even subsolar) in dSph, [$\alpha$/Zn] is frequently supersolar, due to the typically subsolar [Zn/Fe] in dwarfs.  Therefore, the choice of Fe-peak tracer can drastically impact the interpretation of $\alpha$-abundance ratios.

In DLAs, there is a significant scatter in \alphafe{} (both Si and S), although the ratios are almost always supersolar, typically \alphafe{}$\sim0.5\pm0.2$. The uniformly supersolar \alphafe{} in DLAs is either a true enhancement in \alphafe{} or an overestimation due to the depletion of Fe onto dust; providing an upper limit to the true \alphafe{} in DLAs. It is reasonable to assume that \alphafe{} is likely at, or below the Galactic plateau of [S/Fe]$=+0.3$ at a metallicity [Zn/H]$\gtrsim-0.5$ due to dust depletion since the depletion in this regime is likely to be significant.  However, it is impossible to quantify at which metallicity \alphafe{} drops from the plateau.

In an attempt to circumvent the depletion issue in Fe, and determine whether the typical DLA \alphafe{} is consistent with the Galactic data, it is common to consider [$\alpha$/Zn] in DLAs. The bottom panels of Figure \ref{fig:genalphaZn} show [$\alpha$/Zn] in both stars and DLAs. The DLAs show solar or subsolar [S/Zn] and [Si/Zn] at [Zn/H]$\gtrsim-1$ \citep[as previously seen in][]{Centurion00,Prochaska02II,Nissen04,DZavadsky06,Rafelski12}, with a possible increase in [Si/Zn] and [S/Zn] at [Zn/H]$\lesssim -1.5$. The addition of the MRDLA data shows that solar [Si/Zn] and [S/Zn] extend to higher metallicities; despite Si possibly being mildly depleted. The salient point in Figure \ref{fig:genalphaZn} is that [$\alpha$/Zn] in DLAs is different from [$\alpha$/Zn] in either Galactic or dSph stars.

Regardless of whether [Zn/Fe] in DLAs intrinsically matches [Zn/Fe] in dSphs or the Milky Way,  there is no plausible explanation of observed low [$\alpha$/Zn] in DLAs other than low quantities of $\alpha$-elements (relative to the Galaxy and its satellites). However, the [$\alpha$/Zn] in DLAs is apparently in agreement neither with local dSphs, nor the Milky Way. Therefore, whilst the low [$\alpha$/Zn] seen in DLAs is qualitatively similar to the generally low $\alpha$-enhancements seen in dSphs (in \alphafe{}), we conclude that DLAs \emph{do not show identical patterns to either the dwarfs, or the Milky Way, and hence are not analogous to any one single component of the Local Group.}

\subsection{Mn}
\label{sec:Mn}

In addition to $\alpha$-elements, Mn can provide another constraint on the role of both Type Ia (SNe Ia) and II SNe (SNe II) in the chemical enrichment of a galaxy (see Appendix \ref{sec:AppMn} for more details). Previous studies of [Mn/Fe] in DLAs have presented opposing views, with \cite{Pettini00} and \cite{DLAcat30} finding a constant, subsolar  [Mn/Fe] in five DLAs while \cite{Ledoux02} show an increasing [Mn/Fe] with increasing metallicity with the addition of 15 DLAs. However, much of this discrepancy has been attributed to how Mn is corrected for dust depletion as it is somewhat depleted onto dust.

\begin{figure}
\begin{center}
\includegraphics[width=0.5\textwidth]{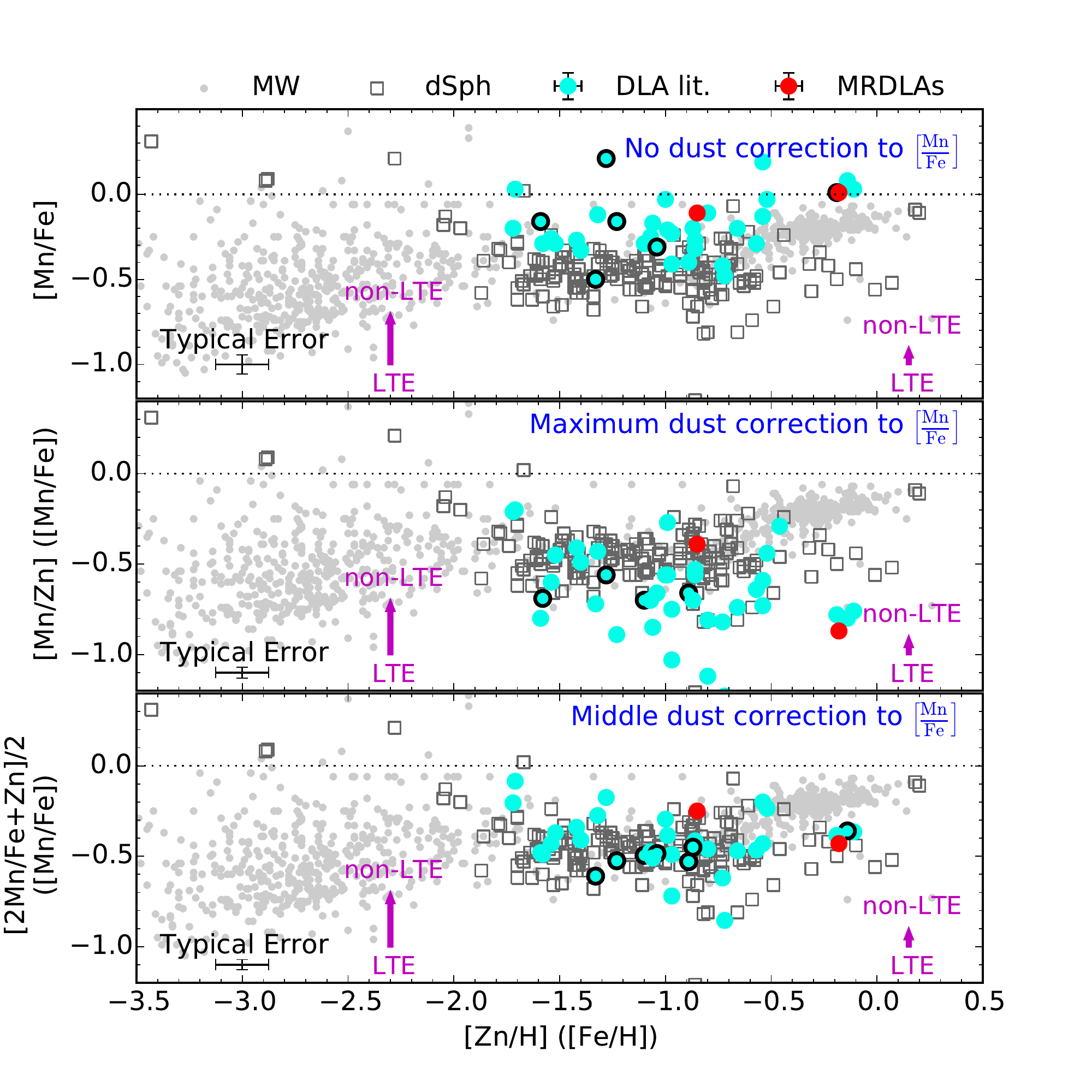}
\caption[{[Mn/Fe-peak]} as a function of metallicity in stars and DLAs]{Mn abundances relative to the Fe-peak elements. To account for dust depletion in metallicity, Zn is used in place of Fe for DLAs. [Mn/Fe] (top panel), and [Mn/Zn] (middle panel) show the upper and lower limits (respectively) of the Mn to Fe-peak ratio to account for dust in the literature DLA (blue) and MRDLA (red) sample. The bottom panel DLA points represent the best estimate of the relative ratios to include dust depletion effects (i.e. the mid-point of [Mn/Fe] and [Mn/Zn]; [2Mn/Fe+Zn]/2). Relative to the stellar data ([Mn/Fe] in all three panels), it is apparent that \mnfe{} in DLAs remains constantly solar (like dSphs) at high metallicities. The magenta arrows show the approximate non-LTE corrections for the Milky Way data \citep{Battistini15}.}
\label{fig:genMnFe}
\end{center}
\end{figure} 

Figure \ref{fig:genMnFe} shows the evolution of Mn for the MRDLA, literature DLA, Milky Way, and satellite samples. [Zn/H] is only adopted as the metallicity indicator in DLAs to minimize dust effects, whereas [Fe/H] is used for the stellar data. In the Milky Way (once non-LTE corrections have been included), [Mn/Fe] remains solar at all metallicities. However, dSphs stars show a relatively flat [Mn/Fe] for [Fe/H]$\gtrsim-2$. Irrespective of non-LTE effects, the dSphs and Milky Way data overlap at [Fe/H]$\lesssim-0.5$. However at [Fe/H]$\gtrsim-0.5$, there is a discrepancy between the two populations, where the Milky Way [Mn/Fe] is higher than the dSphs which remain at [Mn/Fe]$\sim-0.5$ dex (larger than the typical non-LTE correction at this metallicity) due to the relatively low contribution from SNe Ia in dSphs \citep{North12}.

One way of circumventing the issue of the unknown dust depletion of [Mn/Fe] could be to compare to an Fe-peak element with a similar dust depletion factor to Mn, which would effectively remove the relative depletion in the measured [Mn/Fe]. Fe, Cr, and Zn are the commonly measured Fe-peak tracers in DLAs. However the depletion factor of [Mn/H] in the local ISM lies roughly half way between the factors of [Zn/H] and [Fe/H] or [Cr/H] \citep{Savage96}. Therefore [Mn/Fe] (top panel of Figure \ref{fig:genMnFe}) and [Mn/Zn] (assuming Zn traces the Fe-peak for [Zn/H]$\geq-1$; middle panel) should reflect the upper and lower bounds assuming no dust correction ([Mn/Fe]) and some maximum dust correction ([Mn/Zn]). To make a more reasonable correction, the bottom panel shows the middle value between [Mn/Zn] and [Mn/Fe]  to reflect a reasonable dust correction (i.e. [2Mn/(Fe+Zn)]/2; further referred to as \mnfe{} to reflect the dust corrected [Mn/Fe]\footnote{[2Mn/(Fe+Zn)]/2 $=$ \mnfe{} $=$ ($2\times$[Mn/H] $-$ [Fe/H] $-$ [Zn/H])$/2$.}). As noted in Section \ref{sec:ZnFeCr}, [Zn/Fe] is not a perfect tracer of dust depletion. For a DLA with [Zn/Fe] similar to the `average' local dSph ([Zn/Fe]$=-0.3$), the dust correction would be $\sim 0.3$ dex larger than shown in the middle panel of Figure \ref{fig:genMnFe}. This would lead to a downward shift of the DLA points by $\sim 0.3$ dex in the middle panel, and $\sim 0.15$ dex in the lower panel of Figure \ref{fig:genMnFe}, increasing the discrepancy with the Galactic data. The dust-corrected DLA values shown in the lower panel of Figure \ref{fig:genMnFe} are adopted for our discussions on Mn.

Nearly doubling the number of Mn detections in DLAs from both the literature and MRDLA samples (in particular the addition of three high-metallicity sightlines)  further emphasizes the  constant  \mnfe{}$\sim-0.5$ over all metallcities first seen by \cite{Pettini00} and \cite{DLAcat30}, and is similar to the [Mn/Fe]$\sim-0.5$ seen in dSphs \citep{North12}. Upon accounting for any discrepancy from subsolar [Zn/Fe] in DLAs ($\sim-0.15$ dex correction), \mnfe{} would be even lower than the Milky Way value and more similar to dSphs. Based on the chemical evolution models presented in \cite{North12}, lower [Mn/Fe] with increasing metallicity (relative to the Milky Way) implies SNe II are the dominant source of Mn production at high metallicity (although SNe Ia are still contributing). For DLAs, this provides additional similarities to the chemical signatures of dSphs.  

\section{Summary and Conclusions}

Using our sample of 44 MRDLAs from Paper I, and literature compilations of \nldla{} DLAs and \nlstars{} stars, we have extended the comparison between the chemistry of stars and DLAs to understand the chemical evolution of DLAs. With the addition of the MRDLAs, we are able to extend the average chemical enrichment of ISM gas in DLAs to higher metallicities than has previously been possible, \emph{probing, on average, the metallicity regime of the Galactic disc and metal-rich satellites at a redshift of $z\sim2$} (Figure \ref{fig:Festars}). The MRDLAs have reached these relatively high metallicities at a time when the universe was about $\sim3$ Gyr old.

We have presented a simple method of correcting for [Zn/Fe] depletion in DLAs, demonstrating that the majority of DLAs have dust-corrected [Zn/Fe] resembling dSphs, with some systems consistent with the Milky Way (Figure \ref{fig:genZn}). With the large spread in [Zn/Fe] values in dSphs ($-0.9\lesssim$[Zn/Fe]$\lesssim0.2$), it is possible that all DLAs are consistent with dSph [Zn/Fe]. This result emphasizes that \emph{Zn is not a perfect tracer of the Fe-peak in DLAs}, and dust depletion for DLAs needs to be evaluated on an individual basis. As the MRDLA sample pushes to higher metallicity sightlines, we are starting to probe systems with higher amounts of dust depletion. The nature of the enhanced amount of dust depletion of metal-rich DLAs is highlighted by the increased enhancement of [Zn/Fe] at [Zn/H]$\gtrsim-0.7$ (Figure \ref{fig:genZn}). However, the addition of the MRDLAs does not show any significant difference in depletion between [Si/S]  (Figure \ref{fig:genSiS}), suggesting Si is not as depleted as previously expected \citep{Vladilo11}.

Our analysis of the chemistry of individual elements show that the dust-corrected [Mn/Fe] remains constant at $\sim-0.5$ for [Zn/H]$\gtrsim-0.5$, considerably below the values seen in Galactic disc stars (Figure \ref{fig:genMnFe}). In addition, we confirm that DLAs reach solar [S/Zn] and [Si/Zn] at [Zn/H]$\sim-1.5$ (Figure \ref{fig:genalphaZn}). The interpretation of the trends of S, Si, and Mn rely on assumptions about dust depletion or using Zn as a tracer for Fe above [Zn/H]$\gtrsim-1.5$. For DLAs where [Zn/Fe]$\sim0$ similar to the Milky Way at [Zn/H]$>-1.5$ (Figure \ref{fig:genZn}), the observed [$\alpha$/Zn] in DLAs are consistent with the chemical enrichment patterns of [$\alpha$/Fe] dSph stars. In this case however, both [$\alpha$/Zn]$\sim0$ ([Zn/H]$\gtrsim -1.5$; Figure \ref{fig:genalphaZn}) and [Zn/Fe]$\gtrsim0$ ([Zn/H]$\lesssim-1.3$; Figure \ref{fig:genZn}) in DLAs are at odds with the corresponding measurements in dSphs ([$\alpha$/Zn]$>0$ and [Zn/Fe]$<0$) at the same metallicities. In contrast, DLAs with larger dust corrections such that [Zn/Fe] match the measurements in dSphs, Fe would be underpredicted on average by $\sim0.8$ dex (for [Zn/H]$>-1$). With this correction for Fe, both [Si/Fe] and [S/Fe] would be corrected to solar values at lower metallicities relative to the Milky Way, also suggesting that DLAs have similar \alphafe{} values similar to Local Group dSph galaxies. However, DLA [$\alpha$/Zn] still does not match the values in dSphs. \emph{Although there is no one-to-one mapping of DLAs to Local Group components, the majority of DLA abundance patterns are most similar to the patterns in Local Group dSph galaxies (with the exception of lower [$\alpha$/Zn] in DLAs) rather than in the Milky Way.}

The lower masses of dSphs or the continuity of star formation are plausible explanations of why \alphafe{}, [Mn/Fe], and [Zn/Fe] appear to not follow the Milky Way trends \citep[e.g.][]{Shetrone03,Tolstoy09,North12,Venn12,McWilliam13}. As a result, products from high-mass stars are not retained or produced in dSphs, and SNe Ia contribute at lower metallicities than the Milky Way. If the MRDLA sample and other metal-rich DLAs are massive galaxies \citep[as inferred by the $\Delta V_{90}$ statistic;][MRDLA $\Delta V_{90}$ values provided in Table 36 of Paper I]{Prochaska97,Ledoux06,Neeleman13,Christensen14}, we might expect the yields from the high mass stars will be retained. This is supported by [Zn/Fe] matching the Galaxy in most low-metallicity DLAs (assuming modest dust depletion of Fe at low metallicities). However, both simulations and imaging observations show that DLAs are commonly associated with low-mass galaxies and low star formation rates \citep[e.g.][]{Krogager13,Rahmati14,Fumagalli15};  which is consistent with the observed \alphafe{} and [Mn/Fe] trends presented in this work. Therefore DLAs appear to show snapshots of chemical enrichment of ISM gas related to dwarf galaxies, although [$\alpha$/Zn] is suggestive that these dwarf galaxies are not identical to the Local Group dSphs.

\section*{Acknowledgements}
We thank M.~Rafelski, R.~Cooke, V.~D'Odorico, M.~Pettini, J.~Norris, and the anonymous referee for their useful comments on earlier versions of this manuscript. TAMB thanks F.~Herwig, C.~Ritter, M.~Pignatari, and S.~Woosley for their useful discussions on the nucleosynthetic origin of various elements. SLE gratefully acknowledges the receipt of an NSERC Discovery Grant that funded this research. JXP is partially supported by NSF grant AST-1109452.

\bibliography{bibref}

\begin{thebibliography}{228}
\expandafter\ifx\csname natexlab\endcsname\relax\def\natexlab#1{#1}\fi

\bibitem[{{Akerman} {et~al}\mbox{.}(2004){Akerman}, {Carigi}, {Nissen},
  {Pettini}, \& {Asplund}}]{Akerman04}
{Akerman} C.~J., {Carigi} L., {Nissen} P.~E., {Pettini} M., {Asplund} M., 2004,
  \aap, 414, 931

\bibitem[{{Akerman} {et~al}\mbox{.}(2005){Akerman}, {Ellison}, {Pettini}, \&
  {Steidel}}]{Akerman05}
{Akerman} C.~J., {Ellison} S.~L., {Pettini} M., {Steidel} C.~C., 2005, \aap,
  440, 499

\bibitem[{{Aoki} {et~al}\mbox{.}(2009){Aoki}, {Arimoto}, {Sadakane}, {Tolstoy},
  {Battaglia}, {Jablonka}, {Shetrone}, {Letarte}, {Irwin}, {Hill}, {Francois},
  {Venn}, {Primas}, {Helmi}, {Kaufer}, {Tafelmeyer}, {Szeifert}, \&
  {Babusiaux}}]{Aoki09}
{Aoki} W. {et~al.}, 2009, \aap, 502, 569

\bibitem[{{Aoki} {et~al}\mbox{.}(2013){Aoki}, {Beers}, {Lee}, {Honda}, {Ito},
  {Takada-Hidai}, {Frebel}, {Suda}, {Fujimoto}, {Carollo}, \&
  {Sivarani}}]{Aoki13}
{Aoki} W. {et~al.}, 2013, \aj, 145, 13

\bibitem[{{Asplund} {et~al}\mbox{.}(2009){Asplund}, {Grevesse}, {Sauval}, \&
  {Scott}}]{Asplund09}
{Asplund} M., {Grevesse} N., {Sauval} A.~J., {Scott} P., 2009, \araa, 47, 481

\bibitem[{{Asplund} {et~al}\mbox{.}(1997){Asplund}, {Gustafsson}, {Kiselman},
  \& {Eriksson}}]{Asplund97}
{Asplund} M., {Gustafsson} B., {Kiselman} D., {Eriksson} K., 1997, \aap, 318,
  521

\bibitem[{{Badenes} {et~al}\mbox{.}(2008){Badenes}, {Bravo}, \&
  {Hughes}}]{Badenes08}
{Badenes} C., {Bravo} E., {Hughes} J.~P., 2008, \apjl, 680, L33

\bibitem[{{Barbuy} {et~al}\mbox{.}(2015){Barbuy}, {Friaca}, {da Silveira},
  {Hill}, {Zoccali}, {Minniti}, {Renzini}, {Ortolani}, \& {Gomez}}]{Barbuy15}
{Barbuy} B. {et~al.}, 2015, ArXiv e-print 1506.01612

\bibitem[{{Battisti} {et~al}\mbox{.}(2012){Battisti}, {Meiring}, {Tripp},
  {Prochaska}, {Werk}, {Jenkins}, {Lehner}, {Tumlinson}, \&
  {Thom}}]{Battisti12}
{Battisti} A.~J. {et~al.}, 2012, \apj, 744, 93

\bibitem[{{Battistini} \& {Bensby}(2015)}]{Battistini15}
{Battistini} C., {Bensby} T., 2015, ArXiv e-print 1502.01152

\bibitem[{{Bensby} {et~al}\mbox{.}(2005){Bensby}, {Feltzing}, {Lundstr{\"o}m},
  \& {Ilyin}}]{Bensby05}
{Bensby} T., {Feltzing} S., {Lundstr{\"o}m} I., {Ilyin} I., 2005, \aap, 433,
  185

\bibitem[{{Bensby} {et~al}\mbox{.}(2014){Bensby}, {Feltzing}, \&
  {Oey}}]{Bensby14}
{Bensby} T., {Feltzing} S., {Oey} M.~S., 2014, \aap, 562, A71

\bibitem[{{Berg} {et~al}\mbox{.}(2013){Berg}, {Ellison}, {Venn}, \&
  {Prochaska}}]{Berg13}
{Berg} T.~A.~M., {Ellison} S.~L., {Venn} K.~A., {Prochaska} J.~X., 2013,
  \mnras, 434, 2892

\bibitem[{{Berg} {et~al}\mbox{.}(2015){Berg}, {Neeleman}, {Prochaska},
  {Ellison}, \& {Wolfe}}]{Berg15}
{Berg} T.~A.~M., {Neeleman} M., {Prochaska} J.~X., {Ellison} S.~L., {Wolfe}
  A.~M., 2015, \pasp, 127, 167

\bibitem[{{Bergemann} \& {Cescutti}(2010)}]{Bergemann10}
{Bergemann} M., {Cescutti} G., 2010, \aap, 522, A9

\bibitem[{{Bergeron} \& {Boiss{\'e}}(1991)}]{Bergeron91}
{Bergeron} J., {Boiss{\'e}} P., 1991, \aap, 243, 344

\bibitem[{{Boisse} {et~al}\mbox{.}(1998){Boisse}, {Le Brun}, {Bergeron}, \&
  {Deharveng}}]{DLAcat9}
{Boisse} P., {Le Brun} V., {Bergeron} J., {Deharveng} J.-M., 1998, \aap, 333,
  841

\bibitem[{{Bonifacio} {et~al}\mbox{.}(2009){Bonifacio}, {Spite}, {Cayrel},
  {Hill}, {Spite}, {Fran{\c c}ois}, {Plez}, {Ludwig}, {Caffau}, {Molaro},
  {Depagne}, {Andersen}, {Barbuy}, {Beers}, {Nordstr{\"o}m}, \&
  {Primas}}]{Bonifacio09}
{Bonifacio} P. {et~al.}, 2009, \aap, 501, 519

\bibitem[{{Bowen} {et~al}\mbox{.}(2005){Bowen}, {Jenkins}, {Pettini}, \&
  {Tripp}}]{DLAcat52}
{Bowen} D.~V., {Jenkins} E.~B., {Pettini} M., {Tripp} T.~M., 2005, \apj, 635,
  880

\bibitem[{{Caffau} {et~al}\mbox{.}(2005){Caffau}, {Bonifacio}, {Faraggiana},
  {Fran{\c c}ois}, {Gratton}, \& {Barbieri}}]{Caffau05}
{Caffau} E., {Bonifacio} P., {Faraggiana} R., {Fran{\c c}ois} P., {Gratton}
  R.~G., {Barbieri} M., 2005, \aap, 441, 533

\bibitem[{{Caffau} {et~al}\mbox{.}(2014){Caffau}, {Monaco}, {Spite},
  {Bonifacio}, {Carraro}, {Ludwig}, {Villanova}, {Beletsky}, \&
  {Sbordone}}]{Caffau14}
{Caffau} E. {et~al.}, 2014, \aap, 568, A29

\bibitem[{{Carretta} {et~al}\mbox{.}(2010){Carretta}, {Bragaglia}, {Gratton},
  {Lucatello}, {Bellazzini}, {Catanzaro}, {Leone}, {Momany}, {Piotto}, \&
  {D'Orazi}}]{Carretta10}
{Carretta} E. {et~al.}, 2010, \aap, 520, A95

\bibitem[{{Carswell} {et~al}\mbox{.}(2012){Carswell}, {Becker}, {Jorgenson},
  {Murphy}, \& {Wolfe}}]{DLAcat90}
{Carswell} R.~F., {Becker} G.~D., {Jorgenson} R.~A., {Murphy} M.~T., {Wolfe}
  A.~M., 2012, \mnras, 422, 1700

\bibitem[{{Carswell} {et~al}\mbox{.}(1987){Carswell}, {Webb}, {Baldwin}, \&
  {Atwood}}]{Carswell87}
{Carswell} R.~F., {Webb} J.~K., {Baldwin} J.~A., {Atwood} B., 1987, \apj, 319,
  709

\bibitem[{{Cayrel} {et~al}\mbox{.}(2004){Cayrel}, {Depagne}, {Spite}, {Hill},
  {Spite}, {Fran{\c c}ois}, {Plez}, {Beers}, {Primas}, {Andersen}, {Barbuy},
  {Bonifacio}, {Molaro}, \& {Nordstr{\"o}m}}]{Cayrel04}
{Cayrel} R. {et~al.}, 2004, \aap, 416, 1117

\bibitem[{{Centuri{\'o}n} {et~al}\mbox{.}(2000){Centuri{\'o}n}, {Bonifacio},
  {Molaro}, \& {Vladilo}}]{Centurion00}
{Centuri{\'o}n} M., {Bonifacio} P., {Molaro} P., {Vladilo} G., 2000, \apj, 536,
  540

\bibitem[{{Centuri{\'o}n} {et~al}\mbox{.}(2003){Centuri{\'o}n}, {Molaro},
  {Vladilo}, {P{\'e}roux}, {Levshakov}, \& {D'Odorico}}]{DLAcat40}
{Centuri{\'o}n} M., {Molaro} P., {Vladilo} G., {P{\'e}roux} C., {Levshakov}
  S.~A., {D'Odorico} V., 2003, \aap, 403, 55

\bibitem[{{Cescutti} {et~al}\mbox{.}(2008){Cescutti}, {Matteucci},
  {Lanfranchi}, \& {McWilliam}}]{Cescutti08}
{Cescutti} G., {Matteucci} F., {Lanfranchi} G.~A., {McWilliam} A., 2008, \aap,
  491, 401

\bibitem[{{Chen} {et~al}\mbox{.}(2005){Chen}, {Kennicutt}, \&
  {Rauch}}]{DLAcat83}
{Chen} H.-W., {Kennicutt}, Jr. R.~C., {Rauch} M., 2005, \apj, 620, 703

\bibitem[{{Chen} {et~al}\mbox{.}(2004){Chen}, {Nissen}, \& {Zhao}}]{Chen04}
{Chen} Y.~Q., {Nissen} P.~E., {Zhao} G., 2004, \aap, 425, 697

\bibitem[{{Chen} {et~al}\mbox{.}(2002){Chen}, {Nissen}, {Zhao}, \&
  {Asplund}}]{Chen02}
{Chen} Y.~Q., {Nissen} P.~E., {Zhao} G., {Asplund} M., 2002, \aap, 390, 225

\bibitem[{{Christensen} {et~al}\mbox{.}(2014){Christensen}, {M{\o}ller},
  {Fynbo}, \& {Zafar}}]{Christensen14}
{Christensen} L., {M{\o}ller} P., {Fynbo} J.~P.~U., {Zafar} T., 2014, \mnras,
  445, 225

\bibitem[{{Churchill} {et~al}\mbox{.}(2000){Churchill}, {Mellon}, {Charlton},
  {Jannuzi}, {Kirhakos}, {Steidel}, \& {Schneider}}]{DLAcat15}
{Churchill} C.~W., {Mellon} R.~R., {Charlton} J.~C., {Jannuzi} B.~T.,
  {Kirhakos} S., {Steidel} C.~C., {Schneider} D.~P., 2000, \apj, 543, 577

\bibitem[{{Clayton}(2003)}]{Clayton_iso}
{Clayton} D., 2003, {Handbook of Isotopes in the Cosmos}, {Clayton, D.}, ed.

\bibitem[{{Cohen} {et~al}\mbox{.}(2013){Cohen}, {Christlieb}, {Thompson},
  {McWilliam}, {Shectman}, {Reimers}, {Wisotzki}, \& {Kirby}}]{Cohen13}
{Cohen} J.~G., {Christlieb} N., {Thompson} I., {McWilliam} A., {Shectman} S.,
  {Reimers} D., {Wisotzki} L., {Kirby} E., 2013, \apj, 778, 56

\bibitem[{{Cooke} {et~al}\mbox{.}(2012){Cooke}, {Pettini}, \&
  {Murphy}}]{DLAcat96}
{Cooke} R., {Pettini} M., {Murphy} M.~T., 2012, \mnras, 425, 347

\bibitem[{{Cooke} {et~al}\mbox{.}(2010){Cooke}, {Pettini}, {Steidel}, {King},
  {Rudie}, \& {Rakic}}]{DLAcat77}
{Cooke} R., {Pettini} M., {Steidel} C.~C., {King} L.~J., {Rudie} G.~C., {Rakic}
  O., 2010, \mnras, 409, 679

\bibitem[{{Cooke} {et~al}\mbox{.}(2011{\natexlab{a}}){Cooke}, {Pettini},
  {Steidel}, {Rudie}, \& {Jorgenson}}]{DLAcat76}
{Cooke} R., {Pettini} M., {Steidel} C.~C., {Rudie} G.~C., {Jorgenson} R.~A.,
  2011{\natexlab{a}}, \mnras, 412, 1047

\bibitem[{{Cooke} {et~al}\mbox{.}(2011{\natexlab{b}}){Cooke}, {Pettini},
  {Steidel}, {Rudie}, \& {Nissen}}]{Cooke11}
{Cooke} R., {Pettini} M., {Steidel} C.~C., {Rudie} G.~C., {Nissen} P.~E.,
  2011{\natexlab{b}}, \mnras, 417, 1534

\bibitem[{{Cooke} \& {Madau}(2014)}]{Cooke14}
{Cooke} R.~J., {Madau} P., 2014, \apj, 791, 116

\bibitem[{{de la Varga} {et~al}\mbox{.}(2000){de la Varga}, {Reimers},
  {Tytler}, {Barlow}, \& {Burles}}]{DLAcat16}
{de la Varga} A., {Reimers} D., {Tytler} D., {Barlow} T., {Burles} S., 2000,
  \aap, 363, 69

\bibitem[{{Dessauges-Zavadsky} {et~al}\mbox{.}(2004){Dessauges-Zavadsky},
  {Calura}, {Prochaska}, {D'Odorico}, \& {Matteucci}}]{DZavadsky04}
{Dessauges-Zavadsky} M., {Calura} F., {Prochaska} J.~X., {D'Odorico} S.,
  {Matteucci} F., 2004, \aap, 416, 79

\bibitem[{{Dessauges-Zavadsky} {et~al}\mbox{.}(2007){Dessauges-Zavadsky},
  {Calura}, {Prochaska}, {D'Odorico}, \& {Matteucci}}]{Dessauges07}
{Dessauges-Zavadsky} M., {Calura} F., {Prochaska} J.~X., {D'Odorico} S.,
  {Matteucci} F., 2007, \aap, 470, 431

\bibitem[{{Dessauges-Zavadsky} {et~al}\mbox{.}(2001){Dessauges-Zavadsky},
  {D'Odorico}, {McMahon}, {Molaro}, {Ledoux}, {P{\'e}roux}, \&
  {Storrie-Lombardi}}]{DLAcat28}
{Dessauges-Zavadsky} M., {D'Odorico} S., {McMahon} R.~G., {Molaro} P., {Ledoux}
  C., {P{\'e}roux} C., {Storrie-Lombardi} L.~J., 2001, \aap, 370, 426

\bibitem[{{Dessauges-Zavadsky} {et~al}\mbox{.}(2003){Dessauges-Zavadsky},
  {P{\'e}roux}, {Kim}, {D'Odorico}, \& {McMahon}}]{DLAcat45}
{Dessauges-Zavadsky} M., {P{\'e}roux} C., {Kim} T.-S., {D'Odorico} S.,
  {McMahon} R.~G., 2003, \mnras, 345, 447

\bibitem[{{Dessauges-Zavadsky} {et~al}\mbox{.}(2002){Dessauges-Zavadsky},
  {Prochaska}, \& {D'Odorico}}]{DLAcat30}
{Dessauges-Zavadsky} M., {Prochaska} J.~X., {D'Odorico} S., 2002, \aap, 391,
  801

\bibitem[{{Dessauges-Zavadsky} {et~al}\mbox{.}(2006){Dessauges-Zavadsky},
  {Prochaska}, {D'Odorico}, {Calura}, \& {Matteucci}}]{DZavadsky06}
{Dessauges-Zavadsky} M., {Prochaska} J.~X., {D'Odorico} S., {Calura} F.,
  {Matteucci} F., 2006, \aap, 445, 93

\bibitem[{{D'Odorico}(2007)}]{Dodorico07}
{D'Odorico} V., 2007, \aap, 470, 523

\bibitem[{{D'Odorico} \& {Molaro}(2004)}]{DLAcat47}
{D'Odorico} V., {Molaro} P., 2004, \aap, 415, 879

\bibitem[{{Dutta} {et~al}\mbox{.}(2014){Dutta}, {Srianand}, {Rahmani},
  {Petitjean}, {Noterdaeme}, \& {Ledoux}}]{DLAcat108}
{Dutta} R., {Srianand} R., {Rahmani} H., {Petitjean} P., {Noterdaeme} P.,
  {Ledoux} C., 2014, \mnras, 440, 307

\bibitem[{{Edvardsson} {et~al}\mbox{.}(1993){Edvardsson}, {Andersen},
  {Gustafsson}, {Lambert}, {Nissen}, \& {Tomkin}}]{Edvardsson93}
{Edvardsson} B., {Andersen} J., {Gustafsson} B., {Lambert} D.~L., {Nissen}
  P.~E., {Tomkin} J., 1993, \aaps, 102, 603

\bibitem[{{Ellison} {et~al}\mbox{.}(2007){Ellison}, {Hennawi}, {Martin}, \&
  {Sommer-Larsen}}]{DLAcat68}
{Ellison} S.~L., {Hennawi} J.~F., {Martin} C.~L., {Sommer-Larsen} J., 2007,
  \mnras, 378, 801

\bibitem[{{Ellison} {et~al}\mbox{.}(2012){Ellison}, {Kanekar}, {Prochaska},
  {Momjian}, \& {Worseck}}]{Ellison12}
{Ellison} S.~L., {Kanekar} N., {Prochaska} J.~X., {Momjian} E., {Worseck} G.,
  2012, \mnras, 424, 293

\bibitem[{{Ellison} \& {Lopez}(2001)}]{DLAcat26}
{Ellison} S.~L., {Lopez} S., 2001, \aap, 380, 117

\bibitem[{{Ellison} {et~al}\mbox{.}(2001){Ellison}, {Pettini}, {Steidel}, \&
  {Shapley}}]{DLAcat29}
{Ellison} S.~L., {Pettini} M., {Steidel} C.~C., {Shapley} A.~E., 2001, \apj,
  549, 770

\bibitem[{{Ellison} {et~al}\mbox{.}(2010){Ellison}, {Prochaska}, {Hennawi},
  {Lopez}, {Usher}, {Wolfe}, {Russell}, \& {Benn}}]{Ellison10}
{Ellison} S.~L., {Prochaska} J.~X., {Hennawi} J., {Lopez} S., {Usher} C.,
  {Wolfe} A.~M., {Russell} D.~M., {Benn} C.~R., 2010, \mnras, 406, 1435

\bibitem[{{Ellison} {et~al}\mbox{.}(2008){Ellison}, {York}, {Pettini}, \&
  {Kanekar}}]{DLAcat82}
{Ellison} S.~L., {York} B.~A., {Pettini} M., {Kanekar} N., 2008, \mnras, 388,
  1349

\bibitem[{{Erni} {et~al}\mbox{.}(2006){Erni}, {Richter}, {Ledoux}, \&
  {Petitjean}}]{DLAcat58}
{Erni} P., {Richter} P., {Ledoux} C., {Petitjean} P., 2006, \aap, 451, 19

\bibitem[{{Feltzing} {et~al}\mbox{.}(2007){Feltzing}, {Fohlman}, \&
  {Bensby}}]{Feltzing07}
{Feltzing} S., {Fohlman} M., {Bensby} T., 2007, \aap, 467, 665

\bibitem[{{Francois}(1987)}]{Francois87}
{Francois} P., 1987, \aap, 176, 294

\bibitem[{{Francois}(1988)}]{Francois88}
{Francois} P., 1988, \aap, 195, 226

\bibitem[{{Frebel}(2010)}]{Frebel10}
{Frebel} A., 2010, Astronomische Nachrichten, 331, 474

\bibitem[{{Freeman} \& {Bland-Hawthorn}(2002)}]{Freeman02}
{Freeman} K., {Bland-Hawthorn} J., 2002, \araa, 40, 487

\bibitem[{{Fulbright}(2000)}]{Fulbright00}
{Fulbright} J.~P., 2000, \aj, 120, 1841

\bibitem[{{Fumagalli} {et~al}\mbox{.}(2015){Fumagalli}, {O'Meara}, {Prochaska},
  {Rafelski}, \& {Kanekar}}]{Fumagalli15}
{Fumagalli} M., {O'Meara} J.~M., {Prochaska} J.~X., {Rafelski} M., {Kanekar}
  N., 2015, \mnras, 446, 3178

\bibitem[{{Fynbo} {et~al}\mbox{.}(2013){Fynbo}, {Geier}, {Christensen},
  {Gallazzi}, {Krogager}, {Kr{\"u}hler}, {Ledoux}, {Maund}, {M{\o}ller},
  {Noterdaeme}, {Rivera-Thorsen}, \& {Vestergaard}}]{Fynbo13}
{Fynbo} J.~P.~U. {et~al.}, 2013, \mnras, 436, 361

\bibitem[{{Fynbo} {et~al}\mbox{.}(2011){Fynbo}, {Ledoux}, {Noterdaeme},
  {Christensen}, {M{\o}ller}, {Durgapal}, {Goldoni}, {Kaper}, {Krogager},
  {Laursen}, {Maund}, {Milvang-Jensen}, {Okoshi}, {Rasmussen}, {Thorsen},
  {Toft}, \& {Zafar}}]{DLAcat79}
{Fynbo} J.~P.~U. {et~al.}, 2011, \mnras, 413, 2481

\bibitem[{{Garc{\'{\i}}a P{\'e}rez} {et~al}\mbox{.}(2013){Garc{\'{\i}}a
  P{\'e}rez}, {Cunha}, {Shetrone}, {Majewski}, {Johnson}, {Smith}, {Schiavon},
  {Holtzman}, {Nidever}, {Zasowski}, {Allende Prieto}, {Beers}, {Bizyaev},
  {Ebelke}, {Eisenstein}, {Frinchaboy}, {Girardi}, {Hearty}, {Malanushenko},
  {Malanushenko}, {Meszaros}, {O'Connell}, {Oravetz}, {Pan}, {Robin},
  {Schneider}, {Schultheis}, {Skrutskie}, {Simmonsand}, \& {Wilson}}]{Perez13}
{Garc{\'{\i}}a P{\'e}rez} A.~E. {et~al.}, 2013, \apjl, 767, L9

\bibitem[{{Ge} {et~al}\mbox{.}(2001){Ge}, {Bechtold}, \& {Kulkarni}}]{DLAcat25}
{Ge} J., {Bechtold} J., {Kulkarni} V.~P., 2001, \apjl, 547, L1

\bibitem[{{Geisler} {et~al}\mbox{.}(2005){Geisler}, {Smith}, {Wallerstein},
  {Gonzalez}, \& {Charbonnel}}]{Geisler05}
{Geisler} D., {Smith} V.~V., {Wallerstein} G., {Gonzalez} G., {Charbonnel} C.,
  2005, \aj, 129, 1428

\bibitem[{{Gratton} {et~al}\mbox{.}(2003){Gratton}, {Carretta}, {Claudi},
  {Lucatello}, \& {Barbieri}}]{Gratton03}
{Gratton} R.~G., {Carretta} E., {Claudi} R., {Lucatello} S., {Barbieri} M.,
  2003, \aap, 404, 187

\bibitem[{{Guimar{\~a}es} {et~al}\mbox{.}(2012){Guimar{\~a}es}, {Noterdaeme},
  {Petitjean}, {Ledoux}, {Srianand}, {L{\'o}pez}, \& {Rahmani}}]{DLAcat89}
{Guimar{\~a}es} R., {Noterdaeme} P., {Petitjean} P., {Ledoux} C., {Srianand}
  R., {L{\'o}pez} S., {Rahmani} H., 2012, \aj, 143, 147

\bibitem[{{Gustafsson} {et~al}\mbox{.}(1975){Gustafsson}, {Bell}, {Eriksson},
  \& {Nordlund}}]{MARCS}
{Gustafsson} B., {Bell} R.~A., {Eriksson} K., {Nordlund} A., 1975, \aap, 42,
  407

\bibitem[{{Gustafsson} {et~al}\mbox{.}(2008){Gustafsson}, {Edvardsson},
  {Eriksson}, {J{\o}rgensen}, {Nordlund}, \& {Plez}}]{MARCS2}
{Gustafsson} B., {Edvardsson} B., {Eriksson} K., {J{\o}rgensen} U.~G.,
  {Nordlund} {\AA}., {Plez} B., 2008, \aap, 486, 951

\bibitem[{{Gustafsson} {et~al}\mbox{.}(2003){Gustafsson}, {Edvardsson},
  {Eriksson}, {Mizuno-Wiedner}, {J{\o}rgensen}, \& {Plez}}]{MARCS1}
{Gustafsson} B., {Edvardsson} B., {Eriksson} K., {Mizuno-Wiedner} M.,
  {J{\o}rgensen} U.~G., {Plez} B., 2003, in Astronomical Society of the Pacific
  Conference Series, Vol. 288, Stellar Atmosphere Modeling, {Hubeny} I.,
  {Mihalas} D., {Werner} K., eds., p. 331

\bibitem[{{Hendricks} {et~al}\mbox{.}(2014){Hendricks}, {Koch}, {Lanfranchi},
  {Boeche}, {Walker}, {Johnson}, {Pe{\~n}arrubia}, \& {Gilmore}}]{Hendricks14}
{Hendricks} B., {Koch} A., {Lanfranchi} G.~A., {Boeche} C., {Walker} M.,
  {Johnson} C.~I., {Pe{\~n}arrubia} J., {Gilmore} G., 2014, \apj, 785, 102

\bibitem[{{Henry} \& {Prochaska}(2007)}]{DLAcat65}
{Henry} R.~B.~C., {Prochaska} J.~X., 2007, \pasp, 119, 962

\bibitem[{{Herbert-Fort} {et~al}\mbox{.}(2006){Herbert-Fort}, {Prochaska},
  {Dessauges-Zavadsky}, {Ellison}, {Howk}, {Wolfe}, \&
  {Prochter}}]{HerbertFort06}
{Herbert-Fort} S., {Prochaska} J.~X., {Dessauges-Zavadsky} M., {Ellison} S.~L.,
  {Howk} J.~C., {Wolfe} A.~M., {Prochter} G.~E., 2006, \pasp, 118, 1077

\bibitem[{{Israelian} \& {Rebolo}(2001)}]{Israelian01S}
{Israelian} G., {Rebolo} R., 2001, \apjl, 557, L43

\bibitem[{{Kanekar} {et~al}\mbox{.}(2014){Kanekar}, {Prochaska}, {Smette},
  {Ellison}, {Ryan-Weber}, {Momjian}, {Briggs}, {Lane}, {Chengalur},
  {Delafosse}, {Grave}, {Jacobsen}, \& {de Bruyn}}]{DLAcat101}
{Kanekar} N. {et~al.}, 2014, \mnras

\bibitem[{{Kisielius} {et~al}\mbox{.}(2015){Kisielius}, {Kulkarni}, {Ferland},
  {Bogdanovich}, {Som}, \& {Lykins}}]{Kisielius15}
{Kisielius} R., {Kulkarni} V.~P., {Ferland} G.~J., {Bogdanovich} P., {Som} D.,
  {Lykins} M.~L., 2015, ArXiv e-print 1504.01667

\bibitem[{{Kobayashi} \& {Nomoto}(2009)}]{Kobayashi09}
{Kobayashi} C., {Nomoto} K., 2009, \apj, 707, 1466

\bibitem[{{Krogager} {et~al}\mbox{.}(2013){Krogager}, {Fynbo}, {Ledoux},
  {Christensen}, {Gallazzi}, {Laursen}, {M{\o}ller}, {Noterdaeme},
  {P{\'e}roux}, {Pettini}, \& {Vestergaard}}]{Krogager13}
{Krogager} J.-K. {et~al.}, 2013, \mnras

\bibitem[{{Kulkarni} {et~al}\mbox{.}(2005){Kulkarni}, {Fall}, {Lauroesch},
  {York}, {Welty}, {Khare}, \& {Truran}}]{Kulkarni05}
{Kulkarni} V.~P., {Fall} S.~M., {Lauroesch} J.~T., {York} D.~G., {Welty} D.~E.,
  {Khare} P., {Truran} J.~W., 2005, \apj, 618, 68

\bibitem[{{Kulkarni} {et~al}\mbox{.}(1997){Kulkarni}, {Fall}, \&
  {Truran}}]{Kulkarni97}
{Kulkarni} V.~P., {Fall} S.~M., {Truran} J.~W., 1997, \apjl, 484, L7

\bibitem[{{Kulkarni} {et~al}\mbox{.}(2012){Kulkarni}, {Meiring}, {Som},
  {P{\'e}roux}, {York}, {Khare}, \& {Lauroesch}}]{Kulkarni12}
{Kulkarni} V.~P., {Meiring} J., {Som} D., {P{\'e}roux} C., {York} D.~G.,
  {Khare} P., {Lauroesch} J.~T., 2012, \apj, 749, 176

\bibitem[{{Kurucz}(1993)}]{Kurucz93}
{Kurucz} R.~L., 1993. CD-ROM 13, Smithsonian Astrophysical

\bibitem[{{Kurucz}(1998)}]{ATLAS9}
{Kurucz} R.~L., 1998. \url{http://kurucz.harvard.edu/}

\bibitem[{{Lai} {et~al}\mbox{.}(2008){Lai}, {Bolte}, {Johnson}, {Lucatello},
  {Heger}, \& {Woosley}}]{Lai08}
{Lai} D.~K., {Bolte} M., {Johnson} J.~A., {Lucatello} S., {Heger} A., {Woosley}
  S.~E., 2008, \apj, 681, 1524

\bibitem[{{Lanzetta} {et~al}\mbox{.}(1991){Lanzetta}, {Wolfe}, {Turnshek},
  {Lu}, {McMahon}, \& {Hazard}}]{Lanzetta91}
{Lanzetta} K.~M., {Wolfe} A.~M., {Turnshek} D.~A., {Lu} L., {McMahon} R.~G.,
  {Hazard} C., 1991, \apjs, 77, 1

\bibitem[{{Ledoux} {et~al}\mbox{.}(2002{\natexlab{a}}){Ledoux}, {Bergeron}, \&
  {Petitjean}}]{Ledoux02}
{Ledoux} C., {Bergeron} J., {Petitjean} P., 2002{\natexlab{a}}, \aap, 385, 802

\bibitem[{{Ledoux} {et~al}\mbox{.}(1998){Ledoux}, {Petitjean}, {Bergeron},
  {Wampler}, \& {Srianand}}]{DLAcat10}
{Ledoux} C., {Petitjean} P., {Bergeron} J., {Wampler} E.~J., {Srianand} R.,
  1998, \aap, 337, 51

\bibitem[{{Ledoux} {et~al}\mbox{.}(2006{\natexlab{a}}){Ledoux}, {Petitjean},
  {Fynbo}, {M{\o}ller}, \& {Srianand}}]{Ledoux06}
{Ledoux} C., {Petitjean} P., {Fynbo} J.~P.~U., {M{\o}ller} P., {Srianand} R.,
  2006{\natexlab{a}}, \aap, 457, 71

\bibitem[{{Ledoux} {et~al}\mbox{.}(2006{\natexlab{b}}){Ledoux}, {Petitjean}, \&
  {Srianand}}]{DLAcat57}
{Ledoux} C., {Petitjean} P., {Srianand} R., 2006{\natexlab{b}}, \apjl, 640, L25

\bibitem[{{Ledoux} {et~al}\mbox{.}(2002{\natexlab{b}}){Ledoux}, {Srianand}, \&
  {Petitjean}}]{DLAcat32}
{Ledoux} C., {Srianand} R., {Petitjean} P., 2002{\natexlab{b}}, \aap, 392, 781

\bibitem[{{Lemasle} {et~al}\mbox{.}(2012){Lemasle}, {Hill}, {Tolstoy}, {Venn},
  {Shetrone}, {Irwin}, {de Boer}, {Starkenburg}, \& {Salvadori}}]{Lemasle12}
{Lemasle} B. {et~al.}, 2012, \aap, 538, A100

\bibitem[{{Letarte} {et~al}\mbox{.}(2010){Letarte}, {Hill}, {Tolstoy},
  {Jablonka}, {Shetrone}, {Venn}, {Spite}, {Irwin}, {Battaglia}, {Helmi},
  {Primas}, {Fran{\c c}ois}, {Kaufer}, {Szeifert}, {Arimoto}, \&
  {Sadakane}}]{Letarte10}
{Letarte} B. {et~al.}, 2010, \aap, 523, A17

\bibitem[{{Levshakov} {et~al}\mbox{.}(2002){Levshakov}, {Dessauges-Zavadsky},
  {D'Odorico}, \& {Molaro}}]{DLAcat34}
{Levshakov} S.~A., {Dessauges-Zavadsky} M., {D'Odorico} S., {Molaro} P., 2002,
  \apj, 565, 696

\bibitem[{{Lopez} \& {Ellison}(2003)}]{DLAcat41}
{Lopez} S., {Ellison} S.~L., 2003, \aap, 403, 573

\bibitem[{{Lopez} {et~al}\mbox{.}(2002){Lopez}, {Reimers}, {D'Odorico}, \&
  {Prochaska}}]{Lopez02}
{Lopez} S., {Reimers} D., {D'Odorico} S., {Prochaska} J.~X., 2002, \aap, 385,
  778

\bibitem[{{Lopez} {et~al}\mbox{.}(2005){Lopez}, {Reimers}, {Gregg}, {Wisotzki},
  {Wucknitz}, \& {Guzman}}]{DLAcat49}
{Lopez} S., {Reimers} D., {Gregg} M.~D., {Wisotzki} L., {Wucknitz} O., {Guzman}
  A., 2005, \apj, 626, 767

\bibitem[{{Lopez} {et~al}\mbox{.}(1999){Lopez}, {Reimers}, {Rauch}, {Sargent},
  \& {Smette}}]{DLAcat12}
{Lopez} S., {Reimers} D., {Rauch} M., {Sargent} W.~L.~W., {Smette} A., 1999,
  \apj, 513, 598

\bibitem[{{Lu} {et~al}\mbox{.}(1998){Lu}, {Sargent}, \& {Barlow}}]{Lu98}
{Lu} L., {Sargent} W.~L.~W., {Barlow} T.~A., 1998, \aj, 115, 55

\bibitem[{{Lu} {et~al}\mbox{.}(1996{\natexlab{a}}){Lu}, {Sargent}, {Barlow},
  {Churchill}, \& {Vogt}}]{Lu96}
{Lu} L., {Sargent} W.~L.~W., {Barlow} T.~A., {Churchill} C.~W., {Vogt} S.~S.,
  1996{\natexlab{a}}, \apjs, 107, 475

\bibitem[{{Lu} {et~al}\mbox{.}(1996{\natexlab{b}}){Lu}, {Sargent}, {Womble}, \&
  {Barlow}}]{DLAcat4}
{Lu} L., {Sargent} W.~L.~W., {Womble} D.~S., {Barlow} T.~A.,
  1996{\natexlab{b}}, \apjl, 457, L1

\bibitem[{{Mashonkina} {et~al}\mbox{.}(2011){Mashonkina}, {Gehren}, {Shi},
  {Korn}, \& {Grupp}}]{Mashonkina11}
{Mashonkina} L., {Gehren} T., {Shi} J.-R., {Korn} A.~J., {Grupp} F., 2011,
  \aap, 528, A87

\bibitem[{{McConnachie}(2012)}]{Mcconnachie12}
{McConnachie} A.~W., 2012, \aj, 144, 4

\bibitem[{{McWilliam}(1997)}]{McWilliam97}
{McWilliam} A., 1997, \araa, 35, 503

\bibitem[{{McWilliam} {et~al}\mbox{.}(2003){McWilliam}, {Rich}, \&
  {Smecker-Hane}}]{McWilliam03}
{McWilliam} A., {Rich} R.~M., {Smecker-Hane} T.~A., 2003, \apjl, 592, L21

\bibitem[{{McWilliam} {et~al}\mbox{.}(2013){McWilliam}, {Wallerstein}, \&
  {Mottini}}]{McWilliam13}
{McWilliam} A., {Wallerstein} G., {Mottini} M., 2013, \apj, 778, 149

\bibitem[{{Meiring} {et~al}\mbox{.}(2006){Meiring}, {Kulkarni}, {Khare},
  {Bechtold}, {York}, {Cui}, {Lauroesch}, {Crotts}, \& {Nakamura}}]{DLAcat60}
{Meiring} J.~D. {et~al.}, 2006, \mnras, 370, 43

\bibitem[{{Meiring} {et~al}\mbox{.}(2009){Meiring}, {Kulkarni}, {Lauroesch},
  {P{\'e}roux}, {Khare}, \& {York}}]{DLAcat74}
{Meiring} J.~D., {Kulkarni} V.~P., {Lauroesch} J.~T., {P{\'e}roux} C., {Khare}
  P., {York} D.~G., 2009, \mnras, 393, 1513

\bibitem[{{Meiring} {et~al}\mbox{.}(2007){Meiring}, {Lauroesch}, {Kulkarni},
  {P{\'e}roux}, {Khare}, {York}, \& {Crotts}}]{DLAcat67}
{Meiring} J.~D., {Lauroesch} J.~T., {Kulkarni} V.~P., {P{\'e}roux} C., {Khare}
  P., {York} D.~G., {Crotts} A.~P.~S., 2007, \mnras, 376, 557

\bibitem[{{Meiring} {et~al}\mbox{.}(2011){Meiring}, {Tripp}, {Prochaska},
  {Tumlinson}, {Werk}, {Jenkins}, {Thom}, {O'Meara}, \& {Sembach}}]{DLAcat73}
{Meiring} J.~D. {et~al.}, 2011, \apj, 732, 35

\bibitem[{{Meyer} {et~al}\mbox{.}(1995){Meyer}, {Lanzetta}, \&
  {Wolfe}}]{Meyer95}
{Meyer} D.~M., {Lanzetta} K.~M., {Wolfe} A.~M., 1995, \apjl, 451, L13

\bibitem[{{Meyer} \& {Roth}(1990)}]{Meyer90}
{Meyer} D.~M., {Roth} K.~C., 1990, \apj, 363, 57

\bibitem[{{Meyer} \& {York}(1992)}]{Meyer92}
{Meyer} D.~M., {York} D.~G., 1992, \apjl, 399, L121

\bibitem[{{Molaro} {et~al}\mbox{.}(2000){Molaro}, {Bonifacio}, {Centuri{\'o}n},
  {D'Odorico}, {Vladilo}, {Santin}, \& {Di Marcantonio}}]{DLAcat21}
{Molaro} P., {Bonifacio} P., {Centuri{\'o}n} M., {D'Odorico} S., {Vladilo} G.,
  {Santin} P., {Di Marcantonio} P., 2000, \apj, 541, 54

\bibitem[{{Molaro} {et~al}\mbox{.}(2001){Molaro}, {Levshakov}, {D'Odorico},
  {Bonifacio}, \& {Centuri{\'o}n}}]{Molaro01}
{Molaro} P., {Levshakov} S.~A., {D'Odorico} S., {Bonifacio} P., {Centuri{\'o}n}
  M., 2001, \apj, 549, 90

\bibitem[{{Morton}(2003)}]{Morton03}
{Morton} D.~C., 2003, \apjs, 149, 205

\bibitem[{{Neeleman} {et~al}\mbox{.}(2013){Neeleman}, {Wolfe}, {Prochaska}, \&
  {Rafelski}}]{Neeleman13}
{Neeleman} M., {Wolfe} A.~M., {Prochaska} J.~X., {Rafelski} M., 2013, \apj,
  769, 54

\bibitem[{{Nestor} {et~al}\mbox{.}(2008){Nestor}, {Pettini}, {Hewett}, {Rao},
  \& {Wild}}]{DLAcat78}
{Nestor} D.~B., {Pettini} M., {Hewett} P.~C., {Rao} S., {Wild} V., 2008,
  \mnras, 390, 1670

\bibitem[{{Nissen} {et~al}\mbox{.}(2007){Nissen}, {Akerman}, {Asplund},
  {Fabbian}, {Kerber}, {Kaufl}, \& {Pettini}}]{Nissen07}
{Nissen} P.~E., {Akerman} C., {Asplund} M., {Fabbian} D., {Kerber} F., {Kaufl}
  H.~U., {Pettini} M., 2007, \aap, 469, 319

\bibitem[{{Nissen} {et~al}\mbox{.}(2004){Nissen}, {Chen}, {Asplund}, \&
  {Pettini}}]{Nissen04}
{Nissen} P.~E., {Chen} Y.~Q., {Asplund} M., {Pettini} M., 2004, \aap, 415, 993

\bibitem[{{Nissen} {et~al}\mbox{.}(2000){Nissen}, {Chen}, {Schuster}, \&
  {Zhao}}]{Nissen00}
{Nissen} P.~E., {Chen} Y.~Q., {Schuster} W.~J., {Zhao} G., 2000, \aap, 353, 722

\bibitem[{{Nissen} \& {Schuster}(2010)}]{Nissen10}
{Nissen} P.~E., {Schuster} W.~J., 2010, \aap, 511, L10

\bibitem[{{Nissen} \& {Schuster}(2011)}]{Nissen11}
{Nissen} P.~E., {Schuster} W.~J., 2011, \aap, 530, A15

\bibitem[{{Nomoto} {et~al}\mbox{.}(2013){Nomoto}, {Kobayashi}, \&
  {Tominaga}}]{Nomoto13}
{Nomoto} K., {Kobayashi} C., {Tominaga} N., 2013, \araa, 51, 457

\bibitem[{{North} {et~al}\mbox{.}(2012){North}, {Cescutti}, {Jablonka}, {Hill},
  {Shetrone}, {Letarte}, {Lemasle}, {Venn}, {Battaglia}, {Tolstoy}, {Irwin},
  {Primas}, \& {Fran{\c c}ois}}]{North12}
{North} P. {et~al.}, 2012, \aap, 541, A45

\bibitem[{{Noterdaeme} {et~al}\mbox{.}(2012{\natexlab{a}}){Noterdaeme},
  {Laursen}, {Petitjean}, {Vergani}, {Maureira}, {Ledoux}, {Fynbo},
  {L{\'o}pez}, \& {Srianand}}]{DLAcat92}
{Noterdaeme} P. {et~al.}, 2012{\natexlab{a}}, \aap, 540, A63

\bibitem[{{Noterdaeme} {et~al}\mbox{.}(2007{\natexlab{a}}){Noterdaeme},
  {Ledoux}, {Petitjean}, {Le Petit}, {Srianand}, \& {Smette}}]{DLAcat62}
{Noterdaeme} P., {Ledoux} C., {Petitjean} P., {Le Petit} F., {Srianand} R.,
  {Smette} A., 2007{\natexlab{a}}, \aap, 474, 393

\bibitem[{{Noterdaeme} {et~al}\mbox{.}(2008){Noterdaeme}, {Ledoux},
  {Petitjean}, \& {Srianand}}]{DLAcat70}
{Noterdaeme} P., {Ledoux} C., {Petitjean} P., {Srianand} R., 2008, \aap, 481,
  327

\bibitem[{{Noterdaeme} {et~al}\mbox{.}(2012{\natexlab{b}}){Noterdaeme},
  {L{\'o}pez}, {Dumont}, {Ledoux}, {Molaro}, \& {Petitjean}}]{DLAcat94}
{Noterdaeme} P., {L{\'o}pez} S., {Dumont} V., {Ledoux} C., {Molaro} P.,
  {Petitjean} P., 2012{\natexlab{b}}, \aap, 542, L33

\bibitem[{{Noterdaeme} {et~al}\mbox{.}(2012{\natexlab{c}}){Noterdaeme},
  {Petitjean}, {Carithers}, {P{\^a}ris}, {Font-Ribera}, {Bailey}, {Aubourg},
  {Bizyaev}, {Ebelke}, {Finley}, {Ge}, {Malanushenko}, {Malanushenko},
  {Miralda-Escud{\'e}}, {Myers}, {Oravetz}, {Pan}, {Pieri}, {Ross},
  {Schneider}, {Simmons}, \& {York}}]{Noterdaeme12}
{Noterdaeme} P. {et~al.}, 2012{\natexlab{c}}, \aap, 547, L1

\bibitem[{{Noterdaeme} {et~al}\mbox{.}(2007{\natexlab{b}}){Noterdaeme},
  {Petitjean}, {Srianand}, {Ledoux}, \& {Le Petit}}]{DLAcat63}
{Noterdaeme} P., {Petitjean} P., {Srianand} R., {Ledoux} C., {Le Petit} F.,
  2007{\natexlab{b}}, \aap, 469, 425

\bibitem[{{Penprase} {et~al}\mbox{.}(2010){Penprase}, {Prochaska}, {Sargent},
  {Toro-Martinez}, \& {Beeler}}]{Penprase10}
{Penprase} B.~E., {Prochaska} J.~X., {Sargent} W.~L.~W., {Toro-Martinez} I.,
  {Beeler} D.~J., 2010, \apj, 721, 1

\bibitem[{{P{\'e}roux} {et~al}\mbox{.}(2011){P{\'e}roux}, {Bouch{\'e}},
  {Kulkarni}, {York}, \& {Vladilo}}]{Peroux11}
{P{\'e}roux} C., {Bouch{\'e}} N., {Kulkarni} V.~P., {York} D.~G., {Vladilo} G.,
  2011, \mnras, 410, 2237

\bibitem[{{P{\'e}roux} {et~al}\mbox{.}(2006){P{\'e}roux}, {Meiring},
  {Kulkarni}, {Ferlet}, {Khare}, {Lauroesch}, {Vladilo}, \& {York}}]{DLAcat61}
{P{\'e}roux} C., {Meiring} J.~D., {Kulkarni} V.~P., {Ferlet} R., {Khare} P.,
  {Lauroesch} J.~T., {Vladilo} G., {York} D.~G., 2006, \mnras, 372, 369

\bibitem[{{P{\'e}roux} {et~al}\mbox{.}(2008){P{\'e}roux}, {Meiring},
  {Kulkarni}, {Khare}, {Lauroesch}, {Vladilo}, \& {York}}]{Peroux08}
{P{\'e}roux} C., {Meiring} J.~D., {Kulkarni} V.~P., {Khare} P., {Lauroesch}
  J.~T., {Vladilo} G., {York} D.~G., 2008, \mnras, 386, 2209

\bibitem[{{P{\'e}roux} {et~al}\mbox{.}(2002){P{\'e}roux}, {Petitjean},
  {Aracil}, \& {Srianand}}]{DLAcat36}
{P{\'e}roux} C., {Petitjean} P., {Aracil} B., {Srianand} R., 2002, \na, 7, 577

\bibitem[{{Petitjean} {et~al}\mbox{.}(2008){Petitjean}, {Ledoux}, \&
  {Srianand}}]{Petitjean08}
{Petitjean} P., {Ledoux} C., {Srianand} R., 2008, \aap, 480, 349

\bibitem[{{Petitjean} {et~al}\mbox{.}(2000){Petitjean}, {Srianand}, \&
  {Ledoux}}]{DLAcat18}
{Petitjean} P., {Srianand} R., {Ledoux} C., 2000, \aap, 364, L26

\bibitem[{{Petitjean} {et~al}\mbox{.}(2002){Petitjean}, {Srianand}, \&
  {Ledoux}}]{DLAcat33}
{Petitjean} P., {Srianand} R., {Ledoux} C., 2002, \mnras, 332, 383

\bibitem[{{Pettini} {et~al}\mbox{.}(1990){Pettini}, {Boksenberg}, \&
  {Hunstead}}]{Pettini90}
{Pettini} M., {Boksenberg} A., {Hunstead} R.~W., 1990, \apj, 348, 48

\bibitem[{{Pettini} \& {Cooke}(2012)}]{DLAcat93}
{Pettini} M., {Cooke} R., 2012, \mnras, 425, 2477

\bibitem[{{Pettini} {et~al}\mbox{.}(2002){Pettini}, {Ellison}, {Bergeron}, \&
  {Petitjean}}]{Pettini02}
{Pettini} M., {Ellison} S.~L., {Bergeron} J., {Petitjean} P., 2002, \aap, 391,
  21

\bibitem[{{Pettini} {et~al}\mbox{.}(1999){Pettini}, {Ellison}, {Steidel}, \&
  {Bowen}}]{Pettini99}
{Pettini} M., {Ellison} S.~L., {Steidel} C.~C., {Bowen} D.~V., 1999, \apj, 510,
  576

\bibitem[{{Pettini} {et~al}\mbox{.}(2000){Pettini}, {Ellison}, {Steidel},
  {Shapley}, \& {Bowen}}]{Pettini00}
{Pettini} M., {Ellison} S.~L., {Steidel} C.~C., {Shapley} A.~E., {Bowen} D.~V.,
  2000, \apj, 532, 65

\bibitem[{{Pettini} {et~al}\mbox{.}(1995){Pettini}, {Lipman}, \&
  {Hunstead}}]{Pettini95}
{Pettini} M., {Lipman} K., {Hunstead} R.~W., 1995, \apj, 451, 100

\bibitem[{{Pettini} {et~al}\mbox{.}(1994){Pettini}, {Smith}, {Hunstead}, \&
  {King}}]{Pettini94}
{Pettini} M., {Smith} L.~J., {Hunstead} R.~W., {King} D.~L., 1994, \apj, 426,
  79

\bibitem[{{Pettini} {et~al}\mbox{.}(1997){Pettini}, {Smith}, {King}, \&
  {Hunstead}}]{Pettini97}
{Pettini} M., {Smith} L.~J., {King} D.~L., {Hunstead} R.~W., 1997, \apj, 486,
  665

\bibitem[{{Pettini} {et~al}\mbox{.}(2008){Pettini}, {Zych}, {Steidel}, \&
  {Chaffee}}]{Pettini08}
{Pettini} M., {Zych} B.~J., {Steidel} C.~C., {Chaffee} F.~H., 2008, \mnras,
  385, 2011

\bibitem[{{Pignatari} {et~al}\mbox{.}(2013){Pignatari}, {Herwig}, {Hirschi},
  {Bennett}, {Rockefeller}, {Fryer}, {Timmes}, {Heger}, {Jones}, {Battino},
  {Ritter}, {Dotter}, {Trappitsch}, {Diehl}, {Frischknecht}, {Hungerford},
  {Magkotsios}, {Travaglio}, \& {Young}}]{Pignatari13}
{Pignatari} M. {et~al.}, 2013, ArXiv e-print 1307.6961

\bibitem[{{Pomp{\'e}ia} {et~al}\mbox{.}(2008){Pomp{\'e}ia}, {Hill}, {Spite},
  {Cole}, {Primas}, {Romaniello}, {Pasquini}, {Cioni}, \& {Smecker
  Hane}}]{Pompeia08}
{Pomp{\'e}ia} L. {et~al.}, 2008, \aap, 480, 379

\bibitem[{{Preston} {et~al}\mbox{.}(2006){Preston}, {Sneden}, {Thompson},
  {Shectman}, \& {Burley}}]{Preston06}
{Preston} G.~W., {Sneden} C., {Thompson} I.~B., {Shectman} S.~A., {Burley}
  G.~S., 2006, \aj, 132, 85

\bibitem[{{Pritzl} {et~al}\mbox{.}(2005){Pritzl}, {Venn}, \&
  {Irwin}}]{Pritzl05}
{Pritzl} B.~J., {Venn} K.~A., {Irwin} M., 2005, \aj, 130, 2140

\bibitem[{{Prochaska} {et~al}\mbox{.}(2003{\natexlab{a}}){Prochaska}, {Castro},
  \& {Djorgovski}}]{DLAcat44}
{Prochaska} J.~X., {Castro} S., {Djorgovski} S.~G., 2003{\natexlab{a}}, \apjs,
  148, 317

\bibitem[{{Prochaska} {et~al}\mbox{.}(2001{\natexlab{a}}){Prochaska},
  {Gawiser}, \& {Wolfe}}]{Prochaska01}
{Prochaska} J.~X., {Gawiser} E., {Wolfe} A.~M., 2001{\natexlab{a}}, \apj, 552,
  99

\bibitem[{{Prochaska} {et~al}\mbox{.}(2003{\natexlab{b}}){Prochaska},
  {Gawiser}, {Wolfe}, {Castro}, \& {Djorgovski}}]{Prochaska03ApJ595}
{Prochaska} J.~X., {Gawiser} E., {Wolfe} A.~M., {Castro} S., {Djorgovski}
  S.~G., 2003{\natexlab{b}}, \apjl, 595, L9

\bibitem[{{Prochaska} {et~al}\mbox{.}(2003{\natexlab{c}}){Prochaska},
  {Gawiser}, {Wolfe}, {Cooke}, \& {Gelino}}]{Prochaska03ApJS147}
{Prochaska} J.~X., {Gawiser} E., {Wolfe} A.~M., {Cooke} J., {Gelino} D.,
  2003{\natexlab{c}}, \apjs, 147, 227

\bibitem[{{Prochaska} {et~al}\mbox{.}(2002){Prochaska}, {Henry}, {O'Meara},
  {Tytler}, {Wolfe}, {Kirkman}, {Lubin}, \& {Suzuki}}]{DLAcat35}
{Prochaska} J.~X., {Henry} R.~B.~C., {O'Meara} J.~M., {Tytler} D., {Wolfe}
  A.~M., {Kirkman} D., {Lubin} D., {Suzuki} N., 2002, \pasp, 114, 933

\bibitem[{{Prochaska} {et~al}\mbox{.}(2003{\natexlab{d}}){Prochaska}, {Howk},
  \& {Wolfe}}]{Prochaska03}
{Prochaska} J.~X., {Howk} J.~C., {Wolfe} A.~M., 2003{\natexlab{d}}, \nat, 423,
  57

\bibitem[{{Prochaska} \& {McWilliam}(2000)}]{Prochaska00Mn}
{Prochaska} J.~X., {McWilliam} A., 2000, \apjl, 537, L57

\bibitem[{{Prochaska} \& {Wolfe}(1996)}]{Prochaska96}
{Prochaska} J.~X., {Wolfe} A.~M., 1996, \apj, 470, 403

\bibitem[{{Prochaska} \& {Wolfe}(1997{\natexlab{a}})}]{DLAcat8}
{Prochaska} J.~X., {Wolfe} A.~M., 1997{\natexlab{a}}, \apj, 474, 140

\bibitem[{{Prochaska} \& {Wolfe}(1997{\natexlab{b}})}]{Prochaska97}
{Prochaska} J.~X., {Wolfe} A.~M., 1997{\natexlab{b}}, \apj, 487, 73

\bibitem[{{Prochaska} \& {Wolfe}(1999)}]{Prochaska99}
{Prochaska} J.~X., {Wolfe} A.~M., 1999, \apjs, 121, 369

\bibitem[{{Prochaska} \& {Wolfe}(2000)}]{DLAcat22}
{Prochaska} J.~X., {Wolfe} A.~M., 2000, \apjl, 533, L5

\bibitem[{{Prochaska} \& {Wolfe}(2002)}]{Prochaska02II}
{Prochaska} J.~X., {Wolfe} A.~M., 2002, \apj, 566, 68

\bibitem[{{Prochaska} {et~al}\mbox{.}(2007){Prochaska}, {Wolfe}, {Howk},
  {Gawiser}, {Burles}, \& {Cooke}}]{Prochaska07}
{Prochaska} J.~X., {Wolfe} A.~M., {Howk} J.~C., {Gawiser} E., {Burles} S.~M.,
  {Cooke} J., 2007, \apjs, 171, 29

\bibitem[{{Prochaska} {et~al}\mbox{.}(2001{\natexlab{b}}){Prochaska}, {Wolfe},
  {Tytler}, {Burles}, {Cooke}, {Gawiser}, {Kirkman}, {O'Meara}, \&
  {Storrie-Lombardi}}]{Prochaska01I}
{Prochaska} J.~X. {et~al.}, 2001{\natexlab{b}}, \apjs, 137, 21

\bibitem[{{Pruet} {et~al}\mbox{.}(2005){Pruet}, {Woosley}, {Buras}, {Janka}, \&
  {Hoffman}}]{Pruet05}
{Pruet} J., {Woosley} S.~E., {Buras} R., {Janka} H.-T., {Hoffman} R.~D., 2005,
  \apj, 623, 325

\bibitem[{{Rafelski} {et~al}\mbox{.}(2014){Rafelski}, {Neeleman}, {Fumagalli},
  {Wolfe}, \& {Prochaska}}]{Rafelski14}
{Rafelski} M., {Neeleman} M., {Fumagalli} M., {Wolfe} A.~M., {Prochaska} J.~X.,
  2014, \apjl, 782, L29

\bibitem[{{Rafelski} {et~al}\mbox{.}(2012){Rafelski}, {Wolfe}, {Prochaska},
  {Neeleman}, \& {Mendez}}]{Rafelski12}
{Rafelski} M., {Wolfe} A.~M., {Prochaska} J.~X., {Neeleman} M., {Mendez} A.~J.,
  2012, \apj, 755, 89

\bibitem[{{Rahmati} \& {Schaye}(2014)}]{Rahmati14}
{Rahmati} A., {Schaye} J., 2014, \mnras, 438, 529

\bibitem[{{Rao} {et~al}\mbox{.}(2005){Rao}, {Prochaska}, {Howk}, \&
  {Wolfe}}]{DLAcat53}
{Rao} S.~M., {Prochaska} J.~X., {Howk} J.~C., {Wolfe} A.~M., 2005, \aj, 129, 9

\bibitem[{{Rao} \& {Turnshek}(2000)}]{DLAcat17}
{Rao} S.~M., {Turnshek} D.~A., 2000, \apjs, 130, 1

\bibitem[{{Reddy} {et~al}\mbox{.}(2006){Reddy}, {Lambert}, \& {Allende
  Prieto}}]{Reddy06}
{Reddy} B.~E., {Lambert} D.~L., {Allende Prieto} C., 2006, \mnras, 367, 1329

\bibitem[{{Reddy} {et~al}\mbox{.}(2003){Reddy}, {Tomkin}, {Lambert}, \&
  {Allende Prieto}}]{Reddy03}
{Reddy} B.~E., {Tomkin} J., {Lambert} D.~L., {Allende Prieto} C., 2003, VizieR
  Online Data Catalog, 734, 304

\bibitem[{{Roth} \& {Blades}(1995)}]{Roth95}
{Roth} K.~C., {Blades} J.~C., 1995, \apjl, 445, L95

\bibitem[{{Sargent} {et~al}\mbox{.}(1989){Sargent}, {Steidel}, \&
  {Boksenberg}}]{Sargent89}
{Sargent} W.~L.~W., {Steidel} C.~C., {Boksenberg} A., 1989, \apjs, 69, 703

\bibitem[{{Savage} \& {Sembach}(1991)}]{Savage91}
{Savage} B.~D., {Sembach} K.~R., 1991, \apj, 379, 245

\bibitem[{{Savage} \& {Sembach}(1996)}]{Savage96}
{Savage} B.~D., {Sembach} K.~R., 1996, \araa, 34, 279

\bibitem[{{Savaglio} {et~al}\mbox{.}(1994){Savaglio}, {D'Odorico}, \&
  {Moller}}]{Savaglio94}
{Savaglio} S., {D'Odorico} S., {Moller} P., 1994, \aap, 281, 331

\bibitem[{{Sbordone} {et~al}\mbox{.}(2007){Sbordone}, {Bonifacio}, {Buonanno},
  {Marconi}, {Monaco}, \& {Zaggia}}]{Sbordone07}
{Sbordone} L., {Bonifacio} P., {Buonanno} R., {Marconi} G., {Monaco} L.,
  {Zaggia} S., 2007, \aap, 465, 815

\bibitem[{{Searle} \& {Zinn}(1978)}]{Searle78}
{Searle} L., {Zinn} R., 1978, \apj, 225, 357

\bibitem[{{Seitenzahl} {et~al}\mbox{.}(2013){Seitenzahl},
  {Ciaraldi-Schoolmann}, {R{\"o}pke}, {Fink}, {Hillebrandt}, {Kromer},
  {Pakmor}, {Ruiter}, {Sim}, \& {Taubenberger}}]{Seitenzahl13}
{Seitenzahl} I.~R. {et~al.}, 2013, \mnras, 429, 1156

\bibitem[{{Sembach} {et~al}\mbox{.}(1995){Sembach}, {Steidel}, {Macke}, \&
  {Meyer}}]{Sembach95}
{Sembach} K.~R., {Steidel} C.~C., {Macke} R.~J., {Meyer} D.~M., 1995, \apjl,
  445, L27

\bibitem[{{Shetrone} {et~al}\mbox{.}(2003){Shetrone}, {Venn}, {Tolstoy},
  {Primas}, {Hill}, \& {Kaufer}}]{Shetrone03}
{Shetrone} M., {Venn} K.~A., {Tolstoy} E., {Primas} F., {Hill} V., {Kaufer} A.,
  2003, \aj, 125, 684

\bibitem[{{Shi} {et~al}\mbox{.}(2009){Shi}, {Gehren}, {Mashonkina}, \&
  {Zhao}}]{Shi09}
{Shi} J.~R., {Gehren} T., {Mashonkina} L., {Zhao} G., 2009, \aap, 503, 533

\bibitem[{{Shi} {et~al}\mbox{.}(2012){Shi}, {Takada-Hidai}, {Takeda}, {Tan},
  {Hu}, {Zhao}, \& {Cao}}]{Shi12}
{Shi} J.~R., {Takada-Hidai} M., {Takeda} Y., {Tan} K.~F., {Hu} S.~M., {Zhao}
  G., {Cao} C., 2012, \apj, 755, 36

\bibitem[{{Skuladottir} {et~al}\mbox{.}(2015){Skuladottir}, {Andrievsky},
  {Tolstoy}, {Hill}, {Salvadori}, {Korotin}, \& {Pettini}}]{Skuladottir15}
{Skuladottir} A., {Andrievsky} S.~M., {Tolstoy} E., {Hill} V., {Salvadori} S.,
  {Korotin} S.~A., {Pettini} M., 2015, ArXiv e-print 1505.03155

\bibitem[{{Sneden}(1973)}]{MOOG}
{Sneden} C., 1973, \apj, 184, 839

\bibitem[{{Sneden} \& {Crocker}(1988)}]{Sneden88}
{Sneden} C., {Crocker} D.~A., 1988, \apj, 335, 406

\bibitem[{{Sneden} {et~al}\mbox{.}(1991){Sneden}, {Gratton}, \&
  {Crocker}}]{Sneden91}
{Sneden} C., {Gratton} R.~G., {Crocker} D.~A., 1991, \aap, 246, 354

\bibitem[{{Songaila} \& {Cowie}(2002)}]{DLAcat37}
{Songaila} A., {Cowie} L.~L., 2002, \aj, 123, 2183

\bibitem[{{Spite} {et~al}\mbox{.}(2011){Spite}, {Caffau}, {Andrievsky},
  {Korotin}, {Depagne}, {Spite}, {Bonifacio}, {Ludwig}, {Cayrel}, {Fran{\c
  c}ois}, {Hill}, {Plez}, {Andersen}, {Barbuy}, {Beers}, {Molaro},
  {Nordstr{\"o}m}, \& {Primas}}]{Spite11}
{Spite} M. {et~al.}, 2011, \aap, 528, A9

\bibitem[{{Srianand} {et~al}\mbox{.}(2012){Srianand}, {Gupta}, {Petitjean},
  {Noterdaeme}, {Ledoux}, {Salter}, \& {Saikia}}]{DLAcat97}
{Srianand} R., {Gupta} N., {Petitjean} P., {Noterdaeme} P., {Ledoux} C.,
  {Salter} C.~J., {Saikia} D.~J., 2012, \mnras, 421, 651

\bibitem[{{Srianand} {et~al}\mbox{.}(2000){Srianand}, {Petitjean}, \&
  {Ledoux}}]{DLAcat19}
{Srianand} R., {Petitjean} P., {Ledoux} C., 2000, \nat, 408, 931

\bibitem[{{Srianand} {et~al}\mbox{.}(2005){Srianand}, {Petitjean}, {Ledoux},
  {Ferland}, \& {Shaw}}]{Srianand05}
{Srianand} R., {Petitjean} P., {Ledoux} C., {Ferland} G., {Shaw} G., 2005,
  \mnras, 362, 549

\bibitem[{{Starkenburg} {et~al}\mbox{.}(2013){Starkenburg}, {Hill}, {Tolstoy},
  {Fran{\c c}ois}, {Irwin}, {Boschman}, {Venn}, {de Boer}, {Lemasle},
  {Jablonka}, {Battaglia}, {Groot}, \& {Kaper}}]{Starkenburg13}
{Starkenburg} E. {et~al.}, 2013, \aap, 549, A88

\bibitem[{{Steidel} {et~al}\mbox{.}(1997){Steidel}, {Dickinson}, {Meyer},
  {Adelberger}, \& {Sembach}}]{DLAcat6}
{Steidel} C.~C., {Dickinson} M., {Meyer} D.~M., {Adelberger} K.~L., {Sembach}
  K.~R., 1997, \apj, 480, 568

\bibitem[{{Stephens} \& {Boesgaard}(2002)}]{Stephens02}
{Stephens} A., {Boesgaard} A.~M., 2002, \aj, 123, 1647

\bibitem[{{Tafelmeyer} {et~al}\mbox{.}(2010){Tafelmeyer}, {Jablonka}, {Hill},
  {Shetrone}, {Tolstoy}, {Irwin}, {Battaglia}, {Helmi}, {Starkenburg}, {Venn},
  {Abel}, {Francois}, {Kaufer}, {North}, {Primas}, \&
  {Szeifert}}]{Tafelmeyer10}
{Tafelmeyer} M. {et~al.}, 2010, \aap, 524, A58

\bibitem[{{Takada-Hidai} {et~al}\mbox{.}(2002){Takada-Hidai}, {Takeda}, {Sato},
  {Honda}, {Sadakane}, {Kawanomoto}, {Sargent}, {Lu}, \& {Barlow}}]{Takada02}
{Takada-Hidai} M. {et~al.}, 2002, \apj, 573, 614

\bibitem[{{Takeda} {et~al}\mbox{.}(2005){Takeda}, {Hashimoto}, {Taguchi},
  {Yoshioka}, {Takada-Hidai}, {Saito}, \& {Honda}}]{Takeda05}
{Takeda} Y., {Hashimoto} O., {Taguchi} H., {Yoshioka} K., {Takada-Hidai} M.,
  {Saito} Y., {Honda} S., 2005, \pasj, 57, 751

\bibitem[{{Tchernyshyov} {et~al}\mbox{.}(2015){Tchernyshyov}, {Meixner},
  {Seale}, {Fox}, {Friedman}, {Dwek}, {Galliano}, \&
  {Sembach}}]{Tchernyshyov15}
{Tchernyshyov} K., {Meixner} M., {Seale} J., {Fox} A., {Friedman} S.~D., {Dwek}
  E., {Galliano} F., {Sembach} K., 2015, ArXiv e-print 1503.08852

\bibitem[{{Tolstoy} {et~al}\mbox{.}(2009){Tolstoy}, {Hill}, \&
  {Tosi}}]{Tolstoy09}
{Tolstoy} E., {Hill} V., {Tosi} M., 2009, \araa, 47, 371

\bibitem[{{Travaglio} {et~al}\mbox{.}(2004){Travaglio}, {Hillebrandt},
  {Reinecke}, \& {Thielemann}}]{Travaglio04}
{Travaglio} C., {Hillebrandt} W., {Reinecke} M., {Thielemann} F.-K., 2004,
  \aap, 425, 1029

\bibitem[{{Turnshek} {et~al}\mbox{.}(2004){Turnshek}, {Rao}, {Nestor}, {Vanden
  Berk}, {Belfort-Mihalyi}, \& {Monier}}]{DLAcat48}
{Turnshek} D.~A., {Rao} S.~M., {Nestor} D.~B., {Vanden Berk} D.,
  {Belfort-Mihalyi} M., {Monier} E.~M., 2004, \apjl, 609, L53

\bibitem[{{Umeda} \& {Nomoto}(2002)}]{Umeda02}
{Umeda} H., {Nomoto} K., 2002, \apj, 565, 385

\bibitem[{{Venn} {et~al}\mbox{.}(2004){Venn}, {Irwin}, {Shetrone}, {Tout},
  {Hill}, \& {Tolstoy}}]{Venn04}
{Venn} K.~A., {Irwin} M., {Shetrone} M.~D., {Tout} C.~A., {Hill} V., {Tolstoy}
  E., 2004, \aj, 128, 1177

\bibitem[{{Venn} {et~al}\mbox{.}(2012){Venn}, {Shetrone}, {Irwin}, {Hill},
  {Jablonka}, {Tolstoy}, {Lemasle}, {Divell}, {Starkenburg}, {Letarte},
  {Baldner}, {Battaglia}, {Helmi}, {Kaufer}, \& {Primas}}]{Venn12}
{Venn} K.~A. {et~al.}, 2012, \apj, 751, 102

\bibitem[{{Vladilo}(2002{\natexlab{a}})}]{Vladilo02a}
{Vladilo} G., 2002{\natexlab{a}}, \apj, 569, 295

\bibitem[{{Vladilo}(2002{\natexlab{b}})}]{Vladilo02b}
{Vladilo} G., 2002{\natexlab{b}}, \aap, 391, 407

\bibitem[{{Vladilo} {et~al}\mbox{.}(2011){Vladilo}, {Abate}, {Yin}, {Cescutti},
  \& {Matteucci}}]{Vladilo11}
{Vladilo} G., {Abate} C., {Yin} J., {Cescutti} G., {Matteucci} F., 2011, \aap,
  530, A33

\bibitem[{{Vogt} {et~al}\mbox{.}(1994){Vogt}, {Allen}, {Bigelow}, {Bresee},
  {Brown}, {Cantrall}, {Conrad}, {Couture}, {Delaney}, {Epps}, {Hilyard},
  {Hilyard}, {Horn}, {Jern}, {Kanto}, {Keane}, {Kibrick}, {Lewis}, {Osborne},
  {Pardeilhan}, {Pfister}, {Ricketts}, {Robinson}, {Stover}, {Tucker}, {Ward},
  \& {Wei}}]{Vogt94}
{Vogt} S.~S. {et~al.}, 1994, in Society of Photo-Optical Instrumentation
  Engineers (SPIE) Conference Series, Vol. 2198, Instrumentation in Astronomy
  VIII, {Crawford} D.~L., {Craine} E.~R., eds., p. 362

\bibitem[{{Wolfe} {et~al}\mbox{.}(2005){Wolfe}, {Gawiser}, \&
  {Prochaska}}]{Wolfe05}
{Wolfe} A.~M., {Gawiser} E., {Prochaska} J.~X., 2005, \araa, 43, 861

\bibitem[{{Wolfe} {et~al}\mbox{.}(1995){Wolfe}, {Lanzetta}, {Foltz}, \&
  {Chaffee}}]{Wolfe95}
{Wolfe} A.~M., {Lanzetta} K.~M., {Foltz} C.~B., {Chaffee} F.~H., 1995, \apj,
  454, 698

\bibitem[{{Wolfe} {et~al}\mbox{.}(2003){Wolfe}, {Prochaska}, \&
  {Gawiser}}]{Wolfe03}
{Wolfe} A.~M., {Prochaska} J.~X., {Gawiser} E., 2003, \apj, 593, 215

\bibitem[{{Wolfe} {et~al}\mbox{.}(2008){Wolfe}, {Prochaska}, {Jorgenson}, \&
  {Rafelski}}]{Wolfe08}
{Wolfe} A.~M., {Prochaska} J.~X., {Jorgenson} R.~A., {Rafelski} M., 2008, \apj,
  681, 881

\bibitem[{{Wolfe} {et~al}\mbox{.}(1986){Wolfe}, {Turnshek}, {Smith}, \&
  {Cohen}}]{Wolfe86}
{Wolfe} A.~M., {Turnshek} D.~A., {Smith} H.~E., {Cohen} R.~D., 1986, \apjs, 61,
  249

\bibitem[{{Woosley} \& {Weaver}(1995)}]{Woosley95}
{Woosley} S.~E., {Weaver} T.~A., 1995, \apjs, 101, 181

\bibitem[{{Yong} {et~al}\mbox{.}(2013){Yong}, {Norris}, {Bessell},
  {Christlieb}, {Asplund}, {Beers}, {Barklem}, {Frebel}, \& {Ryan}}]{Yong13}
{Yong} D. {et~al.}, 2013, \apj, 762, 27

\bibitem[{{Zafar} {et~al}\mbox{.}(2014{\natexlab{a}}){Zafar}, {Centuri{\'o}n},
  {P{\'e}roux}, {Molaro}, {D'Odorico}, {Vladilo}, \& {Popping}}]{Zafar14N}
{Zafar} T., {Centuri{\'o}n} M., {P{\'e}roux} C., {Molaro} P., {D'Odorico} V.,
  {Vladilo} G., {Popping} A., 2014{\natexlab{a}}, \mnras, 444, 744

\bibitem[{{Zafar} {et~al}\mbox{.}(2011){Zafar}, {M{\o}ller}, {Ledoux}, {Fynbo},
  {Nilsson}, {Christensen}, {D'Odorico}, {Milvang-Jensen}, {Micha{\l}owski}, \&
  {Ferreira}}]{DLAcat100}
{Zafar} T. {et~al.}, 2011, \aap, 532, A51

\bibitem[{{Zafar} {et~al}\mbox{.}(2014{\natexlab{b}}){Zafar}, {Vladilo},
  {P{\'e}roux}, {Molaro}, {Centuri{\'o}n}, {D'Odorico}, {Abbas}, \&
  {Popping}}]{Zafar14Ar}
{Zafar} T., {Vladilo} G., {P{\'e}roux} C., {Molaro} P., {Centuri{\'o}n} M.,
  {D'Odorico} V., {Abbas} K., {Popping} A., 2014{\natexlab{b}}, \mnras, 445,
  2093

\bibitem[{{Zhang} {et~al}\mbox{.}(2011){Zhang}, {Karlsson}, {Christlieb},
  {Korn}, {Barklem}, \& {Zhao}}]{Zhang11}
{Zhang} L., {Karlsson} T., {Christlieb} N., {Korn} A.~J., {Barklem} P.~S.,
  {Zhao} G., 2011, \aap, 528, A92

\end{thebibliography}
\appendix
\section{Data compilation tables}

\subsection{DLA Literature Sample}
\label{sec:AppDLALit}

The compilation of the literature DLA sample has been taken from literature starting  in 1994 up until the end of 2014. In many cases, the same DLAs have been observed multiple times. For any duplicated column densities between studies, values were checked for consistency with each other by comparing the sum of the errors between two measurements with the difference in the two measured column densities. A preference was given to column densities derived with Voigt profile fitting to avoid any contamination from other lines. In addition, abundances derived with higher resolution instruments are preferentially selected as they are more likely to resolve all clouds, whereas lower resolution observations may contain unseen saturated components. Note that column densities in our catalogue derived from Keck/ESI or VLT/XSHOOTER data may indeed be saturated as both instruments do not quite have sufficient resolution to resolve narrow, saturated components of the absorption feature. All references are included in Table \ref{tab:DLAlit}, even if their derived column density was not adopted as the final value in the compilation.

\begin{figure}
\begin{center}
\includegraphics[width=0.5\textwidth]{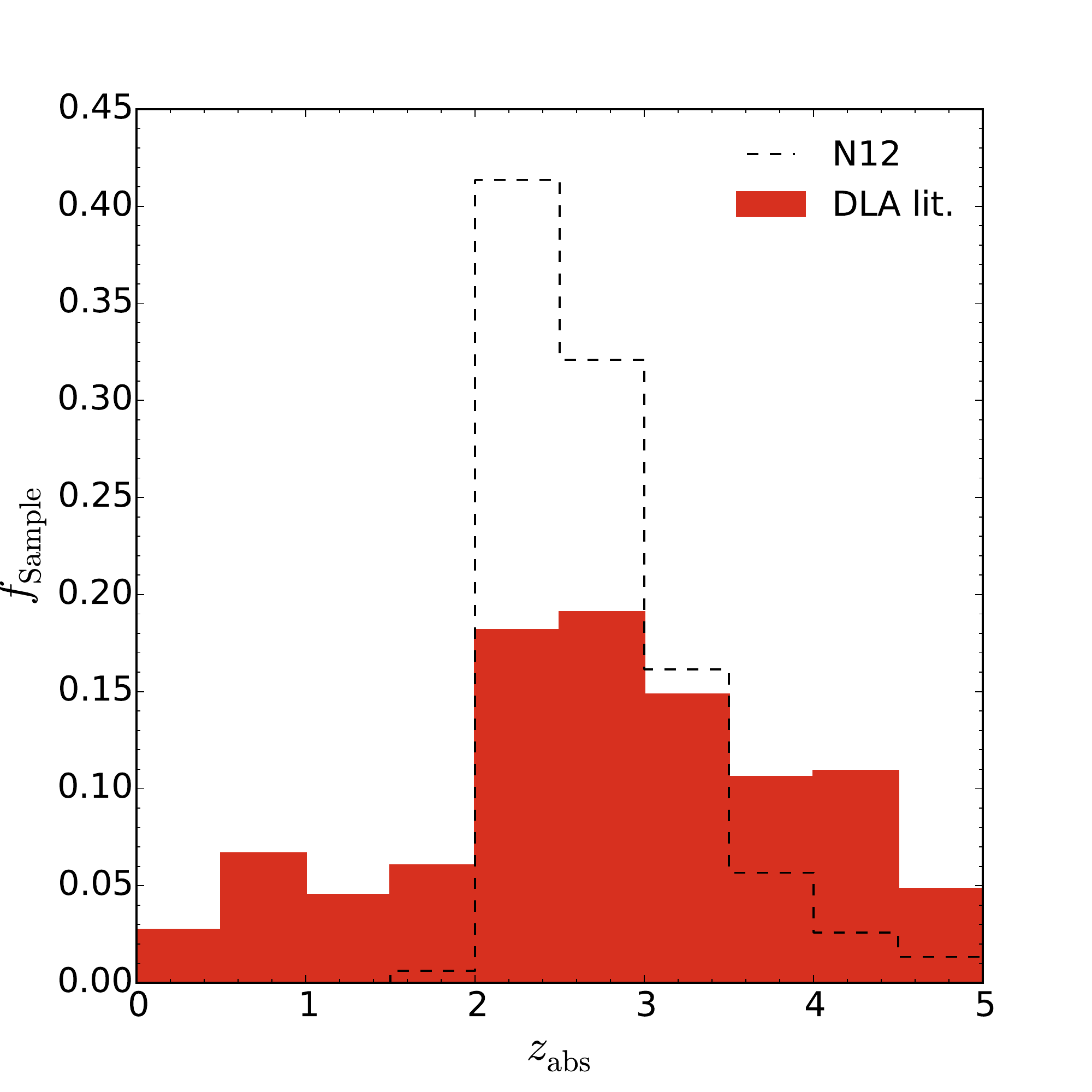}
\caption{The fractional redshift distribution of the DLA literature sample, compared to that of \citet[N12; black dashed line]{Noterdaeme12}. $f_{\rm Sample}$ represents the fraction of the respective sample in each bin. The DLA literature sample spans a large range in redshift from $0\leq$\zabs$\leq 5$, but is slightly biased towards $z\sim2$ due to the selection effects of large surveys such as the SDSS (e.g. N12).}
\label{fig:litzs}
\end{center}
\end{figure}

\begin{figure}
\begin{center}
\includegraphics[width=0.5\textwidth]{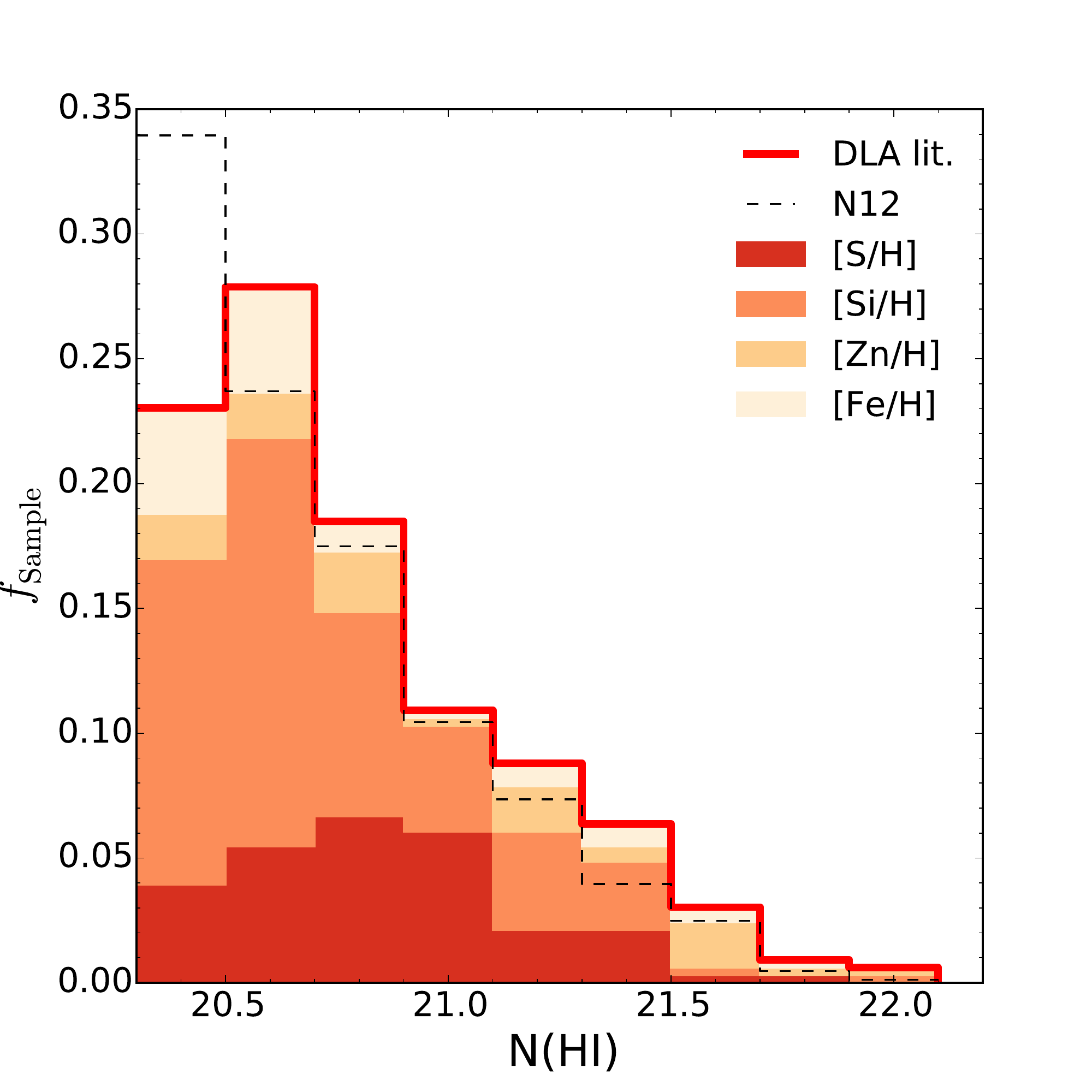}
\caption{The log$N$(\HI{}) distribution for the literature DLA sample (red line). To demonstrate which element is typically used as the metallicity indicator, the log$N$(\HI{}) distribution for the individual elements (S, Si, Zn, and Fe) are stacked, and shown in different shades of red. As in Figure \ref{fig:litzs}, the N12 (dashed line) distributions is also shown. Overall, there is a very good agreement between the distributions of the DLA literature sample and the N12 sample, despite the N12 sample being over an order of magnitude larger. }
\label{fig:litHIs}
\end{center}
\end{figure}

\begin{figure}
\begin{center}
\includegraphics[width=0.5\textwidth]{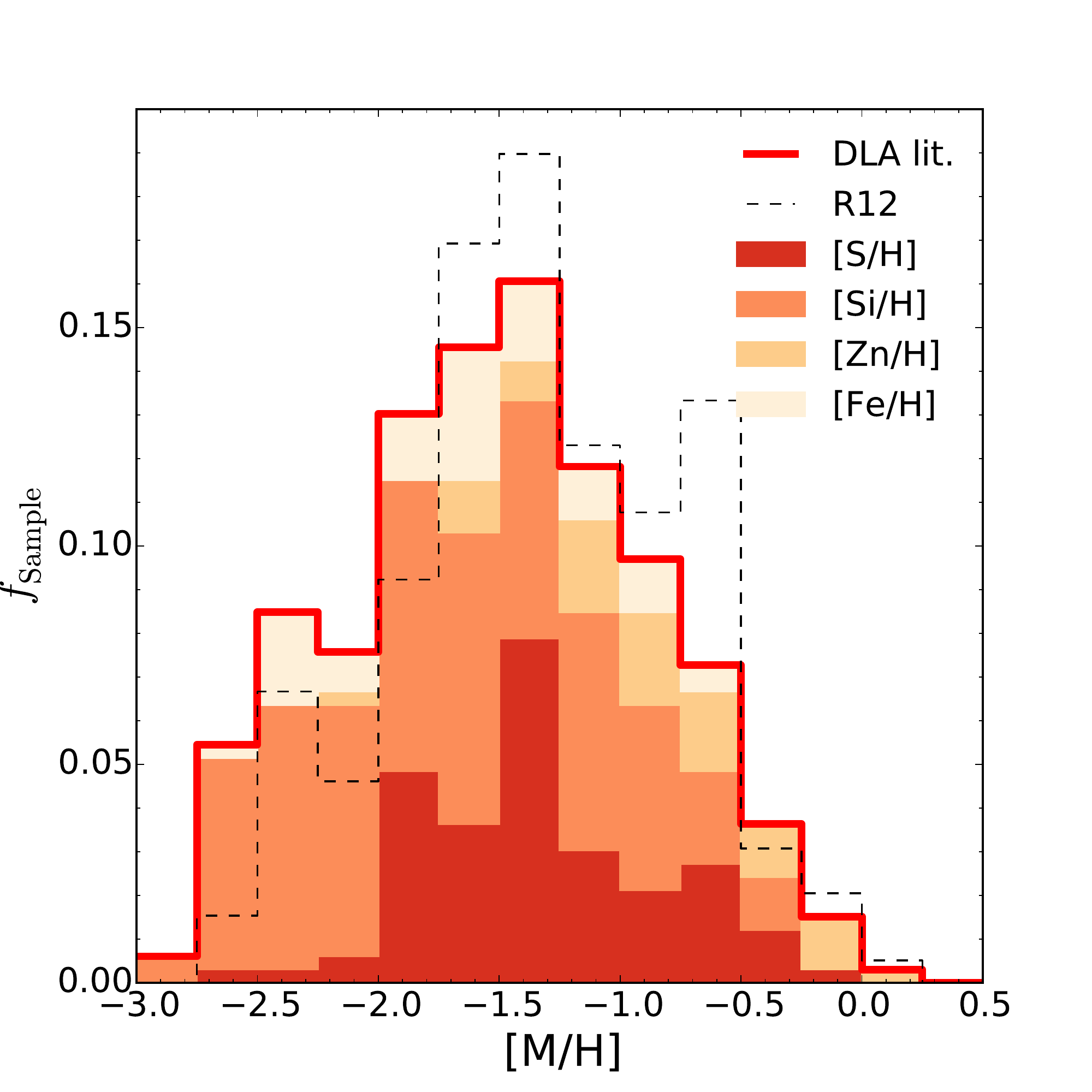}
\caption{The overall metallicity distribution of the DLA literature sample. The DLA literature sample spans a significant range of metallicity between $-3$ and $0.5$, with a median value of [M/H]$=-1.5$. As in \ref{fig:litHIs}, the metallicity distributions for each metallicity indicator (S, Si, Zn, and Fe) are stacked and shown in shades of red. The typical metallicity indicator in DLAs following the \citet{Rafelski12} method is Si, but there is a clear bias to using S and Zn at higher metallicities due to the difficulty of detecting S and Zn in low metal column systems.}
\label{fig:litMs}
\end{center}
\end{figure}

To summarize the properties of the DLA literature sample, Figures \ref{fig:litzs}--\ref{fig:litMs} show the \zabs{}, N(\HI{}), and metallicity distributions (respectively). For comparison, the samples from  \citet[N12; dashed lines in Figures \ref{fig:litzs} and \ref{fig:litHIs}]{Noterdaeme12} and \citet[R12; dashed line in Figure \ref{fig:litMs}]{Rafelski12} are shown, each containing 6839 and 195 DLAs (respectively). The N12 statistical sample ($2\leq z\leq 3.5$) was obtained through DLA identification from SDSS Data Release 9, and has no constraints on metallicity. The R12 literature-only DLA compilation is selected to be unbiased in metallicity, and ignores surveys specifically targeted towards both metal-poor \citep[e.g.][]{Penprase10} and metal-rich \citep[e.g.][]{HerbertFort06} DLAs. We exclude the additional DLAs presented in R12 as they are purposefully selected to be at \zabs{}$>4$. One should note that all but 18 DLAs from the R12 literature-only sample are included in our own DLA literature sample. These 18 DLAs are not included as the original references \citep{Prochaska03ApJ595, Wolfe08} do not provide metal column densities, but rather metallicities.  

Starting with the DLA redshifts (Figure \ref{fig:litzs}), our literature sample spans a large range of redshift, between \zabs{} $\sim 0$ and 5. Most DLAs that are identified have redshifts between $z\sim2$ and 3, which is a selection effect of DLA catalogues derived from large ground-based surveys such as the SDSS (as shown by the N12 sample).  These ground-based surveys (like N12)  are either limited by the UV atmospheric cutoff (3000\AA) or (as in the case of the SDSS) the instrumental efficiency is low at small wavelengths. In addition to being magnitude-limited, these surveys are generally restricted to DLAs at \zabs{}$\gtrsim1.6$. It is shown in Figure \ref{fig:litHIs} that our literature DLA sample spans a large range in \HI{} column densities, and is largely consistent with N12 with the exception at low column densities (logN(\HI{})$\leq 20.7$).

 Figure \ref{fig:litMs} shows the overall metallicity distributions of the literature sample of DLAs in comparison to the R12 sample. The metallicities are obtained for the DLAs following the scheme outlined by \cite{Rafelski12}. The median metallicity of absorbers is around [M/H]$\sim -1.5$ \citep[e.g.][]{Prochaska03ApJ595, Rafelski12}, but they do span a significant range in metallicity (from $-3$ to $0.5$). A closer inspection of which metals are used (represented by the colour scale in Figure \ref{fig:litMs}) shows that there is a slight bias for using certain elements for a given metallicity. S becomes the most common probe at higher metallicities resulting from the S lines (which are typically located in low SNR regions of the spectra due to the Ly$\alpha$ forest) having a higher chance of detection with larger metal contents\footnote{For the typical SNR in the the Ly $\alpha$ forest at the S \sion{} $\lambda 1253$ line  of 5 (or 10), a DLA at \zabs{}$=2$ will have a column density of log$N$(S\sion{})=14.4 (or 14.1) (assuming a typical full width half maximum of 16.5 \kms{}). For a DLA with log$N$(\HI{})$=20.5$; this corresponds to [S/H]$\sim -1.25$ (or $-1.22$) dex. The effect of higher column density systems can be seen in the different shades in Figure \ref{fig:litHIs}; where S is preferentially used in systems with log$N$(\HI{})$\sim21$.}.

\begin{landscape}
\begin{table*}
{\setlength{\tabcolsep}{1pt}
\tiny
\begin{center}
\caption{DLA literature column densities}
\label{tab:DLAlit}
\begin{tabular}{lcccccccccccccccccccc}

\hline
QSO& $z_{\rm em}$& \zabs{}& N(\HI{})& N(N \textsc{i})& N(O \textsc{i})& N(Mg \textsc{i})& N(Mg \textsc{ii})& N(Al \textsc{ii})& N(Al \textsc{iii})& N(Si \textsc{ii})& N(S \textsc{ii})& N(Ca \textsc{ii})& N(Ti \textsc{ii})& N(Cr \textsc{ii})& N(Mn \textsc{ii})& N(Fe \textsc{ii})& N(Co \textsc{ii})& N(Ni \textsc{ii})& N(Zn \textsc{ii})& References\\
\hline
SDSSJ1616+4154& 0.44& 0.3211& $20.60\pm0.20$& $>15.35$& $>15.45$& \nodata{}& \nodata{}& \nodata{}& \nodata{}& $>15.08$& $15.37\pm0.11$& \nodata{}& \nodata{}& \nodata{}& \nodata{}& $15.02\pm0.05$& \nodata{}& \nodata{}& \nodata{}& 98,99\\
SDSSJ1009+0713& 0.46& 0.1140& $20.68\pm0.10$& $15.11\pm0.10$& $>16.00$& \nodata{}& \nodata{}& \nodata{}& \nodata{}& $>15.00$& $15.25\pm0.00$& \nodata{}& $12.70\pm0.03$& \nodata{}& \nodata{}& $15.29\pm0.17$& \nodata{}& $13.93\pm0.18$& \nodata{}& 73\\
SDSSJ1619+3342& 0.47& 0.0963& $20.55\pm0.10$& $14.64\pm0.12$& $>14.80$& $12.49\pm0.14$& \nodata{}& \nodata{}& \nodata{}& $>13.93$& $15.08\pm0.09$& $12.42\pm0.02$& $11.90\pm0.04$& \nodata{}& \nodata{}& $14.38\pm0.15$& \nodata{}& $<13.53$& \nodata{}& 98,99\\
Q0738+313& 0.63& 0.2210& $20.90\pm0.09$& \nodata{}& \nodata{}& $12.00\pm0.10$& $>13.30$& \nodata{}& \nodata{}& \nodata{}& \nodata{}& $11.91\pm0.03$& $<11.48$& $13.11\pm0.24$& \nodata{}& \nodata{}& \nodata{}& \nodata{}& $<12.83$& 60,81,101\\
Q2353-0028& 0.76& 0.6043& $21.54\pm0.15$& \nodata{}& \nodata{}& \nodata{}& \nodata{}& \nodata{}& \nodata{}& \nodata{}& \nodata{}& \nodata{}& \nodata{}& $13.40\pm0.17$& \nodata{}& \nodata{}& \nodata{}& \nodata{}& $13.25\pm0.29$& 78\\
Q1328+307& 0.85& 0.6920& $21.25\pm0.10$& \nodata{}& \nodata{}& \nodata{}& \nodata{}& \nodata{}& \nodata{}& \nodata{}& \nodata{}& $12.50\pm0.10$& \nodata{}& $13.30\pm0.10$& $<12.59$& $14.98\pm0.10$& \nodata{}& \nodata{}& $12.72\pm0.10$& 7,9,31\\
SBS1543+393& 0.87& 0.0090& $20.42\pm0.04$& \nodata{}& $>16.20$& \nodata{}& \nodata{}& \nodata{}& \nodata{}& $>15.10$& $15.19\pm0.04$& \nodata{}& \nodata{}& \nodata{}& \nodata{}& \nodata{}& \nodata{}& $<13.83$& \nodata{}& 52\\
3C336& 0.93& 0.6560& $20.36\pm0.10$& \nodata{}& \nodata{}& \nodata{}& \nodata{}& \nodata{}& \nodata{}& \nodata{}& \nodata{}& \nodata{}& \nodata{}& \nodata{}& $<12.42$& $14.59\pm0.11$& \nodata{}& \nodata{}& \nodata{}& 6,15,31\\
AO0235+164& 0.94& 0.5241& $21.70\pm0.10$& \nodata{}& \nodata{}& \nodata{}& \nodata{}& \nodata{}& \nodata{}& \nodata{}& \nodata{}& \nodata{}& \nodata{}& \nodata{}& \nodata{}& $15.30\pm0.40$& \nodata{}& \nodata{}& \nodata{}& 83\\
Q0827+243& 0.94& 0.5247& $20.30\pm0.05$& \nodata{}& \nodata{}& $12.69\pm0.02$& $>14.59$& \nodata{}& \nodata{}& \nodata{}& \nodata{}& \nodata{}& $<11.76$& $<13.42$& \nodata{}& $14.59\pm0.02$& \nodata{}& \nodata{}& $<12.80$& 60,101\\

\hline
\end{tabular}
Notes. All column densities are given as log[$N$(X)/cm$^-2$]. The table is presented in its entirety in the online version. Only the first 10 entries are shown. 

References-- (1) \cite{Pettini94}. (2) \cite{Meyer95}. (3) \cite{Lu96}. (4) \cite{DLAcat4}. (5) \cite{Prochaska96}. (6) \cite{DLAcat6}. (7) \cite{Pettini97}. (8) \cite{DLAcat8}. (9) \cite{DLAcat9}. (10) \cite{DLAcat10}. (11) \cite{Pettini95}. (12) \cite{DLAcat12}. (13) \cite{Prochaska99}. (14) \cite{Pettini00}. (15) \cite{DLAcat15}. (16) \cite{DLAcat16}. (17) \cite{DLAcat17}. (18) \cite{DLAcat18}. (19) \cite{DLAcat19}. (20) \cite{Centurion00}. (21) \cite{DLAcat21}. (22) \cite{DLAcat22}. (23) \cite{Prochaska01I}. (24) \cite{Prochaska01}. (25) \cite{DLAcat25}. (26) \cite{DLAcat26}. (27) Dessauges-Zavadsky (unpublished).(28) \cite{DLAcat28}. (29) \cite{DLAcat29}. (30) \cite{DLAcat30}. (31) \cite{Ledoux02}. (32) \cite{DLAcat32}. (33) \cite{DLAcat33}. (34) \cite{DLAcat34}. (35) \cite{DLAcat35}. (36) \cite{DLAcat36}. (37) \cite{DLAcat37}. (38) \cite{Lopez02}. (39) \cite{Pettini02}. (40) \cite{DLAcat40}. (41) \cite{DLAcat41}. (42) \cite{Prochaska03}. (43) \cite{Prochaska03ApJS147}. (44) \cite{DLAcat44}. (45) \cite{DLAcat45}. (46) \cite{DZavadsky04}. (47) \cite{DLAcat47}. (48) \cite{DLAcat48}. (49) \cite{DLAcat49}. (50) \cite{Akerman05}. (51) \cite{Srianand05}. (52) \cite{DLAcat52}. (53) \cite{DLAcat53}. (54) \cite{DZavadsky06}. (55) \cite{HerbertFort06}. (56) Dessauges-Zavadsky et al. (unpublished).(57) \cite{DLAcat57}. (58) \cite{DLAcat58}. (59) \cite{Ledoux06}. (60) \cite{DLAcat60}. (61) \cite{DLAcat61}. (62) \cite{DLAcat62}. (63) \cite{DLAcat63}. (64) \cite{Prochaska07}. (65) \cite{DLAcat65}. (66) \cite{Dessauges07}. (67) \cite{DLAcat67}. (68) \cite{DLAcat68}. (69) \cite{Petitjean08}. (70) \cite{DLAcat70}. (71) \cite{Pettini08}. (72) \cite{Peroux08}. (73) \cite{DLAcat73}. (74) \cite{DLAcat74}. (75) \cite{Cooke11}. (76) \cite{DLAcat76}. (77) \cite{DLAcat77}. (78) \cite{DLAcat78}. (79) \cite{DLAcat79}. (80) \cite{Ellison10}. (81) \cite{Kulkarni05}. (82) \cite{DLAcat82}. (83) \cite{DLAcat83}.  (84) \cite{Berg13}.  (85) Paper I. (86) \cite{Penprase10}. (87) \cite{DLAcat82}. (88) \cite{DLAcat89}. (89) \cite{DLAcat90}. (90) \cite{Kulkarni12}. (91) \cite{DLAcat92}. (92) \cite{DLAcat93}. (93) \cite{DLAcat94}. (94) \cite{Ellison12}. (95) \cite{DLAcat96}. (96) \cite{DLAcat97}. (97) \cite{Battisti12}. (98) \cite{DLAcat73}. (99) \cite{DLAcat100}. (100) \cite{DLAcat101}. (101) \cite{Krogager13}. (102) \cite{Vladilo11}. (103) \cite{Rafelski12}. (104) \cite{Fynbo13}. (105) \cite{Zafar14Ar}. (106) \cite{Zafar14N}. (107) \cite{DLAcat108}. (108) \cite{Rafelski14}. (109) This Work
\end{center}
}
\end{table*}

\begin{table*}

\tiny
\begin{center}
\caption{Stellar literature abundances.}
\label{tab:stellar}
\begin{tabular}{lcccccccccccccccc}
\hline
ID& Galaxy& Component& [O/H]& [Mg/H]& [Al/H]& [Si/H]& [S/H]& [Ca/H]& [Ti/H]& [Cr/H]& [Mn/H]& [Fe/H]& [Co/H]& [Ni/H]& [Zn/H]& References\\
\hline
96185\_06& MW& Thick& $0.16$& $-0.33$& $-0.40$& $-0.23$& \nodata{}& $-0.03$& $-0.41$& $-0.61$& $-0.99$& $-0.56$& $-0.40$& \nodata{}& $-0.68$& 4\\
CS22169-035& MW& Halo& $0.05$& $-2.97$& $-3.89$& $-2.71$& \nodata{}& $-2.89$& $-3.07$& $-3.44$& $-3.37$& $-3.04$& $-3.21$& $-3.29$& $-2.90$& 1\\
113357& MW& Thin& \nodata{}& $0.30$& \nodata{}& \nodata{}& \nodata{}& $0.33$& $0.31$& \nodata{}& \nodata{}& $0.25$& \nodata{}& $0.36$& \nodata{}& 2,14,15\\
CS29528-041& MW& Halo& \nodata{}& $-2.92$& $-3.68$& $-3.46$& \nodata{}& $-2.88$& $-2.83$& $-3.44$& $-3.84$& $-3.30$& $-3.37$& $-3.27$& \nodata{}& 1\\
HIP77946& MW& Halo& \nodata{}& $-0.44$& $-1.06$& $-0.48$& \nodata{}& $-0.59$& $-0.55$& $-0.95$& \nodata{}& $-0.95$& \nodata{}& $-0.85$& \nodata{}& 1\\
87533\_06& MW& Thick& $0.05$& $-0.05$& $-0.20$& $-0.06$& \nodata{}& $0.21$& \nodata{}& $-0.18$& $-0.44$& $-0.21$& $-0.19$& $-0.20$& $-0.38$& 4\\
103896& MW& Thin& $-0.16$& $-0.17$& $-0.15$& $-0.19$& \nodata{}& $-0.20$& $-0.20$& $-0.26$& \nodata{}& $-0.24$& \nodata{}& $-0.28$& $-0.35$& 15\\
CS29502-042& MW& Halo& $0.05$& $-2.98$& $-3.98$& $-2.86$& \nodata{}& $-2.95$& $-2.88$& $-3.53$& $-3.79$& $-3.19$& $-2.90$& $-3.22$& $-2.96$& 1\\
HD120559& MW& Halo& $-0.37$& $-0.67$& $-0.62$& $-0.70$& \nodata{}& $-0.76$& $-0.60$& $-0.82$& \nodata{}& $-0.99$& \nodata{}& $-0.92$& \nodata{}& 1\\
ET270& Sculptor& Satellite& \nodata{}& \nodata{}& \nodata{}& \nodata{}& \nodata{}& \nodata{}& \nodata{}& \nodata{}& $-1.86$& $-1.49$& \nodata{}& \nodata{}& \nodata{}& 11\\

\hline
\end{tabular}
The table is presented in its entirety in the online version. Only the first 10 entries are shown. 

\textsc{References} --
        (1) \cite{Frebel10}.
        (2) \cite{Venn04}.
        (3) \cite{Reddy03}.
        (4) \cite{Reddy06}.
        (5) \cite{Letarte10}.
        (6) \cite{Venn12}.
        (7) \cite{Sbordone07}.
        (8) \cite{Pompeia08}.
        (9) \cite{Carretta10}.
        (10) \cite{Aoki09}.
        (11) \cite{North12}.
        (12) \cite{Shetrone03}.
        (13) \cite{Geisler05}.
        (14) \cite{Bensby05}.
        (15) \cite{Bensby14}.
        (16) \cite{Starkenburg13}.
        (17) \cite{Tafelmeyer10}.
        (18) \cite{Fulbright00}.
        (19) \cite{Stephens02}.
        (20) \cite{Edvardsson93}.
        (21) \cite{Hendricks14}.
        (22) \cite{Skuladottir15}.
\end{center}
\end{table*}

\end{landscape}
\subsection{Stellar literature sample}
\label{sec:AppStelLit}

The relevant papers used to acquire the stellar abundances for each Galactic component (in order of increasing metallicity) are summarized below. We have only selected studies which use high-resolution observations (R$>10000$) to obtain abundances accurate to $\pm0.1$ dex. Due to the nature of measuring stellar abundances, there are various assumption required that are not always consistent between studies in the literature. For the simplicity of this work, non-LTE abundance corrections are ignored for all elements as current non-LTE models \citep[such as][for Si, Fe, and Cr; respectively]{Shi09,Bergemann10,Mashonkina11} are only starting to be adopted into model atmosphere codes (see Appendix \ref{sec:AppElem} for typical non-LTE corrections for each element). In addition, no attempt has been made to correct for differences in the atomic data, stellar atmosphere models, or the use of various absorption lines between studies. However, these effects are modest and likely to only introduce systematic differences in stellar abundances of only  $\sim0.1$ dex between studies \citep{Shetrone03,Shi09}.

Most elements within the literature only have abundances derived for one ionization state. However, Fe occasionally has abundances quoted for both the neutral and singly ionized states. For consistency with most of the literature selected, it was assumed that Fe\textsc{i} abundances represent the total Fe abundances in these systems (which generally agree within $\sim0.2$ dex).

\subsubsection*{Galactic Thin and Thick Disc}

The Milky Way disc is decoupled kinematically into a metal-poor thick disc \citep[{$-1\leq$[Fe/H]$\leq -0.4$; scale-height $h\sim$1 kpc) and a metal-rich thin disc ($-0.8\leq$[Fe/H]$\leq 0.2$; $h\sim$0.3 kpc;}][]{Edvardsson93}. The selected thin- and thick-disc stellar samples for our compilation consist of the data provided by \cite{Reddy03} and \cite{Reddy06} (respectively). In addition, the \cite{Venn04} compilation is included to fill the subsample with other literature sources.

Both the Reddy et al.~studies look at nearby dwarf F and G stars using the 2dcoud\'e echelle spectrometer (R$\sim60000$) on the 2.7 m telescope at the McDonald Observatory. As these stars are lower mass main-sequence stars, they provide insight into the chemical composition of the ISM at the time the Milky Way first formed. The combination of both Reddy et al. samples provide a large, homogeneous sample of Milky Way disc stars. Reddy et al.~adopt the ATLAS 9 plane-parallel model \citep{ATLAS9}. LTE is assumed for deriving the abundances, however an empirical model (based on studies of stars with both O \textsc{i} and [O \textsc{i}] lines measured) is used on the O \textsc{i} $\lambda$7771 \AA{} line to derive non-LTE corrections.  When present, the forbidden [O \textsc{i}]$\lambda$6300\AA{} line is preferentially used over the O \textsc{i} $\lambda$7771 \AA{} line as it is more reliable and does not require the non-LTE correction. Hyperfine structure effects were taken into account for the Mn and Cu lines. In \cite{Reddy06}, the probability of the stars being within the thin or thick disc, or halo is calculated; requiring a probability $>70\%$ to determine which population the star belongs. The probability is based on the stars' kinematics being within the expected Gaussian distribution of the population and are weighted by the fraction of stars expected to be within the given population \citep[for more details, see][]{Reddy06}. One should note that  \cite{Reddy03} use a differential analysis\footnote{A differential analysis measures the Sun's abundances simultaneously with the stars. This removes any internal systematic errors between solar and stellar abundance derivations.} for determining stellar abundances. Therefore, their solar abundances are preferentially adopted over \cite{Asplund09} only for the \cite{Reddy03} stars. 

The \cite{Venn04} literature compilation from 14 different sources contains a total of 297 thick-disc stars and 482 thin-disc stars. These stars were identified to be in the thin or thick disc solely based on their kinematics. From their results, 33 stars that were originally classified as thin disc stars \citep{Reddy03} were determined by \cite{Venn04} to reside in the thick disc. Since \cite{Reddy06} adopts a kinematic-based classification similar to \cite{Venn04}, the \cite{Venn04} classification is used in preference over the \cite{Reddy03} description for these 33 stars. All duplicate stars were removed accordingly.

\cite{Bensby05} and \cite{Bensby14} both provide kinematically selected thin and thick disc samples of stars. \cite{Bensby05} used Uppsala MARCS model atmospheres \citep{Asplund97} to derive abundances.  The observations for the 98 disc stars were taken with VLT/UVES, European Space Observatory/Fibre-fed extended range optical spectrograph (ESO/FEROS), and Nordic Optical Telescope/Soviet-Finnish echelle Spectrograph (NOT/SOFIN).  Their study also includes a differential analysis, so their derived solar scale  is adopted in our compilation in preference to the \cite{Asplund09} solar scale.  427 thin disc and 249 thick disc stars were observed by \cite{Bensby14}.  These observation used a combination of VLT/UVES, NOT/SOFIN, NOT/Fibre-fed echelle spectrograph (FIES), ESO/High accuracy radial velocity planet searcher (HARPS), ESO/FEROS, and Magellan/Magellan Inamori Kyocera Echelle (MIKE). Abundances were derived using Uppsala MARCS model atmospheres \citep{Asplund97}. Both Fe \textsc{i} and O lines (from the 7770 \AA{} triplet) were corrected for non-LTE effects. Their abundances were derived in a differential study, so their abundances are not converted to the \cite{Asplund09} scale for our compilation.

\subsubsection*{Stellar Galactic Halo}
\label{sec:AppHalo}
Galactic halo stars are typically characterized by low metallicities ([Fe/H]$\lesssim-1.5$) and an enhanced \alphafe{}, and are believed to be relics of the first mergers to build the Milky Way \citep[e.g.][]{Searle78,Freeman02}. The sample of halo stars is chosen from \citet[80 stars]{Venn04} and \citet[867 stars]{Frebel10}.  Both samples are literature compilations, however the \cite{Venn04} sample has been identified based on the kinematics of nearby stars relative to the Sun whereas \cite{Frebel10} selects halo stars based on their kinematics (when present in the literature) or metallicity (i.e.~stars with metallicities less than the thick disc; [Fe/H]$<-1.5$). Although metal-poor stars typically lie within the halo, there is no guarantee that stars in the \cite{Frebel10} sample are not anomalous metal-poor disc stars with metallicities lower than [Fe/H]$<-1.5$. Other halo studies that have targeted metal-poor stars \citep{Aoki13,Cohen13,Yong13} are also available. However, only a handful of stars have a metallicity that overlaps with the DLA literature sample, and are thus not included in our stellar literature catalogue. In addition to the literature compilations, \cite{Bensby14} also provide a kinematically selected sample of 38 halo stars. See the description above for their observation details.

\subsubsection*{Satellite Galaxies}

All dSph galaxies with abundances for more than 30 stars were selected for our sample. The LMC is also included in this sample as it is another well studied satellite galaxy. Table \ref{tab:samplitdsph} shows a summary of which satellite galaxies were used, the elements available, the total number of stars ($N_{\rm stars}$) within the sample, and the references. Overall, there are \ndsphstar{} stars used from the satellite galaxies. A brief summary follows for each of the relevant literature sources, including any details concerning which lines were used and whether any corrections were adopted.

\begin{table*}
\begin{center}
\caption{Summary of satellite galaxy literature sources}
\label{tab:samplitdsph}
\begin{tabular}{lccc}
\hline
Satellite & Elements & $N_{\rm stars}$ &Reference\\
\hline
Carina & O, Mg, S, Cr, Mn, Fe, Co, Ni, Zn& 37 & 1,2,3\\
Fornax & O, Mg, Si, Cr, Mn, Fe, Ni, Zn & 78 &3,4,5,6\\
LMC & O, Mg, Si, Cr, Fe, Co, Ni & 59&7\\
Sagittarius & O, Mg, Al, Si, Cr, Mn, Fe, Co, Ni, Zn & 39&8,9\\
Sculptor & O, Mg, Cr, Mn, Fe, Co, Ni, Zn& 96&2,3,5,10,11\\

\hline
\end{tabular}

Notes. References --
		(1) \cite{Venn12};
		(2) \cite{Shetrone03};
		(3) \cite{North12};
		(4) \cite{Letarte10};
		(5) \cite{Tafelmeyer10};
		(6) \cite{Hendricks14};
		(7) \cite{Pompeia08};
		(8) \cite{Sbordone07};
		(9) \cite{Carretta10};
		(10) \cite{Geisler05};
		(11) \cite{Starkenburg13}

\end{center}
\end{table*}

\begin{itemize}

\item Fifteen red giant branch (RGB) stars in Sculptor, Fornax, Carina, and Leo \textsc{i} were studied by \cite{Shetrone03} using VLT/UVES (R$\sim40000$). MARCS model atmospheres \citep{MARCS} were adopted, using MOOG \citep[][with LTE assumptions]{MOOG} to determine the abundances. \cite{Shetrone03} compared their abundances  with MARCS/MOOG to those derived from the combination of ATLAS/WIDTH codes \citep{ATLAS9}. By using different model atmosphere and line analysis code, they find that the typical difference in abundances is about 0.1 dex for each species. Hyperfine splitting corrections were adopted for Mn. Most of the oxygen abundances were derived using  [OI] 6300 \AA{}, although some required the use of [OI] 6363 \AA{}.

\item \cite{Geisler05} observed four giant stars in Sculptor using UVES on VLT. Abundances were derived with LTE assumptions using MOOG \citep{MOOG} with MARCS model atmospheres \citep{MARCS}. The determined Mn abundances include corrections for hyperfine splitting. All oxygen abundances were derived using the forbidden [OI] line at 6300 \AA{}.

\item Twelve RGB stars in the Sagittarius dSph were observed with VLT/UVES (R$\sim43000$) for the stellar abundance work by \cite{Sbordone07}. 1D ATLAS model atmospheres \citep{ATLAS9} were used, assuming LTE, for deriving abundances. Hyperfine structure corrections were applied to the Mn, Co, and Cu abundances. O abundances were derived from the [OI] 6300 \AA{} to avoid non-LTE corrections.

\item The study of the LMC by \cite{Pompeia08} looked at 67 RGB stars with the VLT/FLAMES (R$\sim 24000$). Abundances were derived with a MARCS 1D model atmosphere using the ATLAS code \citep{ATLAS9}, assuming LTE. Hyperfine structure corrections were adopted for Cu and Co; the [OI] 6300\AA{} was used to derive the O abundance.

\item \cite{Carretta10} looked at 27 RGB stars in the Sagittarius dSph galaxy using VLT/FLAMES. Kurucz model atmospheres \citep{Kurucz93} were used with LTE assumptions. O abundances were derived from the [OI] 6300 and 6364 \AA{} lines. Hyperfine splitting corrections were adopted for Mn, Sc, Co, and Cu abundances.

\item A large sample of Fornax dSph stars was observed with VLT/FLAMES by \cite{Letarte10}. Abundances were derived for 81 RGB stars using spherical MARCS model atmospheres \citep{MARCS1,MARCS2}, and include a generic correction for hyperfine splitting on Eu and La.

\item \cite{Tafelmeyer10} observed five stars from Sextans, Fornax, and Sculptor with a combination of Hobby--Eberly Telescope/High Resolution Spectrograph and VLT/UVES. Abundances were derived using MARCS model atmospheres \citep{MARCS2}. Oxygen abundances were derived using the [OI] lines.

\item \cite{Venn12} derived abundances for nine RGB stars in the Carina dSph galaxy. Observations were completed with VLT/FLAMES \citep{Lemasle12} and Magellan/MIKE, providing a resolution of R$\sim15000$. Abundances were derived using spherical MARCS model atmospheres \citep{MARCS1,MARCS2}, with hyperfine splitting corrections included on odd-Z elements.

\item \cite{North12} compiled a list of the equivalent widths and stellar parameters from several literature sources to derive Mn in dSphs. Assuming LTE, they repeated the abundance determination under a MARCS \citep{MARCS1,MARCS2} spherical model (apart from Sculptor data, which used plane-parallel models), and redetermined the hyperfine splitting corrections for all the stars under the same model. They concluded that the Mn 5432 \AA{} line is not as reliable as the others due to the differences in the behaviour of the hyperfine splitting correction resulting from the influence of non-LTE effects. Due to the homogeneity of their corrections, the Mn abundances calculated by \cite{North12} are preferentially adopted over other literature sources in our compilation.

\item Seven stars in Sculptor were observed by \cite{Starkenburg13} with VLT/XSHOOTER. Abundances were derived using LTE MARCS model atmospheres \citep{MARCS2}.

\item \cite{Hendricks14} derived abundances of several $\alpha$-elements (Mg, Si, and Ti) for  431 stars in Fornax. Data were taken with VLT/FLAMES, and abundances derived with SPACE (Stellar Parameters and Chemical abundances Estimator).

\item S, Fe, and Mg abundances were derived for 85 stars in Sculptor by \cite{Skuladottir15}. Observations were carried out using VLT/FLAMES and VLT/GIRAFFE. LTE MARCS model atmospheres \citep{MARCS2} were used, but NLTE corrections for the S abundances were derived \citep{Takeda05}. For consistency with other S measurements, we do not include the non-LTE corrections in our sample.

\end{itemize}

\section{ Element details}
\label{sec:AppElem}

\subsection{Fe}
\label{sec:AppFe}
Fe is the most commonly used metallicity indicator in stars as it has many absorption lines present in optical stellar spectra. In addition, Fe is generally the standard for deriving the model atmospheres for abundance measurements. Therefore, Fe is a good benchmark for comparison.

Fe is formed from the decay of $^{56}$Ni produced in the core of SNe Ia \citep{Clayton_iso} and in $\alpha$-rich freezeout in SNe II. Although Fe forms in both types of SNe, much more is produced in SNe Ia \citep{Woosley95}.

Fe is one of the easiest elements to study in stars with its many absorption lines of varying oscillator strength \citep[e.g.~see Table 3 in][]{Reddy03}. As all Fe lines must measure the same overall abundance of Fe, Fe lines are used to derive the parameters of model atmospheres. However, \cite{Mashonkina11} suggest that some Fe \textsc{i} lines in stars suffer from non-LTE effects; which can impact both the derivation of the surface gravity of a star, and the relative contribution of different species towards the overall abundance. Nevertheless, [Fe/H] is used as the metallicity indicator in stars due to its ease of observation.

In DLAs, Fe also has many observable absorption lines \citep[e.g.][]{Morton03}. With a range in oscillator strengths, it is generally possible to measure an unsaturated Fe line to derive an abundance. However, Fe has a condensation temperature of $T_{\rm cond}=1336$ K \citep{Savage96}, making it susceptible to dust depletion. As a result, Fe abundances in DLAs are typically underestimated \citep{Pettini94,Vladilo02a}, and do not provide an \emph{accurate} metallicity benchmark for the comparison to stars. For this reason, Zn is commonly used in DLAs \citep[e.g.][]{Pettini94,Lu96,Prochaska02II}.

\subsection{Zn}
\label{sec:AppZn}
The origin of Zn has remained somewhat of a mystery. The general picture is that Zn is produced primarily in SNe Ia and SNe II \citep{Woosley95, Nomoto13}, however it is unclear what processes actually contribute to the production of Zn. It is thought that most of the Zn is produced by a combination of the s-process and $\alpha$-rich freezeout following nuclear statistical equilibrium \citep{Clayton_iso}. Simulations from \cite{Pignatari13} show that Zn is primarily generated in explosive nucleosynthesis in $\sim10$\Msolar{} stars, but can also be produced from neutrino winds in core-collapse SNe \citep{Pruet05}. However, the yields are dependent on the energy of the explosion, the mass cut, and therefore the amount of fallback. However, [Zn/Fe] can remain roughly solar in SNe Ia production  to within 0.1--0.2 dex \citep{Kobayashi09, Nomoto13}.

Zn measurements in stars usually use one of two multiplets  (further denoted as Mult.; 4722  and 4810  [Mult.~2], and 6362 \AA{} [Mult.~6]). \cite{Chen04} discussed the advantages and disadvantages of using these lines and concluded that, although the 6362 \AA{} line is weaker, it is more reliable as it does not saturate near solar metallicities and does not contain blending from several weak lines in the wings. \cite{Takeda05} tested whether the Zn and S lines required non-LTE corrections by running a grid of 120 model atmospheres with and without non-LTE assumptions, and determined that corrections are typically less than 0.1 dex (the 6362 \AA{} line has nearly negligible corrections) but can be as large as 0.3 dex.

Much of the stellar literature on Zn has been focused on answering the question of whether Zn traces Fe over all metallicities. Part of the motivation of these studies was to determine whether using Zn in DLAs as an Fe-peak tracer is valid or not. With the pioneering work on Zn done by \cite{Sneden88} and \cite{Sneden91}, [Zn/Fe] appeared to be solar for disc stars between metallicities of $-3.0\leq$[Fe/H]$\leq0.0$. This had been confirmed by observations of halo stars \citep{Nissen04}, as well as thin and thick disc stars \citep{Chen04}. However, with the additional observations of more metal-poor stars \citep{Nissen07} and the inclusion of non-LTE effects \citep{Takeda05}, [Zn/Fe] was found to rise with decreasing metallicity below [Fe/H]$\leq-1.5$. \cite{Nissen07} demonstrated that non-LTE effects are responsible for the slight enhancement of [Zn/Fe] in the metallicity range $-2.5\leq$[Fe/H]$\leq -1.5$, whereas the significant rise in [Zn/Fe] seen below [Fe/H]$\leq -2.5$ is likely from the contribution of hypernovae \citep{Nomoto13}. As a result and as discussed in the main body of this paper, \emph{Zn is not a valid tracer of Fe over all metallicities}, however it can be used in place as a valid tracer for Fe above [Fe/H]$\geq -1.5$ in the Milky Way.

Zn is one of the preferred metallicity indicators in DLAs. With a low condensation temperature \citep[$T_{\rm cond}=660$ K;][]{Savage96}, it is considered to be undepleted into dust \citep{Sembach95}. As early stellar studies \citep[e.g.][]{Sneden91} showed that Zn tracked the Fe-peak elements (which are all depleted into dust), Zn became the standard Fe-peak element tracer.  Measuring Zn in DLAs is challenging as the stronger lines are often blended with either the Mg \textsc{i} 2026 or Cr \sion{ii} 2062 lines in low-resolution studies\footnote{Mg\textsc{i} 2026 blending has only been seen in high metallicity systems at high-resolution \citep{DLAcat61}.} \citep{Prochaska02II}, or the DLA contains a small column density of Zn\footnote{Zn is somewhat rare, with a solar abundance of log(Zn/Fe)$_{\odot} \sim-2.8$ \citep{Asplund09}}. Therefore accurate abundance measurements of Zn are typically restricted to observations with high-resolution spectrographs to remove blending \citep{Prochaska02II}, or carefully selected DLAs with lower metallicities where blending is not an issue \citep{DLAcat61}. In addition, \cite{Kisielius15} suggest from atomic modelling that the Zn\sion{} oscillator strengths for Zn\sion{} 2026 and 2062 are underestimated by 0.1 dex; implying that log$N$(Zn\sion{}) in DLAs is overestimated by 0.1 dex. However, this systematic discrepancy is still within the measured error of [Zn/H], and should have little impact on understanding the abundance trends in DLAs where the scatter is generally larger.

Studies of Zn in DLAs are usually used to assess the dust depletion in a given DLA. As Zn is somewhat volatile, it has become a standard of comparison to determine the gas--dust ratio in DLAs. Much of the early work on dust depletion focused on the [Cr/Zn] ratio \citep[cf.][]{Pettini94,Pettini97,Pettini00}. With Zn being relatively undepleted into dust compared to Cr, and assuming that Cr and Zn trace each other (i.e.~[Cr/Zn]$=0$) in a DLA with no dust depletion, [Cr/Zn] would indicate what fraction of Cr was locked into dust. In \cite{Pettini94}, a threshold of [Cr/Zn]$=-0.3$ (i.e. 50\% of Cr locked into dust) was chosen as an indicator of whether there was significant dust depletion in a DLA. Although this study presented a method of selecting DLAs and giving an average gas--dust ratio in DLAs, it did not provide sufficient means to correct for the dust depletion. \cite{Vladilo02a} and \cite{Vladilo02b} presented a scaling relation that would provide a dust correction based on [Fe/Zn] rather than [Cr/Zn]. The scaling law requires two parameters that are known in DLAs: The percent variations in the relative abundance in the dust from changes in either (i) the dust--metal ratio or (ii) the relative abundance of metals in the medium. However, determining these two parameters is quite challenging as it requires assumptions about how similar the ISM in these galaxies is to the ISM in the Milky Way (where these parameters can be derived empirically).

The discrepancy between the apparent nucleosynthetic origin between Zn and Fe at low metallicities in stars combined with the excess of dust depletion of Fe in DLAs has led to difficulty in choosing a tracer of the Fe-peak elements in DLAs. One possibility is adopting abundance corrections  for either Fe (for dust depletion) or Zn (due to non-solar [Zn/Fe] in stars at low metallicities). As dust depletion affects all other Fe-peak elements measured in DLAs \citep[see Figure 4 in][]{Savage96}, and modelling dust depletion in DLAs is complicated and requires an individual understanding of the model parameters in each DLA \citep{Vladilo02a}; it would be more reasonable to check what correction is necessary for the apparent increase in [Zn/Fe] with decreasing metallicity.

\begin{figure}
\begin{center}
\includegraphics[width=0.5\textwidth]{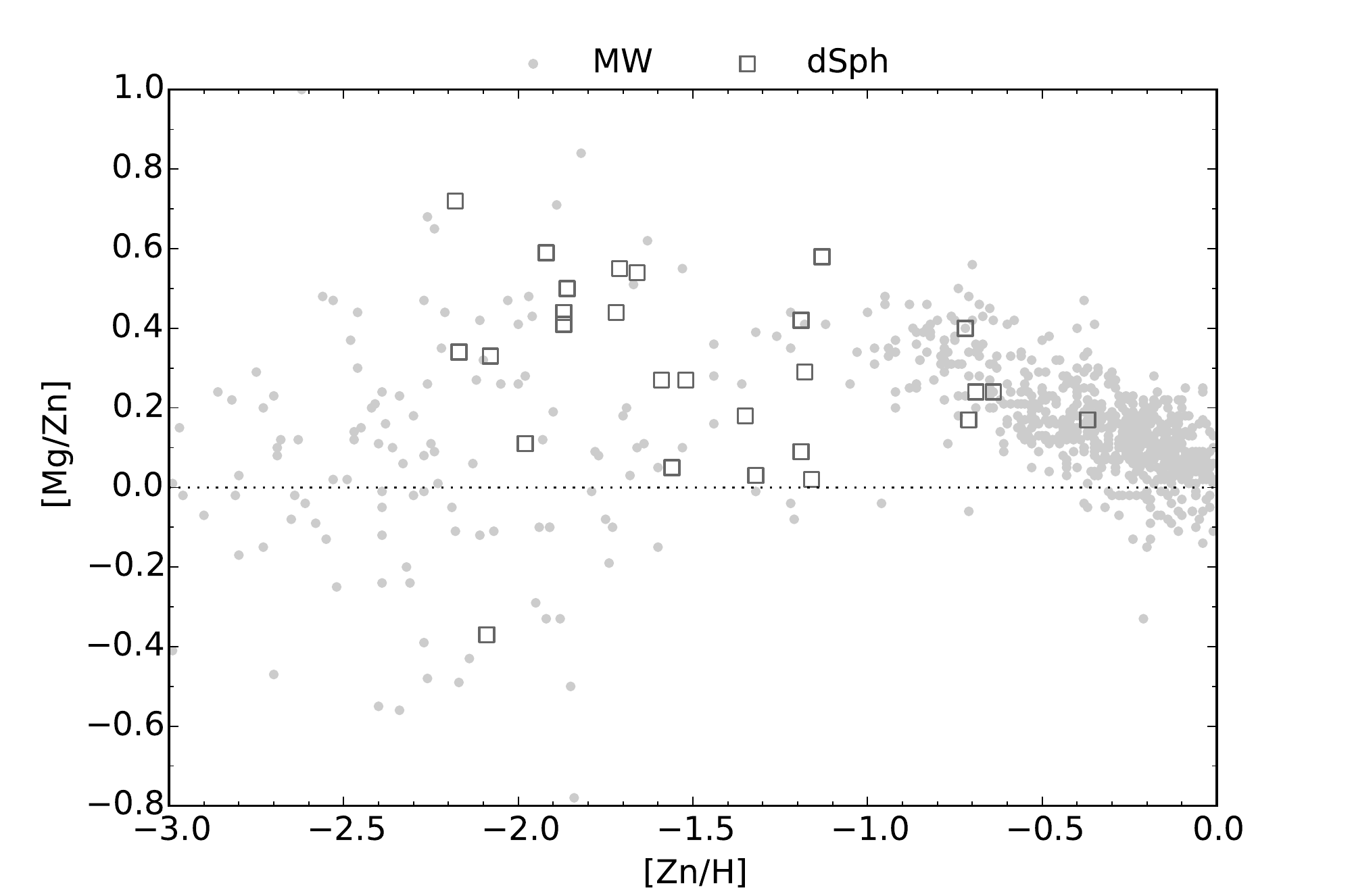}
\caption{[Mg/Zn] as a function of [Zn/H] in the stellar literature sample only. The overall trend for [Zn/H]$\gtrsim-2$ dex suggests that Zn behaves more like an Fe-peak element, as [Mg/Zn] essentially shows the characteristic \alphafe{} curve from $-1\leq$[Zn/H]$\leq0.0$. The discrepancy at low metallicities is likely due to the breakdown in Fe and Zn no longer tracking each other.}
\label{fig:MgZnstars}
\end{center}
\end{figure}

One interesting suggestion that has been made is whether Zn is behaving more like an $\alpha$-element or an Fe-peak element. \cite{Nissen11} first demonstrated by comparing the [Zn/Fe] trend with metallicity for halo stars with or without the typical \alphafe{} enhancement; finding enhancements in [Zn/Fe] of 0.1--0.2 dex when \alphafe{} is also enhanced.  In addition, \cite{Rafelski12} showed with their DLA sample that [$\alpha$/Zn] is constantly solar in DLAs over a large span of metallicity ($-2.1\leq {\rm [Zn/H]}\leq 0.3$), suggesting that Zn behaves more like an $\alpha$-element rather than belonging to the Fe-peak. If this were the case, then a similar trend should be seen in stars. To test this, Figure \ref{fig:MgZnstars} shows [Mg/Zn]\footnote{Mg is chosen as it is one of the most common $\alpha$-elements observed in stars, and is produced in hydrostatic burning of massive stars.} as a function of [Zn/H] in stars. If Zn behaves like an $\alpha$-element, [Mg/Zn] should be solar over all metallicities. However, the trend seen in Figure \ref{fig:MgZnstars}  suggests that Zn behaves like an Fe-peak element down to metallicities of [Zn/H]$\sim-2$ dex, where the trends breaks down due to the nucleosynthetic difference at low metallicities. It is possible that \cite{Rafelski12} were recovering a solar [$\alpha$/Zn] that is seen in DLAs (as discussed in Section \ref{sec:SSi}).

\subsection{Cr}
\label{sec:AppCr}
Cr is typically labelled as an Fe-peak element that is primarily formed as an unstable Fe isotope (such as $^{52}$Fe) that decays into Cr. Although Cr is mainly produced by SNe Ia \citep[e.g.][]{Travaglio04,Seitenzahl13}, SNe II also produce Cr during explosive nucelosynthesis \citep{Pignatari13}. Overall, the total contributions of Cr from both SNe Ia and SNe II are in equal parts, although an individual SNe Ia contribute about five times as much as a single SNe II \citep{Clayton_iso}, therefore SNe Ia dominate the production of Cr \citep[e.g.][]{Bergemann10}. 

For the most part, Cr is seen to trace Fe in Galactic stars ([Fe/Cr]$\sim0$), with a very tight scatter over metallicities \citep[$-1.6\lesssim${[Fe/H]}$\lesssim 0.5$; e.g.][]{Bensby05,Nissen10}. However, at low metallicities there appears to be a increase in [Fe/Cr] with decreasing metallicity starting at [Fe/H]$\sim -2$ \citep{Cayrel04,Lai08, Bonifacio09}. Although this difference was thought to be nucleosynthetic in origin \citep{Cayrel04}, both \cite{Preston06} and \cite{Lai08} found a trend in [Fe/Cr] with surface temperature, suggesting that non-LTE effects may be contributing to the high [Fe/Cr] abundances. \cite{Bergemann10} derived non-LTE models for Cr, and found that the non-LTE effects from their model did account for the $\sim0.35$ dex discrepancy in [Fe/Cr] from solar value. Even at [Fe/H]$\sim0$, a small ($\sim0.1$ dex) correction is often still required on Cr abundances to obtain a solar [Fe/Cr].

Measuring Cr in DLAs relies on three lines, although the Cr\sion{} 2062 line is usually blended with Zn\sion{} 2062 and is not as reliable as the other lines (although the Cr\sion{} 2062 and Zn\sion{} 2062 lines can typically be resolved using R$>5000$ observations). The difficulty with measuring Cr in DLAs, however, results from dust depletion. With its high condensation temperature of $T_{\rm cond}=1277$ K  \citep{Savage96}, Cr is heavily depleted onto dust. As a result, studies of Cr in DLAs have focussed on using Cr as an indicator of dust depletion.

The early work by \cite{Meyer90,Meyer92} and \cite{Pettini90,Pettini94} used [Cr/Zn] to determine the amount of dust depletion in DLAs. Under the assumptions that (i) Zn is a non-refractory element, (ii) Cr and Zn have the same nucleosynthetic origin, and (iii) the solar (Cr/Zn) ratio is identical to (Cr/Zn) in all ISM gas clouds; an underabundance of the [Cr/Zn] ratio along a DLA sightline would be indicative of dust depletion of Cr. All the literature on the [Cr/Zn] ratio in DLAs \citep{Meyer90,Pettini94,Kulkarni05,Akerman05} shows this [Cr/Zn] underabundance with DLAs, and an evolution of the underabundance with metallicity (i.e.~[Zn/H]). \cite{Pettini94} and \cite{Akerman05} have suggested that the metallicity dependence of the underabundance of [Cr/Zn] results from the variation in the amount of dust depletion with metallicity; where the depletion effect is strong at higher metallicities.

With respect to Fe, \cite{Lu96, Prochaska99, Prochaska02II} consistently found a slightly subsolar value of [Fe/Cr] in all DLAs. \cite{Prochaska02II} suggest that this slight enhancement results from dust depletion (i.e.~Cr is slightly less depleted onto dust than Fe) as [Fe/Cr] is also slightly subsolar in the local ISM while no enhancement is seen in stars. \cite{Prochaska99,Prochaska02II} also suggest a slight metallicity dependence in [Fe/Cr], where at low metallicities, [Fe/Cr] is closer to solar. This is contrary to \cite{DZavadsky06}, who suggested that all DLAs exhibit little scatter in [Fe/Cr] and [Si/S] with metallicity, therefore the dust content of DLAs is very uniform.

\subsection{S}
\label{sec:AppS}
S is primarily produced in the oxygen burning layer of massive stars. Si (the end product of oxygen burning) undergoes $\alpha$-rich freezeout during SNe II, producing S \citep{Woosley95}. The combination of SNe II yielding $\sim10\times$ more S than SNe Ia while being $\sim5\times$ more frequent, ensures that SNe II are the dominant source of S in the galaxy \citep{Clayton_iso}.

There are many lines available for observing S in stars, mostly using S \textsc{i}. However, many of these lines produce abundances that are inconsistent with each other \citep{Caffau05,Takeda05}. \cite{Caffau05} were the first to review which lines are useful, and regard Mult.~1, 6, and 8 to be the most reliable. \cite{Caffau05} discuss that Mult.~1 lines (9212.2863, 9212.970, 9228.903, and 9237.538 \AA{}), which are typically strong lines,  are often blended or near other features. As the Mult.~1 lines are the strongest lines, they are often used for measuring S in metal-poor stars. Mult.~6 (8693.931 and 8694.626 \AA{}) are about 10 times weaker than Mult 1., but are not near any other lines that can potentially blend. However, \cite{Caffau05} mention that there are large discrepancies in the oscillator strengths adopted between studies for these lines, making comparisons difficult. The most reliable lines seem to be the Mult.~8 lines which are free from blending and have consistently measured oscillator strengths.  As the lines originate from the same lower level as Mult.~6. their dependence on the effective temperature and surface gravity of the stars are the same, providing consistent abundances between multiplets. Although both \cite{Nissen04} and \cite{Caffau05} claim that non-LTE corrections are small and can be ignored, \cite{Takeda05} suggest the opposite and that non-LTE should be included for Mult.~1 and 6 lines as abundances can be overestimated by up to 0.25 dex (by not including non-LTE corrections). The combination of selecting certain lines and the inclusion of non-LTE corrections has caused difficulty in comparing results between multiple studies of S in the literature.

With the large variety of inconsistent methods used to determine S abundances, it has made understanding the observed nucleosynthetic trends of S very difficult.  The main interest in S has been whether or not S behaves like O (and other $\alpha$-elements) at low metallicities (i.e. [S/Fe] and [S/Zn] $\sim+0.4$ dex plateau at low metallicities [Fe/H]$\leq-1.0$ dex.). The first studies completed by \cite{Francois87,Francois88} suggest [S/Fe] in the halo are in agreement with other $\alpha$-elements such as magnesium and oxygen. However, \cite{Israelian01S} claim that no such plateau exists. Using a small sample of stars and a reanalysis of the \cite{Francois87} data, \cite{Israelian01S} found that [S/Fe] increased with decreasing metallicity to about $\sim+0.7$ dex using the Mult.~6 lines.  This decrease was also confirmed by \cite{Takada02} who did a similar analysis with 67 dwarf and giant stars. Subsequent studies \citep[][using combinations of Mult.~1, 3 with non-LTE corrections, 6, and 8]{Nissen04,Nissen07,Spite11,Caffau14} did find the plateau at low metallicities ([Fe/H]$\sim0.35$). However, \cite{Caffau05} showed that the large scatter can confirm both cases.

In DLAs, S is one of the best $\alpha$-elements to measure. With a low condensation temperature \citep[$T_{\rm cond}=648$ K][]{Savage96}, S is likely not depleted into dust and provides accurate $\alpha$-element measurements. However, only three lines are available to measure S \citep{Morton03}. As the lines are near the Ly $\alpha$ transition,  the absorption lines are often found in the Ly $\alpha$ forest, and require large column densities to be distinguished from blending. As a result, there is a tendency for S to only be found in the higher column density DLAs (see Section \ref{sec:DLAlit}).

\subsection{Si}
\label{sec:AppSi}
Si is an $\alpha$-element formed during the oxygen burning phases in stars. It is created either from the combination of two O nuclei, or He with Mg during explosive nucleosynthesis \citep{Woosley95, Pignatari13}. Although both SNe Ia and SNe II produce equal amounts of Si, SNe II are five times more frequent, thus dominating the production of Si \citep{Clayton_iso}.

Si is typically measured by two lines in stars: the Si \textsc{i} 3905.5 and 4102.9 \AA{}.  Recently, \cite{Shi12} have used several IR lines to derive Si abundances in nearby stars. The trouble with measuring Si in stars is that neither of the Si \textsc{i} lines gives systematically consistent results between studies \citep[][]{Shi09}. Although the variety in stellar atmosphere codes and adopted parameters play a significant role in these discrepancies between studies \citep{Zhang11}, part of the problem is potential blending from other features. The 3905 \AA{} line is often severely blended with a CH line, while the 4102 \AA{} line falls within the wings of the H$\delta$ absorption feature \citep{Cayrel04}. However for metal-poor stars, only the 3905 \AA{} line is available for metallicities below [Fe/H]$<-2.5$. At these low metallicities, there is an enhancement in the carbon abundance \citep[e.g.][]{Akerman04}, greatly increasing the contribution of the CH line. In the past, LTE was generally assumed for the 3905 \AA{} line. However, \cite{Preston06} were the first to notice in their sample of red horizontal branch stars that [Si/Fe] decreased with increasing surface temperature and did not find significant CH blending. The non-LTE models from \cite{Shi09} demonstrated that the decrease in [Si/Fe] was a result of not including non-LTE corrections in their models. The magnitude of the non-LTE corrections for Si varies depending on which line is adopted \citep{Shi09}. Although non-LTE effects should be included, the work by \cite{Shi12} demonstrated that the Si\textsc{i} IR lines can be used with LTE assumptions to determine Si abundances. As with other $\alpha$-elements, Si shows a constant enhancement relative to Fe ([Si/Fe]$\sim0.4$) at low metallicities, with a decrease towards solar values with increasing metallicity starting at [Fe/H]$=-1$ \citep[e.g.][]{Perez13}.

Si is generally the most frequently observed $\alpha$-element in DLAs \citep{Prochaska02II}. With at least four commonly used absorption lines of varying oscillator strengths, there is often at least one line to provide a Si abundance reliably. The difficulty with Si is that it is somewhat depleted into dust \citep[$T_{\rm cond}=1311$ K;][]{Savage96}. Comparing to S abundances in DLAs, \cite{Vladilo11} found an average Si depletion of $0.27\pm0.16$ dex. However, \cite{Rafelski12} point out that [Si/S] is relatively constant over all metallicities, hinting that Si might not be as depleted as previously thought. Despite its affinity to deplete onto dust, Si is still frequently used to study $\alpha$-elements and understand the star formation history in DLAs \citep{Prochaska02II}. 

\subsection{Mn}
\label{sec:AppMn}
Mn is an Fe-peak element formed from explosive Si burning as $^{55}$Co, which decays into $^{55}$Mn. It is produced both in SNE Ia and SNe II, but it is unclear whether either source of Mn has metallicity dependent yields \citep{Woosley95,Cescutti08}. Mn is primarily produced in SNe II explosive burning, similar to Cr \citep{Pignatari13}. The metallicity dependence seen in yields is likely a result of the odd--even effect (i.e. increased number of metals results in an increase in the number of free neutrons that can form $^{55}$Co), where the excess neutrons come from the conversion of N$^{14}$ to O$^{18}$ \citep{Badenes08}.

The trouble with observing Mn (or any odd-Z element) in stars is that hyperfine splitting corrections must be included \cite[e.g.][]{Prochaska00Mn,North12}. \cite{Feltzing07} checked for deviations in LTE effects, but found no effects for the  four Mn \textsc{i} lines used ($\lambda$ 5394, 5492, 6013, and 6016 \AA{}).  \cite{North12} did an extensive comparison of many of the optical Mn lines, deriving [Mn/Fe] for the four Mn \textsc{i} lines. \cite{North12} found a disagreement between [Mn/Fe] derived with Mn \textsc{i} 5432 \AA{} compared to the other three lines  of their study ($\lambda$ 5407, 5420, and  5516\AA{}) for [Fe/H]$>-1.0$. Therefore, \cite{North12} deemed the Mn \textsc{i} 5432 line unreliable for deriving Mn abundances.

To test whether Mn yields do have a metallicity dependence, there have been several observational studies to constrain the nucleosynthesis of Mn. One of the largest studies started with \cite{Nissen00}, where they observed 119 stars in the thin and thick disc, as well as the halo of the Milky Way. They found a steady increase in the [Mn/Fe] ratio with metallicity, supporting a metallicity dependence of Mn.  However, they noticed that below a metallicity of [Fe/H]$<-0.7$ dex,  the slope was much steeper than at higher metallicities. They claimed that the discontinuity mirrors the same discontinuity seen in \alphafe{}, suggesting that SNe II contribute to Mn production in the lower metallicity components of the Milky Way (i.e. the halo and thick disc). This is contrary to what is argued by \cite{Prochaska00Mn}, where they reanalysed the \cite{Nissen00} abundances by including more accurate hyperfine splitting corrections. \cite{Prochaska00Mn} found a shallower slope, which they claim to be consistent with metallicity-independent yields. Although the trend observed by \cite{Nissen00} could be due to selection effects (as thick-disc stars were only chosen to have a metallicity less than [Fe/H]$<-0.7$; thin-disc stars with [Fe/H]$\sim-0.7$ dex), work by \cite{Feltzing07} confirms the observed  change in slope trend by using a kinematically selected sample of stars. \cite{Feltzing07} went one step further by comparing [Mn/O] as it evolves with [O/H], and found that [Mn/O] is a constant for [O/H]$\leq-0.5$, suggesting that Mn and O are produced in balanced amounts by SNe II. For [O/H]$\geq-0.5$ dex, there is a steady increase in Mn with [O/H]. \cite{Feltzing07} claim that because SNe Ia do not contribute until [O/H]=0 that the rise in [Mn/O] must be a result of metallicity-dependent yields from SNe II. Non-LTE effects also have an impact on [Mn/Fe] at low metallicities, bringing [Mn/Fe] up to solar values \citep{Battistini15} for all metallicities in the Milky Way. In dSphs, the sample provided by \cite{North12} shows a clear overlap with the thick-disc and halo stars in the [Mn/Fe] versus [Fe/H] plot. After including models of various star formation histories and predicted nucleosynthetic yields of Mn, \cite{North12} were only able to reproduce the observed amounts of [Mn/Fe] in dSphs if metallicity dependent yields for both SNe Ia and SNe II were adopted.

To measure Mn in DLAs, there are five different absorption lines, all with very similar oscillator strengths, that can be used. The difficulty with Mn is that it has a relatively high condensation temperature ($T_{\rm cond}=1190$ K), making it prone to dust depletion. The main studies of Mn in DLAs have been completed by \cite{Pettini00}, \cite{DLAcat30}, and \cite{Ledoux02}. The work by \cite{Pettini00} and \cite{DLAcat30} shows that [Mn/Fe] is constant with metallicity (or at least a much flatter evolution than in Milky Way stars) for the range of metallicity between $-3\leq$[Fe/H]$\leq0$. However, the \cite{Pettini00} study is based solely on a small sample of five DLAs that show minor dust depletion in sightlines; while \cite{DLAcat30} added and additional three systems. With such a small sample, it is difficult to conclude whether all DLAs demonstrate a metallicity-independent [Mn/Fe], despite the caution of avoiding strong effects of dust depletion. The constant [Mn/Fe] with increasing metallicity in DLAs is somewhat at odds with the sample presented by \cite{Ledoux02}, which shows an increase in [Mn/Fe] as a function of [Zn/H] with a similar slope as the data from stars \citep[e.g.][]{North12}. Although \cite{Ledoux02} were attempting to model dust depletion for Mn, they speculate that the only way to model the depletion correctly would be to include the metallicity dependence of [Mn/Fe].

\end{document}